\documentclass[aps,prd,twocolumn,superscriptaddress,nofootinbib]{revtex4-1}

\usepackage{amsmath,amssymb,bm}
\usepackage{slashed}
\usepackage{graphicx}
\usepackage[T1]{fontenc}
\usepackage[utf8]{inputenc}
\usepackage{gauss}
\usepackage[colorlinks,linkcolor=blue,citecolor=blue,urlcolor=blue]{hyperref}
\usepackage{comment}
\usepackage{multirow}
\usepackage{ytableau}
\usepackage{makecell}
\usepackage{caption}
\usepackage{tabularx}
\usepackage{blkarray}
\usepackage{float}
\usepackage{booktabs}
\usepackage{xcolor}
\usepackage{cancel}
\usepackage{ulem}
\allowdisplaybreaks 

\newcommand{\be}{\begin{equation}}
\newcommand{\ee}{\end{equation}}
\newcommand{\bea}{\begin{eqnarray}}
\newcommand{\eea}{\end{eqnarray}}
\newcommand{\ba}{\begin{eqnarray}}
\newcommand{\ea}{\end{eqnarray}}

\begin{document}

\title{Chiral Structure and Selection Rules in Light-Front Nucleon-Pentaquark Mixing}

\author{Fangcheng He}
\email{fangchenghe123@gmail.com}
\affiliation{Department of Physics, New Mexico State University, Las Cruces, NM 88003, USA}
\affiliation{Nuclear Science Division, Lawrence Berkeley National Laboratory, Berkeley, CA 94720, USA}

\author{Edward Shuryak}
\email{edward.shuryak@stonybrook.edu}
\affiliation{Center for Nuclear Theory, Department of Physics and Astronomy, Stony Brook University, Stony Brook, New York 11794-3800, USA}

\author{Wan Wu}
\email{wuw20@mails.tsinghua.edu.cn}
\affiliation{Center for Nuclear Theory, Department of Physics and Astronomy, Stony Brook University, Stony Brook, New York 11794-3800, USA}
\affiliation{Physics Department, Tsinghua University, Beijing 100084, China}

\author{Ismail Zahed}
\email{ismail.zahed@stonybrook.edu}
\affiliation{Center for Nuclear Theory, Department of Physics and Astronomy, Stony Brook University, Stony Brook, New York 11794-3800, USA}

\begin{abstract}
We present a light-front Hamiltonian analysis of nucleon–pentaquark mixing induced by $\sigma$- and $\pi$-type transition operators in a fully Pauli-consistent five-quark basis. The pentaquark configurations are constructed using a systematic permutation-group classification of orbital, spin-flavor, and color degrees of freedom, and the hyperfine interaction is diagonalized to obtain orthonormal eigenchannels with definite quantum numbers. We compute the mixing coefficients for all 27 positive-parity $P$-wave pentastates and find a highly sparse structure: only 6 channels contribute to the nucleon wave function, while the remaining 21 vanish due to symmetry selection rules. The nonzero contributions are concentrated in a small set of hyperfine eigenchannels, demonstrating a strong dominance pattern. The $\sigma$- and $\pi$-induced amplitudes populate the same subset of states and are related by a fixed phase, reflecting their common chiral structure, which eliminates interference in the normalization. As a result, their contributions add incoherently, yielding a total five-quark probability of about $29\%$, with the remaining $71\%$ residing in the three-quark core. These results show that nucleon-pentaquark mixing is governed primarily by symmetry selection rules and chiral structure, and that the five-quark content is dominated by a small number of dynamically selected channels.
\end{abstract}


\maketitle

\section{Introduction}
\label{sec_introduction}

The internal structure of the nucleon in Quantum Chromodynamics is inherently multi-partonic. While the minimal three-quark picture captures the gross quantum numbers of baryons, it is well established that higher Fock components containing explicit quark-antiquark pairs play an important role in spin, flavor, and momentum distributions, as well as in baryon spectroscopy and transition processes \cite{Brodsky:1997de,Ji:2013dva}. In a light-front description, these effects are naturally encoded in the presence of higher Fock sectors in the baryon wave function, with the lowest nontrivial extension beyond the three-quark sector given by $qqqq\bar q$ configurations.

The importance of five-quark components has been emphasized from several complementary perspectives. Early quark-model studies highlighted the role of $qqqq\bar q$ admixtures in explaining spin observables, flavor asymmetries, and baryon mass splittings \cite{Isgur:1978wd,Glozman:1996wq}. In parallel, chiral soliton and large-$N_c$ approaches demonstrated that exotic baryonic configurations and meson-baryon dressing arise naturally once collective and topological degrees of freedom are taken into account \cite{Diakonov:1997mm,Dashen:1993jt}. More recently, explicit five-quark Fock components have been explored in both phenomenological and lattice-inspired analyses as a mechanism for understanding sea-quark effects and hidden-flavor baryons \cite{An:2012kj,Alexandrou:2020okk}.

A quantitative treatment of such five-quark components requires two essential ingredients. First, one must construct a complete and Pauli-allowed basis for the internal quantum numbers of the pentaquark system. This is nontrivial because the four quarks are identical fermions, so their combined orbital, spin-flavor, and color wave function must be antisymmetric under the permutation group $S_4$, while the full five-body state must be a color singlet. Second, one must specify the microscopic operators that couple the three-quark nucleon sector to the five-quark sector and evaluate the corresponding transition matrix elements in a controlled dynamical framework.

The purpose of this work is to provide both ingredients in a unified and explicit form, and to quantify the resulting nucleon–pentaquark mixing in a light-front Hamiltonian framework. We construct a complete set of Pauli-consistent pentaquark states using a systematic permutation-group classification of orbital, spin-flavor, and color degrees of freedom based on Young tableaux. These states are organized into symmetry-adapted bases that allow for efficient evaluation of matrix elements and a transparent implementation of the Pauli principle. The present work is in the continuation of the work presented recently in~\cite{Miesch:2025ael}, which is aimed at addressing the higher Fock sectors of the nucleon and its tomography~\cite{Shuryak:2026pqt} (and references therein), based on the wealth of multi-particle states discovered by LHCb~\cite{Johnson:2024ExoticLHCb}. 

In particular, we address nucleon–pentaquark mixing induced by two physically motivated operators. The first is a $\sigma$-type $^3P_0$ pair-creation operator, which creates a quark-antiquark pair in a color-singlet, flavor-singlet, spin-triplet state with one unit of relative orbital angular momentum \cite{Micu:1968mk,LeYaouanc:1972vsx}. The second is a $\pi$-type spin-momentum operator that has additional coupling between longitudinal momentum fractions and spin degrees of freedom, in close analogy with pion-induced transitions in light-front Hamiltonian approaches \cite{Miesch:2025ael}. These operators form a chiral pair and impose strong and complementary selection rules on the allowed pentaquark configurations as we will show.

We evaluate the corresponding transition matrix elements in a light-front basis, separating color-spin-flavor recoupling from longitudinal and transverse orbital overlaps. Hyperfine color-spin interactions are included and diagonalized in the pentaquark sector, yielding orthonormal eigenchannels that reorganize the symmetry basis into physically relevant states \cite{DeRujula:1975qlm}. The physical nucleon is then constructed as a dressed state containing admixtures of these eigenchannels, with coefficients determined by the transition matrix elements and energy denominators following \cite{Miesch:2025ael}.

A central result of this work is that nucleon–pentaquark mixing exhibits a highly constrained and sparse structure. Among the $27$ positive-parity $P$-wave pentastates, only $6$ contribute to the nucleon wave function, while the remaining $21$ vanish due to symmetry selection rules. The nonzero contributions are concentrated in a small set of hyperfine eigenchannels, revealing a strong dominance pattern in the five-quark admixture. Furthermore, the $\sigma$- and $\pi$-induced amplitudes populate the same subset of states and are related by a fixed phase, reflecting their common chiral structure. As a result, their interference vanishes in the normalization, and the two contributions add incoherently. After normalization, the nucleon contains a five-quark component of approximately $29\%$, dominated by a small number of dynamically selected channels.

The organization of this paper is as follows. In Sec.~\ref{sec:groupdecomp} we present the permutation-group framework and construct explicit Young-basis states for three- and four-quark systems. In Sec.~\ref{sec:five_quark_decomp} we assemble the Pauli-consistent five-quark basis and classify the allowed P-wave pentaquark configurations, we calculate the color-spin hyperfine interaction based on these pentaquark configurations. In Sec.~\ref{sec:construct_P} we introduce the light-front Hamiltonian and construct the corresponding orbital wave functions. In Sec.~\ref{sec:mixing_consolidated} we define the $\sigma$- and $\pi$-type transition operators and derive the associated selection rules and matrix elements. In Sec.~\ref{sec:Nphys_explicit} we construct the physical nucleon state including five-quark admixtures and analyze the resulting mixing pattern, including the role of chiral structure and interference. Our conclusions are in~\ref{sec_conclusions}. The appendices provide additional technical details: Appendix~\ref{App_3} and Appendix~\ref{App_4} summarize the Young-basis constructions and tensor-product projections, while Appendix~\ref{sec:lf_pwave_consolidated} and \ref{sec:lf_swave_consolidated} detail the light-front eigenbasis and numerical implementation of the P-wave pentaquark states and S-wave three quark states, respectively. In Appendix~\ref{sec:full_pwavefunction} we give the explicit orbital, spin-flavor, and color (OSFC) form of the nucleon and pentastates. The derivation of chiral relation between $\sigma$- and $\pi$-type interactions in the Foldy-Wouthuysen reduction is presented in Appendix~\ref{sec_FW}.

\section{Group decomposition}
\label{sec:groupdecomp}

The central dynamical problem addressed in this work is the construction of pentaquark 
states with well-defined total quantum numbers $(J^P,I,S,\ldots)$ while enforcing the Pauli 
principle for the identical quarks. In constituent or effective descriptions, the dynamics is encoded in Hamiltonian matrix 
elements evaluated in a basis of color$\times$spin$\times$flavor$\times$orbital states. The bottleneck is not only the 
diagonalization but the faithful construction of a basis that is complete, non-redundant, and automatically antisymmetric 
under exchange of identical quarks.

For a $qqqq\bar q$ system with four identical quarks, the Pauli constraint is most conveniently implemented at the level of
the permutation group $S_4$ acting on the four-quark labels. One organizes the four-quark wavefunction into irreducible 
representations (irreps) $[f]$ of $S_4$, with the understanding that the total four-quark state must be antisymmetric under 
any quark exchange. Since the full wavefunction factors into orbital, spin-flavor, and color parts (and sometimes 
additional internal labels), each factor transforms in some $S_4$ irrep; the product of these irreps must contain the 
totally antisymmetric irrep $[1111]$. This section fixes explicit Young-basis states and projection formulas that will be 
repeatedly invoked when coupling subsystems, counting independent states, and evaluating symmetry-constrained matrix 
elements.

Two levels of permutation structure appear in practice. For internal couplings that single out a three-quark cluster 
(e.g.\ when relating a four-quark core to baryon-like substructures or when defining Jacobi coordinates), $S_3$ enters 
naturally. For the full four-quark core, $S_4$ controls the allowed symmetry patterns. Because the subsequent sections 
require explicit coefficients (rather than abstract group-theory statements), we provide normalized Young-basis vectors 
and explicit tensor-product projections. These are the concrete bookkeeping rules behind phrases such 
as ``mixed symmetry'' or ``Pauli-allowed coupling'' and they directly determine which dynamical channels exist.

\subsection{Three-quark Young basis and physical interpretation}
For three quarks, the relevant permutation irreps are the fully symmetric $[3]$ and the mixed-symmetry $[21]$. 
The totally antisymmetric irrep $[111]$ also exists, but for most spin-flavor applications with three quarks it 
enters implicitly through color (e.g.\ a baryon color wavefunction is $[111]$ under $S_3$), leaving the spin-flavor and
orbital factors to be symmetric or mixed such that the total is antisymmetric. The practical lesson is that for any 
three-quark subcluster the exchange symmetry dictates selection rules: operators symmetric under exchanges couple only to
the symmetric components, while exchange-odd structures project onto mixed/antisymmetric components.

To make these statements operational, we adopt an explicit normalized Young basis in a two-state label space $\{i_1,i_2\}$.
This two-state space can be viewed as a pedagogical stand-in for any binary internal label (e.g.\ a reduced flavor doublet, 
or a spin-$\tfrac12$ projection basis) used to build symmetry-adapted combinations. The algebra of $S_3$ does not depend
on the physical meaning of the label; only the action of permutations matters.

A convenient normalized basis is
\begin{equation}
\begin{aligned}
[3] &= |i_1 i_1 i_1\rangle,\\
[21]_\alpha &= \frac{1}{\sqrt{2}}\big(|i_1 i_2\rangle-|i_2 i_1\rangle\big)\,|i_1\rangle,\\
[21]_\beta &= \frac{1}{\sqrt{6}}\big(2|i_1 i_1 i_2\rangle-|i_1 i_2 i_1\rangle-|i_2 i_1 i_1\rangle\big).
\end{aligned}
\end{equation}
The two mixed-symmetry vectors correspond to the two standard Young tableaux of $[21]$,
\begin{equation}
\begin{aligned}
[21]_\alpha=\ytableaushort{1 3 , 2},
\qquad
[21]_\beta=\ytableaushort{1 2 , 3}.
\end{aligned}
\end{equation}
Physically, the distinction between $\alpha$ and $\beta$ is a choice of coupling order and basis 
within the same irrep; the subscript keeps track of which Young operator has been applied. In later 
matrix elements, this label becomes important because intermediate couplings (e.g.\ coupling quarks $1,2$ 
first versus $2,3$ first) lead to different recoupling coefficients.

The normalization conventions above are chosen so that inner products are orthonormal within each irrep, 
and so that the subsequent tensor-product projection coefficients are simple square roots. These conventions are also
convenient when the same $S_3$ irrep appears in different physical sectors (spin, flavor, orbital), because they allow one 
to reuse identical projection formulae. More details can be found in Appendix~\ref{App_3}.

\subsection{Four-quark Young basis and why it matters for pentaquarks}
For four identical quarks, the permutation group is $S_4$ and the relevant irreps are
\begin{equation}
[4],\qquad [31],\qquad [22],\qquad [211],\qquad [1111].
\end{equation}
In pentaquark construction, $[1111]$ plays the role of the Pauli target: the total four-quark wavefunction must 
transform as $[1111]$ under $S_4$. The four-quark wavefunction factorizes into
\bea
\Psi_{4q}=\Psi_{\rm orb}\,\Psi_{\rm spin\!-\!flavor}\,\Psi_{\rm color},
\eea
so that the allowed symmetry types are those for which
\begin{equation}
[f_{\rm orb}]\otimes[f_{\rm spin\!-\!flavor}]\otimes[f_{\rm color}] \supset [1111].
\end{equation}
This single inclusion condition is the group-theoretic statement of the Pauli principle. It enforces, 
for example, that if the orbital state is symmetric $[4]$ (ground-state $S$-wave for the four-quark core), 
then the combined spin$\otimes$flavor$\otimes$color must be $[1111]$. If instead the orbital part is mixed 
(e.g.\ a single $P$-wave excitation corresponds typically to $[31]$), then spin$\otimes$flavor$\otimes$color must 
compensate accordingly. Thus the decomposition tables below are not ornamental: they specify which orbital 
excitations can coexist with which color-spin-flavor patterns and therefore which physical channels exist.

As in the $S_3$ case, we present explicit normalized Young-basis vectors in a two-state label space $\{i_1,i_2\}$. 
Again this is a concrete representation used to fix phases and normalizations of basis states. In actual applications 
the labels will correspond to spin projections, isospin components, strange/nonstrange flavor indices, or orbital 
single-particle labels; the $S_4$ symmetry algebra is the same.

A normalized basis for the $S_4$ irreps $[4]$, $[31]$, and $[22]$ is
\begin{widetext}
\begin{equation}
\begin{aligned}
[4] &= |i_1 i_1 i_1 i_1\rangle,\\
[31]_\alpha &= \frac{1}{\sqrt{2}}\big(|i_1 i_2\rangle-|i_2 i_1\rangle\big)\,|i_1 i_1\rangle,\\
[31]_\beta &= \frac{1}{\sqrt{6}}\big(2|i_1 i_1 i_2\rangle-|i_1 i_2 i_1\rangle-|i_2 i_1 i_1\rangle\big)\,|i_1\rangle,\\
[31]_\gamma &= \frac{1}{\sqrt{12}}\big(3|i_1 i_1 i_1 i_2\rangle-|i_2 i_1 i_1 i_1\rangle-|i_1 i_2 i_1 i_1\rangle-|i_1 i_1 i_2 i_1\rangle\big),\\
[22]_\alpha &= \sqrt{\frac{1}{4}}\big(|i_1 i_2 i_1 i_2\rangle-|i_2 i_1 i_1 i_2\rangle-|i_1 i_2 i_2 i_1\rangle+|i_2 i_1 i_2 i_1\rangle\big),\\
[22]_\beta &= \sqrt{\frac{1}{12}}\big(2|i_1 i_1 i_2 i_2\rangle-|i_1 i_2 i_1 i_2\rangle-|i_2 i_1 i_1 i_2\rangle-|i_1 i_2 i_2 i_1\rangle-|i_2 i_1 i_2 i_1\rangle+2|i_2 i_2 i_1 i_1\rangle\big).
\end{aligned}
\end{equation}
The three basis vectors of the $[31]$ irrep are labeled by $\alpha,\beta,\gamma$ and 
are defined by the standard Young tableaux; they enter naturally when coupling mixed-symmetry 
factors to produce an overall antisymmetric four-quark core. The corresponding Young tableaux are
\begin{equation}
\begin{aligned}
[31]_\alpha=\ytableaushort{1 3 4, 2},
\qquad
[31]_\beta=\ytableaushort{1 2 4, 3},
\qquad
[31]_\gamma=\ytableaushort{1 2 3, 4},
\end{aligned}
\end{equation}
\begin{equation}
\begin{aligned}
[22]_\alpha=\ytableaushort{1 3, 2 4},
\qquad
[22]_\beta=\ytableaushort{1 2, 3 4}.
\end{aligned}
\end{equation}
\end{widetext}
The irrep $[31]$ is the archetype of ``one unit of mixed symmetry'': 
individual basis states transform into each other under permutation.
In orbital language, this irrep is the typical symmetry of a single $P$-wave 
excitation of an otherwise symmetric four-body configuration; in spin-flavor language it 
corresponds to a core where three quarks are in a mixed symmetric pattern and one quark plays a 
distinguished role. The irrep $[22]$ is ``pairwise'' symmetry and is especially common in 
diquark-like decompositions where $(12)$ and $(34)$ are correlated pairs.

For $[211]$ we label the three standard basis vectors by $\alpha,\beta,\gamma$ through their tableaux
\begin{equation}
\begin{aligned}
[211]_\alpha=\ytableaushort{1 3, 2, 4},
\qquad
[211]_\beta=\ytableaushort{1 2, 3, 4},
\qquad
[211]_\gamma=\ytableaushort{1 4, 2, 3}.
\end{aligned}
\end{equation}
In later sections $[211]$ appears naturally when coupling a mixed-symmetry factor to another 
mixed-symmetry factor and demanding an overall antisymmetric four-quark core. More details 
can be found in Appendix~\ref{App_4}.

\subsection{Explicit tensor-product projections}

Once a symmetry-adapted basis is chosen, two tasks recur throughout the pentaquark analysis.

First, one must decide which combinations of orbital, spin-flavor, and color irreps can yield 
a Pauli-allowed four-quark state. This is a tensor-product question in $S_4$ and is solved by decomposition rules such as
\begin{equation}
[31]\otimes[31]=[4]\oplus[31]\oplus[22]\oplus[211].
\end{equation}
Second, having identified the allowed irreps, one must actually build normalized states 
and compute matrix elements. This requires explicit projection coefficients that map 
products of basis vectors into a definite irrep, because Hamiltonians are typically 
sums of two-body operators symmetric under particle relabeling, and their reduced 
matrix elements depend sensitively on these projections.

The formulas below provide exactly these projections. The notation
\begin{equation}
\big([f]_{AB}:[f_A]_A\otimes[f_B]_B\big)
\end{equation}
denotes the normalized projection of a product state transforming in irrep $f_A$ of physical factor A and irrep $f_B$ of physical factor B onto 
the irrep $[f]$ in the combined $AB$ space.(for example, $A$ might denote spin  symmetry, $B$ might denote color symmetry and $AB$ is the spin-color space.)

The explicit Young-basis states and projectors assembled here and in Appendix~\ref{App_3}-\ref{App_4} 
serve three concrete purposes in the remainder of the manuscript. They provide a systematic enumeration 
of Pauli-allowed $qqqq$ cores for a given orbital symmetry (most crucially, distinguishing $S$-wave $[4]$ 
from $P$-wave $[31]$ excitations). They fix the relative phases between different coupling schemes, 
which is essential when assembling total states with the antiquark and when comparing to alternative 
clusterizations such as baryon-meson versus diquark-triquark. They allow us to evaluate matrix elements 
of permutation-symmetric two-body operators by reducing them to reduced matrix elements within definite 
$S_4$ irreps, thereby separating dynamics from symmetry.

In the next section we will use these projections to construct explicit pentaquark basis states 
with definite color singlet structure, and to tabulate the resulting $(S,I)$ multiplets and their 
parity assignments for both $S$-wave and $P$-wave cores.

\section{Five-Quark Decomposition}
\label{sec:five_quark_decomp}

This section assembles the explicit $|qqqq\bar q\rangle$ basis used throughout the
dynamical analysis of nucleon-pentaquark mixing on the light front.
The goal is not to re-derive the permutation-group machinery already fixed in
Sec.~\ref{sec:groupdecomp}, but to show how that machinery becomes a practical ``interface''
between (i) Pauli consistency for the $q^4$ core, (ii) hyperfine diagonalization
in the positive-parity $P$-wave sector, and (iii) the transition operators that
connect $|qqq\rangle$ and $|qqqq\bar q\rangle$ Fock sectors.

We build the pentaquark state from a four-quark core $(q^4)$ coupled to an
antiquark $\bar q$.  The four quarks are identical fermions, so the complete
four-quark wave function must be antisymmetric under $S_4$ permutations:
the product of orbital ($L$), spin-flavor ($SF$), and color ($C$) symmetry
types must contain the fully antisymmetric irrep $[1111]$.
The explicit Young-basis conventions and all required $S_4$ tensor-product
projections are fixed once and for all in this section; we use them here without
repetition.

For the positive-parity pentaquarks of interest, the orbital part carries one
unit of relative angular momentum.  In the four-quark sector this implies
orbital permutation symmetry $L[4]$ or $L[31]$, depending on which Jacobi
coordinate carries the $P$-wave excitation.
Color is chosen as $C[211]$ for the $q^4$ core so that coupling to the antiquark
color $\bar{\mathbf 3}$ yields an overall color singlet.  With these ingredients,
the remaining task is to enumerate and write the canonical OSFC basis states
in a form that makes both hyperfine matrix elements and transition operators
straightforward to evaluate.

\subsection{Pauli-consistent OSFC bookkeeping for the $q^4$ core}

The key reason to work in an $S_4$-irrep basis is that the operators that drive
both spectroscopy and mixing are built from pairwise structures acting inside
the $q^4$ core.  Hyperfine interactions are sums of two-body color-spin
operators (schematically $\lambda_i\!\cdot\!\lambda_j\,\sigma_i\!\cdot\!\sigma_j$),
while the transition operators that connect $|qqq\rangle$ to $|qqqq\bar q\rangle$
select very specific spin and orbital components of the five-quark wave
function.  In an $S_4$-adapted basis, permutations $P_{ij}$ act as sparse linear
maps organized by $[f]$ labels, so the symmetry reduction is performed once
(through the Sec.~\ref{sec:five_quark_decomp} projectors) and then reused uniformly in every matrix
element.

Concretely, we label a four-quark configuration by its orbital symmetry $[f]_L$,
spin-flavor symmetry $[f']_{SF}$, and color symmetry $C[211]$, and we couple
these sectors so that the resulting $q^4$ state is Pauli allowed.  The
corresponding five-quark state is obtained by coupling the antiquark color to
make a singlet, and the antiquark spin-flavor quantum numbers are appended in
the usual way.  This construction keeps the orbital dependence explicit through
$P$-wave functions $\varphi_{m_L}$ or $\tilde{\varphi}_{1m_L}$ carrying the magnetic
projection $m_L$, while the internal OSFC structure is handled algebraically by
the $S_4$ projectors.

\subsection{Counting of $P$-wave pentaquarks}

The allowed $P$-wave configurations are determined by which $(L\otimes SF)$
symmetry types can combine with color $C[211]$ to produce an antisymmetric
four-quark core.  For the present construction we focus on total isospin
$I=\tfrac12$ and spins $S=\tfrac12,\tfrac32,\tfrac52$.
A compact bookkeeping is to list the admissible $(L[f]\otimes SF[f'])$ families
and count the multiplicities in each $(I,S)$ channel.

The resulting number of $P$-wave states is summarized in
Table~\ref{tab:Pwave_counts}.

\begin{table}[H]
    \centering
    \begin{tabular}{cccc}
        \toprule
        $num$ & $I=\frac{1}{2}\ S=\frac{1}{2}$ & $I=\frac{1}{2}\ S=\frac{3}{2}$ & $I=\frac{1}{2}\ S=\frac{5}{2}$ \\
        \midrule
        $L[4] \otimes SF[31]$ & 3 & 3 & 1 \\
        $L[31] \otimes SF[31]$ & 3 & 3 & 1 \\
        $L[31] \otimes SF[22]$ & 2 & 2 & 1 \\
        $L[31] \otimes SF[4]$  & 2 & 1 & 0 \\
        $L[31] \otimes SF[211]$ & 3 & 2 & 0 \\
        \midrule
        $Total$ & 13 & 11 & 3\\
        \bottomrule
    \end{tabular}
    \caption{The number of $P$-wave states in the pentaquark basis for $I=\tfrac12$ and the listed spins $S$.}
    \label{tab:Pwave_counts}
\end{table}

It is also useful to indicate representative decompositions of five-quark
spin-flavor configurations into separate flavor and spin irreps.  For the
channels emphasized here, the dominant decompositions are listed in
Table~\ref{tab:FS_decomp}.

\begin{table}[H]
    \centering
    \begin{tabular}{ccc}
        \toprule
        $I=\frac{1}{2}\ S=\frac{1}{2}$ & $I=\frac{1}{2}\ S=\frac{3}{2}$ & $I=\frac{1}{2}\ S=\frac{5}{2}$ \\
        \midrule
        $F[22]\ S[22]$ & $F[22]\ S[31]$ & $F[22]\ S[4]$  \\
        $F[31]\ S[31]$ & $F[31]\ S[4]$  & $F[31]\ S[4]$ \\
        \bottomrule
    \end{tabular}
    \caption{Representative decompositions of five-quark spin-flavor configurations into separate flavor and spin irreps.}
    \label{tab:FS_decomp}
\end{table}

\subsection{Canonical $P$-wave pentaquark wave functions}

We now write the canonical $P$-wave pentaquark states in the OSFC basis, with
a four-quark color core in $C[211]$ coupled to the antiquark to form an overall
color singlet.  The internal OSFC structure is fixed by the requirement that the
four-quark core be antisymmetric under $S_4$ once orbital, spin-flavor, and color
are combined. 

We first define the $L[4]\otimes SF[31]$ family:
\begin{widetext}
\begin{equation}
\label{eq:psiA}
\Psi_{P m_L}^A
=
\frac{1}{\sqrt{3}}
\left(
\left[
C[211]_\beta\left(L[4]SF[31]\right)_\alpha
-
C[211]_\alpha\left(L[4]SF[31]\right)_\beta
+
C[211]_\gamma\left(L[4]SF[31]\right)_\gamma
\right]
\, C[11]\, S F[1]
\right).
\end{equation}
The spin-flavor substructures needed in later projections can be organized as
\begin{equation}
\begin{aligned}
&L[4]\left( SF[31]:F[31] \otimes S[31]\right)_{\alpha, \beta, \gamma}, \\
&L[4]\left(SF[31]:F[22] \otimes S[31]\right)_{\alpha, \beta, \gamma}, \\
&L[4]\left(SF[31]:F[31] \otimes S[22]\right)_{\alpha, \beta, \gamma}, \\
&L[4]\left(SF[31]:F[31] \otimes S[4]\right)_{\alpha, \beta, \gamma}, \\
\end{aligned}
\end{equation}
The last one is for $S=5/2$ and $S=3/2$ cases.
Next, for the $L[31]\otimes SF[4]$ family we define
\begin{equation}
\Psi_P^P
=
\frac{1}{\sqrt{3}}
\left(
\left[
C[211]_\beta\left(L[31]SF[4]\right)_\alpha
-
C[211]_\alpha\left(L[31]SF[4]\right)_\beta
+
C[211]_\gamma\left(L[31]SF[4]\right)_\gamma
\right]
\, C[11]\, S F[1]
\right),
\end{equation}
with the corresponding $SF$ couplings
\begin{equation}
\begin{aligned}
& \left(L[31]\left(SF[4]:F[31] S[31]\right)\right)_{\alpha, \beta, \gamma}, \\
& \left(L[31]\left(SF[4]:F[22]S[22]\right)\right)_{\alpha, \beta, \gamma}.
\end{aligned}
\end{equation}

For the $L[31]\otimes SF[31]$ family we define
\begin{equation}
\Psi_{P 1}^P
=
\frac{1}{\sqrt{3}}
\left(
\left[
C[211]_\beta\left(L[31]SF[31]\right)_\alpha
-
C[211]_\alpha\left(L[31]SF[31]\right)_\beta
+
C[211]_\gamma\left(L[31]SF[31]\right)_\gamma
\right]
\, C[11]\, S F[1]
\right),
\end{equation}
with 
\begin{equation}
\begin{aligned}
& \left(L[31]\left( SF[31]:F[31]S[31]\right)\right)_{\alpha, \beta, \gamma}, \\
& \left(L[31]\left( SF[31]:F[22]S[31]\right)\right)_{\alpha, \beta, \gamma}, \\
& \left(L[31]\left( SF[31]:F[31]S[22]\right)\right)_{\alpha, \beta, \gamma}, \\
& \left(L[31]\left( SF[31]:F[31]S[4]\right)\right)_{\alpha, \beta, \gamma}.
\end{aligned}
\end{equation}

Finally, the remaining $L[31]\otimes SF[22]$ and $L[31]\otimes SF[211]$ families are
\begin{equation}
\Psi_{P 2}^P
=
\frac{1}{\sqrt{3}}
\left(
\left[
C[211]_\beta\left(L[31]SF[22]\right)_\alpha
-
C[211]_\alpha\left(L[31]SF[22]\right)_\beta
+
C[211]_\gamma\left(L[31]SF[22]\right)_\gamma
\right]
\, C[11]\, S F[1]
\right),
\end{equation}
\begin{equation}
\Psi_{P 3}^P
=
\frac{1}{\sqrt{3}}
\left(
\left[
C[211]_\beta\left(L[31]SF[211]\right)_\alpha
-
C[211]_\alpha\left(L[31]SF[211]\right)_\beta
+
C[211]_\gamma\left(L[31]SF[211]\right)_\gamma
\right]
\, C[11]\, S F[1]
\right).
\end{equation}

with 
\begin{equation}
\begin{aligned}
& \left(L[31]\left( SF[22]:F[31]S[31]\right)\right)_{\alpha, \beta, \gamma}, \\
& \left(L[31]\left( SF[22]:F[22]S[22]\right)\right)_{\alpha, \beta, \gamma}, \\
& \left(L[31]\left( SF[22]:F[22]S[4]\right)\right)_{\alpha, \beta, \gamma},
\end{aligned}
\end{equation}
and
\begin{equation}
\begin{aligned}
& \left(L[31]\left( SF[211]:F[31]S[31]\right)\right)_{\alpha, \beta, \gamma}, \\
& \left(L[31]\left( SF[211]:F[22]S[31]\right)\right)_{\alpha, \beta, \gamma}, \\
& \left(L[31]\left( SF[211]:F[31]S[22]\right)\right)_{\alpha, \beta, \gamma}, \\
\end{aligned}
\end{equation}
In summary, the 13 states with $(I=\frac{1}{2},S=\frac{1}{2})$ can be labeled as
\bea
LSF=&&L[4]SF[31]_\text{a} :  S[31]\otimes F[31],~~~~~~ L[4]SF[31]_\text{b} :  S[31]\otimes F[22],~~~~~~ L[4]SF[31]_\text{c} :  S[22]\otimes F[31]  \nonumber\\
&&L[31]SF[31]_\text{a} :  S[31]\otimes F[31],~~~~~~ L[31]SF[31]_\text{b} :  S[31] \otimes F[22],~~~~~~ L[31]SF[31]_\text{c} :  S[22]\otimes   F[31]    \nonumber\\
&& L[31]SF[22]_\text{a} :  S[31]\otimes F[31],~~~~~~  L[31]SF[22]_\text{b} :  S[22] \otimes F[22],   \nonumber\\
&&L[31]SF[4]_\text{a} :  S[31]\otimes F[31],~~~~~~ L[31]SF[4]_\text{b} :  S[22] \otimes F[22],    \nonumber\\
&&L[31]SF[211]_\text{a} :  S[31]\otimes F[31],~~~~~~ L[31]SF[211]_\text{b} :  S[31] \otimes F[22],~~~~~~ L[31]SF[211]_\text{c} :  S[22]\otimes   F[31] 
\eea
The 11 states with $(I=\frac{1}{2},S=\frac{3}{2})$ can be labeled as
\bea
LSF=&&L[4]SF[31]_\text{a} :  S[31]\otimes F[31],~~~~~~ L[4]SF[31]_\text{b} :  S[31]\otimes F[22],~~~~~~ L[4]SF[31]_\text{c} :  S[4]\otimes F[31]  \nonumber\\
&&L[31]SF[31]_\text{a} :  S[31]\otimes F[31],~~~~~~ L[31]SF[31]_\text{b} :  S[31] \otimes F[22],~~~~~~ L[31]SF[31]_\text{c} :  S[4]\otimes   F[31]    \nonumber\\
&& L[31]SF[22]_\text{a} :  S[31]\otimes F[31],~~~~~~  L[31]SF[22]_\text{b} :  S[4] \otimes F[22],   \nonumber\\
&&L[31]SF[4] :  S[31]\otimes F[31],    \nonumber\\
&&L[31]SF[211]_\text{a} :  S[31]\otimes F[31],~~~~~~ L[31]SF[211]_\text{b} :  S[31] \otimes F[22]
\eea
The three states with $(I=\frac{1}{2},S=\frac{5}{2})$ can be labeled as
\bea
LSF=&&L[4]SF[31]:  S[4]\otimes F[31]\nonumber\\
&&L[31]SF[31] :  S[4]\otimes F[31]\nonumber\\
&& L[31]SF[22]:  S[4]\otimes F[22],
\eea
\end{widetext}
The labels $L[4]$ and $L[31]$ represent the light-cone P wave state, they are the functions of longitudinal momentum $x_\xi$ and transverse momentum $k_{\xi,\perp}$ in Jacobi coordinates, we will show later how to
construct them using the eigenstates of light cone Hamiltonian. The full pentaquark wavefunction can be found in Appendix~\ref{sec:OSFC}.


\subsection{Color-spin hyperfine interaction including antiquark effects}
\label{sec:colorspin_full}

The short-range dynamics of the pentaquark is governed by the color-spin
hyperfine interaction induced by one-gluon exchange. For a genuine five-body
system this interaction acts not only among the four identical quarks but also
between each quark and the antiquark. The full hyperfine Hamiltonian therefore
takes the form
\begin{widetext}
\begin{equation}
H_{\rm CS}
=-\sum_{i<j\leq 5}\frac{V_{1g}(r_{ij})}{m_Q^2}\lambda_{ij}\Sigma_{ij}=
- \frac{V_{1g}(r_{ij})}{m_Q^2}\left(
\sum_{i<j}^{4}
\boldsymbol{\lambda}_i \cdot \boldsymbol{\lambda}_j\,
\boldsymbol{\sigma}_i \cdot \boldsymbol{\sigma}_j
+
\sum_{i=1}^{4}
\boldsymbol{\lambda}_i \cdot \boldsymbol{\lambda}_{\bar q}\,
\boldsymbol{\sigma}_i \cdot \boldsymbol{\sigma}_{\bar q}
\right),
\label{eq:Hcs_full}
\end{equation}
\end{widetext}
where $\boldsymbol{\lambda}_i$ and $\boldsymbol{\sigma}_i$ act on the color and
spin of quark $i$, while $\boldsymbol{\lambda}_{\bar q}=-\boldsymbol{\lambda}^*$
and $\boldsymbol{\sigma}_{\bar q}=-\boldsymbol{\sigma}^*$ act on the antiquark.

Assuming the $\Delta$--nucleon mass splitting to be saturated by the color-spin
interaction, we have
\begin{equation}
M_\Delta-M_N=\frac{16V_{1g}}{m_Q^2}\rightarrow (1232-939)\,{\rm MeV}.
\end{equation}

The first term in Eq.~\eqref{eq:Hcs_full} acts entirely within the $q^4$
subsystem. Since it is fully symmetric under permutations of the four quarks,
its matrix elements are block diagonal in irreducible representations of the
permutation group $S_4$. This implies that it does not mix states belonging to
different $S_4$ irreps, although off-diagonal matrix elements may appear
between different basis vectors within the same irrep.

The second term involves the antiquark and couples each quark individually to
the antiquark. Although the total color operator satisfies
\begin{equation}
\sum_{i=1}^{4}\boldsymbol{\lambda}_i \cdot \boldsymbol{\lambda}_{\bar q}
=
-\frac{16}{3}
\end{equation}
in a color-singlet $q^4\bar q$ state, the operator
\begin{equation}
\sum_{i=1}^{4}
(\boldsymbol{\lambda}_i \cdot \boldsymbol{\lambda}_{\bar q})
(\boldsymbol{\sigma}_i \cdot \boldsymbol{\sigma}_{\bar q})
\end{equation}
does not in general factorize into a universal color coefficient times a spin
factor depending only on total spins. Its matrix elements depend on the full
color-spin recoupling structure of the basis states.

We evaluate the hyperfine interaction in the OSFC basis, in which the four-quark
subsystem is first coupled to definite color $C[211]$, spin $S_{q^4}$, and
spin-flavor symmetry, and then combined with the antiquark to form a total
color singlet and total spin $S$. The four-quark core spin takes integer values
\begin{equation}
S_{q^4}=0,1,2,
\end{equation}
which couple with the antiquark spin $1/2$ to give total pentaquark spin
\begin{equation}
S=\frac{1}{2},\frac{3}{2},\frac{5}{2}.
\end{equation}

In this basis, both the quark-quark and quark-antiquark parts of the hyperfine
interaction are evaluated directly using standard color and spin recoupling.
For the quark-quark term,
\begin{equation}
H_{qq}=-\frac{V_{1g}(r_{ij})}{m_Q^2}\sum_{i<j}^4
(\boldsymbol{\lambda}_i \cdot \boldsymbol{\lambda}_j)\,
(\boldsymbol{\sigma}_i \cdot \boldsymbol{\sigma}_j),
\label{eq:H_qq}
\end{equation}
the operator is invariant under permutations of the four quarks and therefore
preserves the $S_4$ symmetry of the OSFC basis. Off-diagonal matrix elements
arise from recoupling between different Young basis vectors within the same
symmetry sector.

For the quark-antiquark term,
\begin{equation}
H_{q\bar q}=-\frac{V_{1g}(r_{ij})}{m_Q^2}\sum_{i=1}^{4}
(\boldsymbol{\lambda}_i \cdot \boldsymbol{\lambda}_{\bar q})\,
(\boldsymbol{\sigma}_i \cdot \boldsymbol{\sigma}_{\bar q}),
\label{eq:H_qqbar}
\end{equation}
each quark contributes separately, and the resulting matrix elements depend on
the detailed color-spin structure of the OSFC basis states. In the present
basis, this operator is diagonal or block diagonal, but its diagonal values are
state-dependent and cannot be expressed solely in terms of $(S_{q^4},S)$.

The full hyperfine matrices are therefore constructed directly in the OSFC
basis for each $(I,S)$ sector. The explicit matrices for isospin
$I=\tfrac{1}{2}$ and total spins $S=\tfrac{1}{2},\tfrac{3}{2},\tfrac{5}{2}$ are
given in Tables~\ref{tab:CS12},~\ref{tab:CS32} and \ref{tab:CS52} below. These
matrices include both diagonal contributions and off-diagonal mixings arising
from quark-quark recoupling.

Diagonalization of these matrices yields orthonormal hyperfine eigenchannels,
which provide the physical pentaquark basis used in the subsequent nucleon
dressing and mixing analysis.


\begin{table*}[htbp]
    \centering
    \begin{tabular}{|c|c|c|c|c|c|c|c|c|c|c|c|c|c|} \hline
  & $[4][31]_\text{a}$ & $[4][31]_\text{b}$ & $[4][31]_\text{c}$ & $[31][31]_\text{a}$ & $[31][31]_\text{b}$  & $[31][31]_\text{c}$ & $[31][22]_\text{a}$ & $[31][22]_\text{b}$ & $[31][4]_\text{a}$ & $[31][4]_\text{b}$& $[31][211]_\text{a}$ &$[31][211]_\text{b}$ &$[31][211]_\text{c}$ \\  \hline
  $[4][31]_\text{a}$ & $-\frac{16}{3}$  &     & 8 &  & & & & & &  & & & \\  \hline
  $[4][31]_\text{b}$ &   & $\frac{56}{3}$   &   &  & & & & & &  & & & \\  \hline
  $[4][31]_\text{c}$ & 8  &     & 0  &  & & & & & &  & & & \\  \hline
  $[31][31]_\text{a}$  &   &     &   & $-\frac{4}{3}$ &  & 4  & $-8 \sqrt{\frac{2}{3}}$ & &  $-\frac{32}{\sqrt{3}}$ &  & $-12$ & & $4 \sqrt{3}$\\  \hline
  $[31][31]_\text{b}$ &   &     &   &  & $\frac{56}{3}$  &  & &  $4 \sqrt{2}$& & $-8 \sqrt{2}$  & &  $-8 \sqrt{3}$& \\  \hline
  $[31][31]_\text{c}$ &   &     &   & 4 &    & 12 & $8 \sqrt{\frac{2}{3}}$ & & $-\frac{16}{\sqrt{3}}$ &  &  $-4$ & & $4 \sqrt{3}$\\  \hline
  $[31][22]_\text{a}$ & & & &  $-8 \sqrt{\frac{2}{3}}$  &     & $8 \sqrt{\frac{2}{3}}$  & $\frac{16}{3}$ & &$-\frac{8 \sqrt{2}}{3} $ & & $-8 \sqrt{6}$  &  &  \\  \hline
  $[31][22]_\text{b}$ &   &     &   &  & $4 \sqrt{2}$ & & & 8 &  & $-8$ & & $-4 \sqrt{6}$ & \\  \hline
  $[31][4]_\text{a}$ &   &     &   & $-\frac{32}{\sqrt{3}}$ & & $-\frac{16}{\sqrt{3}}$ & $-\frac{8 \sqrt{2}}{3}$ & &  8 & & &  & \\  \hline
  $[31][4]_\text{b}$ &   &     &   &  & $-8 \sqrt{2}$ & & & $-8$ & & 8 &  & & \\  \hline
  $[31][211]_\text{a}$ &   &     &   & $-12$ & &  $-4$ & $-8 \sqrt{6}$ & & &  &  $-\frac{4}{3}$& & $-4 \sqrt{3}$\\  \hline
  $[31][211]_\text{b}$ &   &     &   &  & $-8 \sqrt{3}$& & &$-4 \sqrt{6}$ & &  & & $\frac{8}{3}$ & \\  \hline
  $[31][211]_\text{c}$ &   &     &   & $4 \sqrt{3}$ & & $4 \sqrt{3}$ & & & &  & $-4 \sqrt{3}$ & & 4\\  \hline
    \end{tabular}
    \caption{Color-spin hyperfine matrix $\sum_{i<j\leq5}\lambda_{ij} \Sigma_{ij}$ for $I=\frac{1}{2}$ and $S=\frac{1}{2}$. The first number in the bracket denotes the representation for the orbital part, while the second corresponds to the spin-flavor representation. For example, $[4][31]_a$ represents the state $L[4]SF[31]_a$. }
    \label{tab:CS12}
\end{table*}

\begin{table*}[htbp]
    \centering
    \begin{tabular}{|c|c|c|c|c|c|c|c|c|c|c|c|} \hline
  & $[4][31]_\text{a}$ & $[4][31]_\text{b}$ & $[4][31]_\text{c}$ & $[31][31]_\text{a}$ & $[31][31]_\text{b}$  & $[31][31]_\text{c}$ & $[31][22]_\text{a}$ & $[31][22]_\text{b}$ & $[31][4]$ & $[31][211]_\text{a}$ &$[31][211]_\text{b}$ \\  \hline
  $[4][31]_\text{a}$ & $-\frac{4}{3}$  &     &  $4 \sqrt{10}$ &  & & & & & &  & \\  \hline
  $[4][31]_\text{b}$ &   & $-\frac{4}{3}$   &   &  & & & & & &  & \\  \hline
  $[4][31]_\text{c}$ &  $4 \sqrt{10}$ &     &  0 &  &  & &  & & &  & \\  \hline
  $[31][31]_\text{a}$  &   &     &   & $-\frac{10}{3}$ & & $2 \sqrt{10}$  & $-2 \sqrt{\frac{2}{3}}$ & & $-\frac{8}{\sqrt{3}}$ & $-6$ & \\  \hline
  $[31][31]_\text{b}$ &   &     &   & & $\frac{14}{3}$ &  & & $-2 \sqrt{10}$ & &  & $-2 \sqrt{3}$ \\  \hline
  $[31][31]_\text{c}$ &   &     &   & $2 \sqrt{10}$ &    &  & $-4 \sqrt{\frac{5}{3}}$& & $4 \sqrt{\frac{10}{3}}$ & $-2 \sqrt{10}$  & \\  \hline
  $[31][22]_\text{a}$ &   &     &   & $-2 \sqrt{\frac{2}{3}}$ & & $-4 \sqrt{\frac{5}{3}}$ &  $-\frac{8}{3}$ &   & $-\frac{8 \sqrt{2}}{3}$ & $-2 \sqrt{6}$ & \\  \hline
  $[31][22]_\text{b}$ &   &     &   &  &$-2 \sqrt{10}$& & &  &  &  & $-2 \sqrt{30}$\\  \hline
  $[31][4]$ &   &     &   & $-\frac{8}{\sqrt{3}}$ & &$4 \sqrt{\frac{10}{3}}$ & $-\frac{8 \sqrt{2}}{3}$ &  &  &  & \\  \hline
  $[31][211]_\text{a}$ &   &     &   & $-6$ & & $-2 \sqrt{10}$&$-2 \sqrt{6}$ &  & & $-\frac{10}{3}$ & \\  \hline
  $[31][211]_\text{b}$ &   &     &   &  & $-2 \sqrt{3}$ & & & $-2 \sqrt{30}$ & & & $\frac{2}{3}$ \\  \hline
    \end{tabular}
    \caption{Color-spin hyperfine matrix for $I=\frac{1}{2}$ and $S=\frac{3}{2}$}
    \label{tab:CS32}
\end{table*}

\begin{table*}[htbp]
    \centering
    \begin{tabular}{|c|c|c|c|} \hline
  & $[4][31]$ & $[31][31]$ & $[31][22]$ \\  \hline
  $[4][31]$ & $-\frac{40}{3}$  &    &    \\  \hline
  $[31][31]$  &   & $-\frac{40}{3}$  &  \\  \hline
  $[31][22]$ &   &     &  $-\frac{40}{3}$\\  \hline
    \end{tabular}
    \caption{Color-spin hyperfine matrix for $I=\frac{1}{2}$ and $S=\frac{5}{2}$}
    \label{tab:CS52}
\end{table*}

\subsection{Connection to Young tableaux}
\label{subsec:CS_young_example}

The computation of the explicit color-spin matrices is most transparent when
formulated in the Young-tableaux language used throughout the OSFC
construction. The central point is that the four identical quarks furnish a
representation of the permutation group $S_4$, and the Pauli principle requires
the $q^4$ wavefunction to be antisymmetric, i.e.\ to transform as $[1111]$ under
$S_4$. In the OSFC basis each factor---orbital, spin-flavor, and color---is
assigned an $S_4$ Young pattern, and the allowed pentaquark basis vectors are
those for which the tensor product
\begin{equation}
[f]_L \otimes [f]_{\rm SF} \otimes [f]_C \supset [1111]
\end{equation}
contains the fully antisymmetric irrep. With color fixed to $[211]_C$ in the
$q^4$ sector so that $[211]_C \otimes \bar{\mathbf{3}}_C \to \mathbf{1}_C$, the
orbital patterns relevant for positive-parity $P$-wave pentaquarks are
$L[4]$ and $L[31]$, and the spin-flavor Young patterns $[f]_{\rm SF}$ are
constrained accordingly.

Once the OSFC basis is fixed, the hyperfine operator is evaluated using
explicit Young basis vectors. For the quark-quark part~\eqref{eq:H_qq},
one exploits the fact that the operator is invariant under $S_4$ permutations
of the quark labels. Therefore it can only connect states within the same
$S_4$ irrep and is block diagonal in the $S_4$-adapted OSFC basis. The actual
matrix elements are obtained by evaluating the two-body operator in a basis
where a chosen pair, is first coupled in color and spin, and then
recoupled back to the chosen OSFC-Young basis.

Concretely, the color factor for a quark pair depends only on whether the pair
is in the $\bar{\mathbf{3}}_C$ or $\mathbf{6}_C$ channel,
\begin{equation}
\boldsymbol{\lambda}_i\cdot\boldsymbol{\lambda}_j =
\begin{cases}
-\frac{8}{3}, & (qq)_{\bar{\mathbf{3}}_C},\\
+\frac{4}{3}, & (qq)_{\mathbf{6}_C},
\end{cases}
\end{equation}
while the spin factor depends only on whether the pair is in $s_{ij}=0$ or
$s_{ij}=1$,
\begin{equation}
\boldsymbol{\sigma}_i\cdot\boldsymbol{\sigma}_j =
\begin{cases}
-3, & s_{ij}=0,\\
+1, & s_{ij}=1.
\end{cases}
\end{equation}

Thus, for any OSFC state $|\Psi_\alpha\rangle$, the contribution of a given pair
$(ij)$ is fixed once the amplitudes for finding that pair in the allowed
color-spin channels are known. These amplitudes are determined by the
Young-tableaux recoupling coefficients (isoscalar factors) relating the
pair-coupled basis to the chosen OSFC basis. Summing over all six pairs yields
the diagonal entries and, in sectors where multiple Young basis vectors exist
with the same quantum numbers, the off-diagonal mixings that appear in the
explicit matrices.

The quark-antiquark term in~\eqref{eq:H_qqbar}
is evaluated in the same OSFC basis. Although the total color contraction is
fixed in a color-singlet $q^4\bar q$ state,
\begin{equation}
\left(\sum_{i=1}^4 \boldsymbol{\lambda}_i + \boldsymbol{\lambda}_{\bar q}\right)
|\mathbf{1}_C\rangle = 0,
\end{equation}
the operator itself acts on each quark separately, and its matrix elements
depend on the detailed color-spin structure of the basis states.

The spin dependence may be expressed through angular-momentum algebra as
\begin{equation}
\sum_{i=1}^4 \boldsymbol{\sigma}_i\cdot\boldsymbol{\sigma}_{\bar q}
=
2\left[S(S+1)-S_{q^4}(S_{q^4}+1)-\frac{3}{4}\right],
\end{equation}
but this relation alone does not determine the full matrix elements, since the
color-spin operator involves correlated pairwise contractions. As a result,
the quark-antiquark contribution is diagonal or block diagonal in the OSFC
basis but is generally state-dependent.

To illustrate how an explicit table entry is obtained, one evaluates the
matrix element of each pair operator
$\boldsymbol{\lambda}_i\cdot\boldsymbol{\lambda}_j\,
\boldsymbol{\sigma}_i\cdot\boldsymbol{\sigma}_j$ by decomposing the state into
pair-coupled channels, multiplying by the corresponding eigenvalues, and
summing over all six pairs. The quark-antiquark term is treated analogously,
with each quark contributing separately.

\section{The construction of Light-front $P$-wave }\label{sec:construct_P}
The light front Hamiltonian for the pentaquark state includes the kinetic and confining terms
\bea
H'_{LF}&=&\sum_{i=1}^5\Bigg(\frac{k^2_{i\perp}+m_Q^2}{x_i}
\nonumber\\
&+&2\sigma_T((i\partial/\partial x_i)^2+M^2 r^2_{i\perp})^{1/2}\Bigg)
\eea
Here $x_i$ are momentum fractions
for quarks, $i=1..5$ and $r_{i\perp}$ are transverse coordinates.  The the string tension $\sigma_T=m_\rho^2/\pi\rightarrow (0.44\,\rm{GeV})^{2}$ is  fixed from meson Reggion trajectories, and the mass is approximated as $M\approx 5m_Q$, with $m_Q$ the constituent quark mass, which is taken to be 383MeV~\cite{Kock:2020frx}. 

To eliminate the motion of center mass, we use the following Jacobi coordinates for the longitudinal part,
\bea
x_\alpha&=&\frac{x_1-x_2}{\sqrt{2}},\nonumber\\
x_\beta&=&\frac{x_1+x_2-2x_3}{\sqrt{6}} ,\nonumber\\
x_\gamma&=&\frac{x_1+x_2+x_3-3x_4}{\sqrt{12}} ,\nonumber\\
x_\delta&=&\frac{x_1+x_2+x_3+x_4-4x_5}{\sqrt{20}}
\eea
Similarly for the transverse part. The light front Hamiltonian, free of CM, can be expressed using Jacobi coordinates~\cite{He:2025dik}
\bea\label{eq:LFHamiJob}
H_{LF}&\equiv&\sum_{i=1}^5\frac{k^2_{i\perp}+m_Q^2}{x_i} + 5\sigma_Ta 
\nonumber\\
&-&\frac{\sigma_T}{a}\sum_{\xi=\alpha,\beta,\gamma,\delta}((\partial/\partial x_\xi)^2+M^2 (\partial/\partial_{\vec{k}_{\xi\perp}})^2) \nonumber\\
\eea

The parameter a is chosen to be $a=7.59$, which has been determined by minimizing the ground state mass~\cite{He:2025dik}.
For the penta quark state, P wave state can be obtained by diagonalizing the full Hamiltonian for five quark states. The longitudinal and transverse eigenbasis used to diagonalize the full Hamiltonian are described in Appendix~.\ref{sec:lf_pwave_consolidated}.  The P stats for different orbital representation satisfy the following orthogonality conditions
\bea\label{eq:ortho}
&&\int [d^2\vec{k}_{\perp,_\xi}][dx_\xi ]\psi^{\mp 1}_{L[4]}(x,k_\perp)e^{\pm i\phi_{\hat{\delta}}}=C^\perp_{L[4]}\delta_{m,\mp 1}, 
\nonumber\\
&&\int [d^2\vec{k}_{\perp,_\xi}][dx_\xi ]\psi^{0}_{L[4]}(x,k_\perp)x_j=C^{||}_{L[4]}\delta_{j,\hat{\delta}}, 
\nonumber\\
&&\int [d^2\vec{k}_{\perp,_\xi}][dx_\xi ]\psi^{0}_{L[31]_i}(x,k_\perp)x_j=C^{||}_{L[31]}\delta_{ij}
\eea
where $d{x_\xi}=dx_\alpha dx_\beta dx_\gamma dx_\delta$ and similarly for $d^2 \vec{k}_{\xi\perp}$. $\psi^{n}_{L[4]}(x,k_\perp)$ and $\psi^{n}_{L[31]}(x,k_\perp)$ represent light-cone P-wave states for different group representation. Orthogonality conditions are imposed for P-wave states with transverse direction and longitudinal direction, respectively. 
The coefficients $C^\perp_{L[4]}$,  $C^{||}_{L[4]}$ and $C^{||}_{L[31]}$ can be obtained once the corresponding P-state states are constructed using the orthogonality conditions described above.
To construct P-wave state, we first calculate the following projections:
\bea\label{eq:proj}
C^{n,\pm 1}_{L[4]}&=&\int [d^2\vec{k}_{\perp,\xi}][dx_\xi ] \psi_n(x,k_\perp)e^{\pm i\phi_{\hat{\delta}}}\nonumber\\
C^{n,0}_{L[31]_\alpha}&=&\int [d^2\vec{k}_{\perp,\xi}][dx_\xi ] \psi_n(x,k_\perp)x_\alpha\nonumber\\
C^{n,0}_{L[31]_\beta}&=&\int [d^2\vec{k}_{\perp,\xi}][dx_\xi ] \psi_n(x,k_\perp)x_\beta\nonumber\\
C^{n,0}_{L[31]_\gamma}&=&\int [d^2\vec{k}_{\perp,\xi}][dx_\xi ] \psi_n(x,k_\perp)x_\gamma\nonumber\\
C^{n,0}_{L[31]_\delta}&=&\int [d^2\vec{k}_{\perp,\xi}][dx_\xi ] \psi_n(x,k_\perp)x_\delta
\eea
Here, $\psi_n(x,k_\perp)$ are the eigenstates of the full five quark Hamiltonian. We consider the lowest 11 states, their corresponding energy and projections onto different P-wave components are listed in Table.~\ref{tab:Mix}. Using the results in Table.~\ref{tab:Mix}, together with orthogonality conditions in Eq.~(\ref{eq:ortho}), we obtain
\bea\label{eq:L4pm1}
\psi^{+1}_{L[4]}(x,k_\perp)&=&(0.394653 +0.29994i) \psi_2(x,k_\perp)
\nonumber\\
&+&(-0.24334 + 0.931223i) \psi_3(x,k_\perp)\nonumber\\
\psi^{-1}_{L[4]}(x,k_\perp)&=&(0.641508 -0.618858i)  \psi_2(x,k_\perp)
\nonumber\\
&-&(0.334356+0.045252i) \psi_3(x,k_\perp)
\eea
One can then calculate the coefficient $C^\perp_{L[4]}$, defined in Eq.~(\ref{eq:ortho}), as
\bea
C^\perp_{L[4]}=\int [d^2\vec{k}_{\perp,_\xi}][dx_\xi ]\psi^{-1}_{L[4]}(x,k_\perp)e^{i\phi_{\hat{\delta}}}&=&0.469,
\eea

The state corresponding to the representation $L[31]_\alpha$ is contributed solely by a single eigenstate,
\bea
\psi^{0}_{L[31]_\alpha}(x,k_\perp)=\psi_8(x,k_\perp)
\eea
Accordingly, the coefficient $C^{||}_{L[31]}$ can be obtained as
\bea
C^{||}_{L[31]}&=&\int [d^2\vec{k}_{\perp,_\xi}][dx_\xi ]\psi^{0}_{L[31]_\alpha}(x,k_\perp)x_\alpha
\nonumber\\
&=&-0.0137+0.0465i
\eea
On the other hand, one can verify the following relations
\bea
&&|C^{8,0}_{L[31]_\alpha}|^2\approx |C^{6,0}_{L[31]_\beta}|^2+|C^{9,0}_{L[31]_\beta}|^2
\nonumber\\
&&\approx |C^{5,0}_{L[31]_\gamma}|^2+|C^{10,0}_{L[31]_\gamma}|^2\approx |C^{4,0}_{L[4]_\delta}|^2+|C^{11,0}_{L[4]_\delta}|^2
\eea
Therefore, the state corresponding to the representation $L[31]_\beta$ can be expressed as
\bea
\psi^{0}_{L[31]_\beta}(x,k_\perp)=a_1\psi_6(x,k_\perp)+a_2\psi_9(x,k_\perp)
\eea
where the coefficients $a_1$ and $a_2$ are determined by 
\bea
a_1C^{6,0}_{L[31]_\beta}+a_2C^{9,0}_{L[31]_\beta}=C^{8,0}_{L[31]_\alpha}, |a_1|^2+|a_2|^2=1
\eea
Finally, the state corresponding to the representation $L[31]_\beta$, $L[31]_\gamma$ and $L[4]_\delta$ are given by
\bea
\psi^{0}_{L[31]_\beta}&=&(0.053i+0.526)\psi_6(x,k_\perp)
\nonumber\\
&+&(-0.580+0.620i)\psi_9(x,k_\perp) \nonumber\\
\psi^{0}_{L[31]_\gamma}&=&(0.475-0.372i)\psi_5(x,k_\perp)
\nonumber\\
&+&(0.291i-0.742)\psi_{10}(x,k_\perp) \nonumber\\
\psi^{0}_{L[4]_\delta}&=&(-0.521+0.352i)\psi_4(x,k_\perp)
\nonumber\\
&+&(0.725+0.280i)\psi_{11}(x,k_\perp) \nonumber\\
\eea
Thus we obtain
\bea
C^{||}_{L[4]}&=&\int [d^2\vec{k}_{\perp,_\xi}][dx_\xi ]\psi^{0}_{L[4]}(x,k_\perp)x_\delta
\nonumber\\
&=&-0.0135+0.0457i
\eea
This value is very close to $C^{||}_{L[31]}$, indicating the longitudinal P-wave exhibits good symmetry across different directions in the Jacobi coordinates.
\begin{table*}[htbp]
    \centering
    \begin{tabular}{|c|c|c|c|c|c|c|c|c|} \hline
  n&$E_n$(GeV)& $C^{n,+1}_{L[4]}$ & $C^{n,-1}_{L[4]}$ & $C^{n,0}_{L[31]_\alpha}$ & $C^{n,0}_{L[31]_\beta}$ & $C^{n,0}_{L[31]_\gamma}$ & $C^{n,0}_{L[4]_\delta}$   \\  \hline
  1& $4.71$ & 0  &  0   &  0  & 0 & 0 & 0  \\  \hline
  2& $4.86$ & $0.305+0.321i$  & $0.0988 -0.1196i$  & 0  & 0 & 0 & 0 \\  \hline
  3& $4.86$ & $-0.213+0.0813i$ & $-0.0859-0.4007i$  & 0  & 0& 0 & 0 \\  \hline
  4& $4.99$ & 0 &  0  &  0  & 0  & 0 & $0.023\, -0.019 i$    \\  \hline
  5& $5.00$ & 0  & 0   &  0 & 0 & $-0.023\, +0.017i$ & 0   \\  \hline
  6& $5.00$ & 0  & 0  &  0 & $-0.005+0.025i$ & 0 & 0   \\  \hline
  7& $5.00$ & 0  & 0  &  0  & 0 & 0 & 0 \\  \hline
  8& $5.00$ & 0  & 0    &   $-0.0137+0.0465i$ & 0 & 0 & 0   \\  \hline
  9& $5.05$ & 0  & 0    & 0 & $0.037\, -0.018i$ & 0 & 0   \\  \hline
  10& $5.05$ & 0  & 0    &  0 & 0.0 & $-0.038+0.003i$ & 0 \\  \hline
  11& $5.06$ & 0  & 0    &  0 & 0.0 & 0 & $0.003\, +0.037 i$  \\  \hline 
    \end{tabular}
    \caption{The projection of lowest 11 states of 5 quark full Hamiltonian. The second column is the energy. The third to eighth columns represent the projections defined in Eq.~(\ref{eq:proj}). 
    The ground state shows zero overlap with these P wave projections.}
    \label{tab:Mix}
\end{table*}
Once the above states are obtained, they can be combined with the spin flavor and color wave functions to compute the hyperfine splitting, the final energy should be $$E_n+{\rm hyperfine\, splitting}$$.

\subsection{Physical interpretation}

The expressions above are symmetry-adapted wave functions, but their role is
dynamical: they are the minimal scaffolding that allows one to compute and
interpret the observables that matter for nucleon-pentaquark mixing in a
light-front Hamiltonian framework.

The first ingredient is the hyperfine interaction within the $q^4$ core.
Because the hyperfine Hamiltonian is a sum of two-body operators acting on pairs
of quarks, its matrix elements are controlled by how orbital, spin, flavor, and
color rearrange under quark exchanges.  In an $[f]$-labeled $S_4$ basis these
exchange operations are organized and sparse, so the full hyperfine matrix is
assembled from algebraic recoupling factors fixed by the Sec.~\ref{sec:five_quark_decomp} projectors.
This makes the subsequent diagonalization physically transparent: the resulting
eigenchannels correspond to definite $(I,S)$ multiplets built from Pauli-allowed
OSFC components.

The second ingredient is the set of transition operators that connect the
three-quark nucleon to the five-quark sector.  In the present application these operators include a $\sigma$-type ${}^3P_0$ pair-creation vertex and a $\pi$-type including an additional spin-momentum operator.  Both impose sharp selection rules.  The ${}^3P_0$
vertex creates a spin-triplet pair with one unit of relative orbital angular
momentum. In a light-front formulation, longitudinal and transverse dynamics
separate in a particularly clean way; as a result, only specific $P$-wave orbital
components contribute to overlap integrals.  Writing the five-quark states in an
explicit $([f]_L[f']_{SF})$ basis isolates those contributing components without
ambiguity and separates the purely group-theoretic recoupling from the
longitudinal/transverse integrals.

The third ingredient is comparison and matching to existing dynamical templates.
Constructions in the style of Miesch-Shuryak-Zahed~\cite{Miesch:2025ael} rely on controlled orbital
truncations while retaining exact spin-flavor-color combinatorics.  The basis
given here is designed to meet that standard: the orbital content is explicit
and minimal, the Pauli principle is implemented exactly at the $q^4$ level, and
the resulting states are immediately usable for hyperfine splittings and for
$|qqq\rangle \leftrightarrow |qqqq\bar q\rangle$ transition matrix elements.

\section{Nucleon-pentaquark $\sigma, \pi$ mixing operators}
\label{sec:mixing_consolidated}

In this section we consolidate the explicit results obtained for the transition (mixing) between the
three-quark nucleon state and the positive-parity $P$-wave pentaquark states generated by the pair-creation operator.
The presentation follows the same logic used in the Miesch-Shuryak-Zahed treatment~\cite{Miesch:2025ael}: we separate
the calculation into (i) the internal color-spin-flavor matrix elements (group theory and
recoupling), and (ii) the orbital overlaps (longitudinal and transverse) dictated by the light-front
basis and by the vertex structure. The end product is a set of transition matrix elements
$\langle N|T|P\rangle$ resolved by $(I,S)$ and by the five-quark orbital-spin-flavor symmetry
labels, together with the corresponding hyperfine eigenvectors and their mixing strengths.

\subsection{Definition of the transition matrix element and basis}
\label{subsec:mix_def}

We define the nucleon state in the minimal three-quark Fock sector as
\begin{equation}
|N\rangle \;=\; |L[3]\rangle \otimes |SF[3]\rangle \otimes |C[111]\rangle,
\label{eq:40}
\end{equation}
with $L[3]$ the totally symmetric $S$-wave orbital component and $C[111]$ the color singlet.
The pentaquark state is taken in the minimal $qqqq\bar q$ sector, with a four-quark core in color
$[211]$ coupled to the antiquark color $[11]$ to form a color singlet. The orbital part is a
$P$-wave excitation in the five-body Jacobi basis (or equivalently the light-front
$\alpha,\beta,\gamma,\delta$ basis), and the spin-flavor part is organized by $S_4$ Young shapes
$[31],[22],[4],[211]$ as detailed in the group decomposition sections.

The generic mixing matrix element of interest is of the form
\begin{equation}
\mathcal{M}_{N\leftrightarrow P}
\;=\;
\langle N|\, \mathcal{O}_{45}\, |P\rangle,
\end{equation}
where $\mathcal{O}_{45}$ acts on the created quark-antiquark pair labeled $(4,5)$ and includes
the appropriate color, flavor and spin projectors
\begin{equation}
\chi^C_{45},\qquad \chi^F_{45},\qquad \chi^S_{45},
\label{eq:42}
\end{equation}
with $\chi^C_{45}$ projecting onto a color singlet, $\chi^F_{45}$ onto a flavor singlet, and
$\chi^S_{45}$ onto the spin channel dictated by the chosen vertex (triplet for $\sigma$-type
$^3P_0$, and also triplet for the $\pi$-type operator used below).

The pentaquark orbital-spin-flavor basis states are labeled as in the tables,
e.g.\ $L[4]\,SF[31]_{\rm a,b,c}$, $L[31]\,SF[31]_{\rm a,b,c}$, $L[31]\,SF[22]_{\rm a,b}$, etc.,
with the explicit spin-flavor factorization
$SF=\;S[\cdot]\otimes F[\cdot]$ indicated in the captions.

\subsection{$\sigma$-type and $\pi$-type mixing operators ($^3P_0$ model) and selection rules}
\label{subsec:sigma_operator}

The $\sigma$-type  and $\pi$-type mixings are induced by a $^3P_0$ pair-creation vertex. In momentum space, a
standard form is
\begin{widetext}
\begin{equation}
\begin{aligned}
T_{\sigma,\pi}
=
-3 \gamma_0
\int d\vec{p}_4\, d\vec{p}_5\;
\delta\!\left(\vec{p}_4+\vec{p}_5\right)\,
e^{-r_\sigma^2(\vec{p}_4-\vec{p}_5)^2/6}\,
\mathcal{T}_{\sigma,\pi}\chi_{45}^{F}\chi_{45}^{C} ,
\label{eq:43}
\end{aligned}
\end{equation}
where the pair is
created in a color singlet, flavor singlet and spin triplet configuration. The sigma and pion operator as chiral partners, can be written explicitly as~\cite{Miesch:2025ael}
\bea
{\rm sigma:}\qquad &&\mathbf S_{4\bar q}.(\hat{\mathbf z}\times \mathbf K_{4\bar q})\nonumber\\
{\rm pion^a:}\qquad &&\mathbf S_{4\bar q}. \mathbf K_{4\bar q}\,\mathbf \tau^a
\eea
as we also detail in Appendix~\ref{sec_FW}, with both carrying the same dimensions and the correct parity assignments. Projecting these structures onto the intrinsic $\delta$ coordinate amounts to identifying the transferred momentum with the momentum conjugate to that Jacobi transfer variable, $\mathbf K_{4\bar{q}} \rightarrow \frac{4}{\sqrt{5}}\mathbf k_\delta$.  The effective
transition operators are therefore taken in the form
\begin{equation}
\mathcal{T}_\sigma
=
\frac{1}{2i}
\left[
\left(S^{(4)}_{+}+S^{(\bar q)}_{+}\right) \frac{4}{\sqrt{5}}k_{\delta,-}\frac{1}{\sqrt{3}}\chi^{Sz=+1}_{45}
-
\left(S^{(4)}_{-}+S^{(\bar q)}_{-}\right) \frac{4}{\sqrt{5}}k_{\delta,+}\frac{1}{\sqrt{3}}\chi^{Sz=-1}_{45}
\right],
\label{eq:Tsigma_delta}
\end{equation}
and 
\bea
\mathcal{T}_\pi
&=&
\Bigg[
\left(S^{(4)}_z+S^{(\bar q)}_z\right) m_N\frac{(\sqrt{5}x_\sigma-\sqrt{3}x_\gamma)}{2}\frac{-1}{\sqrt{3}}\chi^{Sz=0}_{45}
\nonumber\\
&+&\frac{1}{2}
\left(
\left(S^{(4)}_{+}+S^{(\bar q)}_{+}\right)\frac{4}{\sqrt{5}}k_{\delta,-}\frac{1}{\sqrt{3}}\chi^{Sz=+1}_{45}
+
\left(S^{(4)}_{-}+S^{(\bar q)}_{-}\right)\frac{4}{\sqrt{5}}k_{\delta,+}\frac{1}{\sqrt{3}}\chi^{Sz=-1}_{45}
\right)
\Bigg] \tau^a ,
\label{eq:Tpi_delta}
\eea
where $k_{\delta,\pm}=k_{\delta,x}\pm i k_{\delta,y}=k_\delta e^{\pm i\phi_\delta}$ are the transverse circular components in the $\delta$ mode. Note that the spin wave function $\chi_{45}$ has been included in the above equation to ensure that the quark and anti-quark pair has zero total angular momentum.
Equation~\eqref{eq:Tsigma_delta} is the intrinsic realization of the
light-front scalar spin-orbit structure
$\mathbf S\!\cdot(\hat z\times\mathbf k)$ and therefore excites purely transverse orbital components ($m_L=\pm1$).  In contrast,
Eq.~\eqref{eq:Tpi_delta} is the intrinsic form of
$\mathbf S\!\cdot\mathbf k\,\tau^a$ and contains both transverse
($m_L=\pm1$) and longitudinal ($m_L=0$) components. However, the operator $S^{(4)}_z+S^{(\bar{q})}_z$ vanish when projected to the zero spin state, so the longitudinal ($m_L=0$) components in Eq.~\eqref{eq:Tpi_delta} is zero.

On the light front, it is convenient to express the operator in the Jacobi transverse coordinates.
Using the $\gamma$-$\delta$ variables introduced previously, the transverse $\sigma$-type vertex
used in the explicit numerical evaluation is written as
\begin{equation}
\begin{aligned}
T_{\sigma,\pi}(\vec{k}_{\xi,\perp},\vec{k}_{\delta,\perp})
&=
-3\gamma_0\,
\delta^2(\vec{k}_{4,\perp}+\vec{k}_{5,\perp})\,
e^{-r_q^2(\vec{k}_{4,\perp}-\vec{k}_{5,\perp})^2/6}\mathcal{T}_{\sigma,\pi}\,
\chi^F_{45}\chi^C_{45}
\\
&=
-3\gamma_0\,
\delta^2\!\left(\frac{5\sqrt{3}\,\vec{k}_{\gamma,\perp}+3\sqrt{5}\,\vec{k}_{\delta,\perp}}{10}\right)\,
e^{-r_q^2(\sqrt{5}\vec{k}_{\delta,\perp}-\sqrt{3}\vec{k}_{\gamma,\perp})^2/24}\,
\mathcal{T}_{\sigma,\pi}\,
\chi^F_{45}\chi^C_{45}
\\
&=-4\gamma_0\,
e^{-8r_q^2\vec{k}_{\delta,\perp}^2/15}\,
\mathcal{T}_{\sigma,\pi}\,
\chi^F_{45}\chi^C_{45}\,
\delta^2\!\left(\vec{k}_{\gamma,\perp}+\sqrt{\frac{3}{5}}\vec{k}_{\delta,\perp}\right),
\end{aligned}
\label{eq:T1_sigma}
\end{equation}
with $\gamma_0=2.6$ and $r_q=0.3\,{\rm fm}$ in the numerical examples~\cite{Capstick:1993kb}.

The key selection rule relevant to the transverse $\sigma$-type and $\pi$-type mixings are that the orbital part must
supply a $P$-wave in the $\delta$ direction in order to saturate the $e^{\pm i\phi_\delta}$ factor.
This is why, in the explicit evaluation of the transverse mixing matrix element, only the pentaquark
orbital representation with $L[4]$ (transverse $m_\delta=\pm 1$ components) contributes at leading
order, while the $L[31]_{\alpha,\beta,\gamma}$ longitudinal components do not contribute to the
same transverse overlap. {Therefore, the only pentastate with non-zero mixing with three quark state is $\Psi_{P m_L}^A$ defined in Eq.~(\ref{eq:psiA}).
The transition matrix element between N and pentastate $\Psi_{P m_L}^A$ can be expressed as}
\bea\label{eq:trans}
&&\langle N(S_z=\frac{1}{2})|T_{\sigma,\pi}|P(S_z=\frac{1}{2})\rangle=\int [d^2\vec{k}_{\perp,_\xi}][dx_\xi ]\psi^\dagger_S(x'_\alpha,x'_\beta;k_{\alpha,\perp},k_{\beta,\perp})\psi^{1}_{L[4]}(x_\alpha,x_\beta,x_\sigma,x_\delta;k_{\alpha,\perp},k_{\beta,\perp},k_{\gamma,\perp},k_{\delta,\perp})
\nonumber\\
&\times&\frac{1}{\sqrt{3}}\langle SF[3]C[111]~|T_{\sigma,\pi}(\vec{k}_{\xi,\perp},\vec{k}_{\delta,\perp})|~SF[31]
\left(
\left[
C[211]_\beta SF[31]_\alpha
-
C[211]_\alpha SF[31]_\beta
+
C[211]_\gamma SF[31]_\gamma
\right]
\, C[11]\, S F[1]
\right)\rangle
\nonumber\\
\eea
where $\psi_S(x;\vec{k}_\perp)$ represent the light from S wave three quark state obtained by diagonalizing the Hamiltonian given in Appendix~\ref{sec:lf_swave_consolidated}, and $x'_\alpha$ and $x'_\beta$ are defined as
\bea\label{eq:xpab}
x'_{\alpha,\beta}&=&\frac{x_{\alpha,\beta}}{1-x_4-x_5}=\frac{10 x_{\alpha,\beta }}{5 \sqrt{3} x_{\gamma} +3 \sqrt{5}  x_{\delta} +6}
\eea
The three effective light-front fractions \(x'_i(x)\) entering
the nucleon wavefunction are obtained from the five-body fractions \(x_i\) by
removing the total plus momentum carried by the created sigma-pair~\footnote{Clearly if $X_\sigma=0$ the overlap would be zero. The created pair carries non-zero longitudinal momentum.},
\begin{equation}
X_\sigma \equiv x_4 + x_5,
\qquad
0 < X_\sigma < 1,
\end{equation}
and rescaling the remaining three-quark fractions to unit sum
\begin{equation}
x'_1(x) = \frac{x_1}{1-X_\sigma},\qquad
x'_2(x) = \frac{x_2}{1-X_\sigma},\qquad
x'_3(x) = \frac{x_3}{1-X_\sigma}.
\end{equation}
Then one can obtain the corresponding Jacobi coordinates $x'_\alpha$ and $x'_\beta$ as shown in Eq.~(\ref{eq:xpab}). The wave function $\psi_S(x'_\alpha,x'_\beta;k_{\alpha,\perp},k_{\beta,\perp})$ expressed in terms of the new variables $x'$, is normalized to unity.
\end{widetext}

\subsection{$\sigma$-induced and $\pi$-induced mixing matrix elements by symmetry channel}
\label{subsec:pi_matrix_elements}

For $(I,S)=(\tfrac{1}{2},\tfrac{1}{2})$ the reduced transition elements
$\langle N(\tfrac{1}{2})|T_\pi|P(\tfrac{1}{2})\rangle$ and $\langle N(\tfrac{1}{2})|T_\sigma|P(\tfrac{1}{2})\rangle$in the symmetry basis are listed in
Table~\ref{tab:T_S12}, and for $(I,S)=(\tfrac{1}{2},\tfrac{3}{2})$ they are listed in
Table~\ref{tab:T_S32}, and the matrix elements for $(I,S)=(\tfrac{1}{2},\tfrac{5}{2})$ are zero. These values are the direct counterparts of the $\sigma$-mixing table, but
with a different selection pattern: many channels vanish, and the dominant nonzero entries appear in
the symmetry sectors that carry the appropriate longitudinal $P$-wave content.

\begin{table*}[htbp]
    \centering
    \begin{tabular}{|c|c|c|c|} \hline
   & $\langle N,\frac{1}{2}|T_\pi|v_n,\frac{1}{2}\rangle$   & $\langle N,\frac{1}{2}|T_\sigma|v_n,\frac{1}{2}\rangle$  \\  \hline
     $[4][31]_\text{a}$ & $0.0721846 -0.0416758 i$   & $0.0416758+0.0721846 i$\\  \hline
    $[4][31]_\text{b}$ & $0.0721846 -0.0416758 i$  & $0.0416758+0.0721846 i$\\  \hline 
    $[4][31]_\text{c}$ & $-0.216554+0.125027 i$  & $-0.125027 -0.216554 i$\\  \hline  
    $[31][31]_\text{a}$ & 0   & $0$\\  \hline 
    $[31][31]_\text{b}$ & 0  & $0$\\  \hline 
    $[31][31]_\text{c}$ & 0   & $0$\\  \hline 
    $[31][22]_\text{a}$ & 0  & $0$\\  \hline 
    $[31][22]_\text{b}$ & 0  & $0$\\  \hline 
    $[31][4]_\text{a}$ & 0  & $0$\\  \hline 
   $[31][4]_\text{b}$ &  0  & $0$\\  \hline 
   $[31][211]_\text{a}$ &0  & $0$\\  \hline 
   $[31][211]_\text{b}$ &  0 & $0$\\  \hline 
    $[31][211]_\text{c}$& 0  & $0$\\  \hline 
    \end{tabular}
    \caption{The transition matrix element between the pentastate with $S=\frac{1}{2}$, $I=\frac{1}{2}$ and nucleon state induced by the pion and sigma-type interactions. The numbers in the first and second brackets represent the representation of the orbital angular momentum and spin-flavor. For example, the label $[4][31]_\text{a}$ represent the state $L[4]SF[31]_a$ with spin $S=1/2$ and $I=1/2$.}
    \label{tab:T_S12}
\end{table*}

\begin{table*}[htbp]
    \centering
    \begin{tabular}{|c|c|c|} \hline
   & $\langle N,\frac{1}{2}|T_\pi|v_n,\frac{1}{2}\rangle$    & $\langle N,\frac{1}{2}|T_\sigma|v_n,\frac{1}{2}\rangle$  \\  \hline
     $[4][31]_\text{a}$ & $0.204169 -0.117877 i$  & $0.117877 +0.204169 i$\\  \hline
    $[4][31]_\text{b}$ & $0.204169 -0.117877 i$  & $0.117877 +0.204169 i$\\  \hline 
    $[4][31]_\text{c}$ & 0  & 0 \\  \hline  
    $[31][31]_\text{a}$ & 0   & $0$\\  \hline 
    $[31][31]_\text{b}$ & 0 & 0 \\  \hline 
    $[31][31]_\text{c}$ & 0  & $0$\\  \hline 
    $[31][22]_\text{a}$ & 0  & $0$\\  \hline 
    $[31][22]_\text{b}$ & 0  & $0$\\  \hline 
   $[31][4]$ &  0  & $0$\\  \hline 
   $[31][211]_\text{a}$ &0  & $0$\\  \hline 
   $[31][211]_\text{b}$ & 0  & $0$\\  \hline 
   \end{tabular}
    \caption{The transition matrix element between the pentastate with $S=\frac{3}{2}$, $I=\frac{1}{2}$ and nucleon state induced by the pion and sigma-type interactions.}
    \label{tab:T_S32}
\end{table*}

\begin{table*}[htbp]
\centering
\begin{tabular}{|c|c|c|c|c|} 
\hline
 & \makecell{Spectrum1 (color-spin)} & $\langle N,\tfrac{1}{2}|T_\pi|P,\tfrac{1}{2}\rangle$ & $\langle N,\tfrac{1}{2}|T_\sigma|P,\tfrac{1}{2}\rangle$  \\ \hline
\multirow{13}{*}{$\left(\tfrac{1}{2}\right)^+$}
& $c^p_0-662.6$  & 0 &0\\
& $c^p_0-662.6$  & 0 &0\\
& $c^p_0-341.8$  & 0 &0\\      
& $c^p_0-341.8$  & $0.0721846-0.0416758i$ &$0.0416758 +0.0721846 i$\\     
& $c^p_0-118.8$  & 0 &0 \\      
& $c^p_0-118.8$  & 0 &0 \\ 
& $c^p_0-105.6$  & 0 &0 \\      
& $c^p_0-105.6$  & 0 &0 \\    
& $c^p_0-105.6$  & $-0.13347+0.0770592 i$  &$-0.0770592-0.13347 i$\\    
& $c^p_0+203.3$  & $-0.18518 +0.106914i$  &$-0.106914-0.18518 i$\\ 
& $c^p_0+203.3$  & $0$ &0 \\ 
& $c^p_0+203.3$  & $0$ &0 \\ 
& $c^p_0+537.2$  & $0$ &0 \\
\hline
\multirow{11}{*}{$\left(\tfrac{3}{2}\right)^+$}
& $c^p_0-219.8$  & 0 & 0 \\
& $c^p_0-219.8$  & 0 & 0 \\
& $c^p_0-219.8$  & $0.140519 -0.0811285i$  & $0.0811285 +0.140519 i$\\      
& $c^p_0-122.1$  & 0 & 0 \\     
& $c^p_0-122.1$  & 0 & 0 \\       
& $c^p_0+24.4$   & 0 & 0 \\ 
& $c^p_0+24.4$   & $0.204169-0.117877i$ &$0.117877 +0.204169 i$\\      
& $c^p_0+244.2$  & $-0.14812 +0.0855169i$ &$-0.0855169-0.14812 i$ \\    
& $c^p_0+244.2$  & 0 & 0  \\
& $c^p_0+244.2$  & 0  & 0 \\
& $c^p_0+244.2$  & 0  & 0 \\
\hline
\multirow{3}{*}{$\left(\tfrac{5}{2}\right)^+$}
& $c^p_0+244.2$  & $0$ & 0 \\
& $c^p_0+244.2$  & $0$ & 0\\
& $c^p_0+244.2$  & $0$ & 0\\
\hline
\end{tabular}
\caption{Spectrum lines and transition matrix elements between the nucleon state and $P$-wave pentaquark
eigenstates with $I=\tfrac{1}{2}$ and spin assignments $\tfrac{1}{2}^+$, $\tfrac{3}{2}^+$, $\tfrac{5}{2}^+$, which are obtained by diagonalizing the matrix of spin color interaction shown in Table~\ref{tab:CS12},~\ref{tab:CS32} and \ref{tab:CS52}. $c_0^p=c_0+150$(MeV) with $c_0=1756$(MeV) is has been determined in our pervious work for S-wave analysis, 150MeV is the mass gap between the first excited state and ground state of light cone Hamiltonian, as shown in Table~.\ref{tab:Mix}.   
}
\label{tab:specanT_consolidated}
\end{table*}

\subsection{$\sigma, \pi$ mixing summary}
\label{subsec:mix_interpretation}

The $\sigma$-type $^3P_0$ vertex enforces a strong transverse selection rule: it couples most
efficiently to pentaquark configurations that contain an explicit transverse $P$-wave component in the
$\delta$ coordinate, thereby isolating the $L[4]$ orbital representation as the dominant contributor
to the transverse overlap. After hyperfine diagonalization, this translates into the set of eigenchannels with distinct mixing strengths shown in Table~\ref{tab:specanT_consolidated}.

Although the $\pi$-type vertex includes an additional longitudinal coupling, it vanishes due to the spin projection applied to the fourth and fifth quarks. As a result, the corresponding transition matrix elements are very similar to those arising from the $\sigma$-type interaction.  The vanishing of the $(I,S)=(\tfrac{1}{2},\tfrac{5}{2})$
entries is consistent with the spin constraints built into the operator and the required pair quantum numbers.

These matrix elements enter as off-diagonal blocks that
mix the lowest nucleon eigenstate with a tower of pentaquark eigenchannels. The relative importance of the
$\sigma$- and $\pi$-induced admixtures then depends on the interplay between: (i) the hyperfine-split
pentaquark spectrum, (ii) the orbital overlap suppression factors (transverse for both $\sigma$ and $\pi$),
and (iii) the effective couplings $\gamma_0$ and the corresponding pion-sector coupling normalization.
These ingredients together determine whether a small but non-negligible $qqqq\bar q$ component is generated in the
physical nucleon state, and which pentaquark symmetry channels dominate that admixture,
as we now discuss.

\section{Mixed nucleon state from $\sigma, \pi$ transitions}
\label{sec:Nphys_explicit}

We now write the physical nucleon eigenstate in the presence of explicit $qqqq\bar q$ admixtures
generated by the $\sigma$-type ($^3P_0$) and $\pi$-type (spin-momentum) transition operators.
Following the MSZ logic, we treat the mixing in leading-order degenerate (or quasi-degenerate)
perturbation theory: the bare $|N\rangle$ in the three-quark sector mixes with a tower of
orthonormal pentaquark eigenchannels $|P_n\rangle$ obtained after hyperfine diagonalization in the
five-quark sector. The resulting physical nucleon is then written as a normalized superposition
\begin{widetext}
\begin{equation}
\begin{aligned}
|N_{\rm phys},S_z\rangle
&=
\sqrt{Z_N}\,
\Bigg[
|N,S_z\rangle
+
\sum_{n\in{\cal P}}
C^{(\sigma)}_{n}(S_z)\,|P_n,S_z\rangle
+
\sum_{n\in{\cal P}}
C^{(\pi)}_{n}(S_z)\,|P_n,S_z\rangle
\Bigg],
\end{aligned}
\label{eq:Nphys_general_expansion}
\end{equation}
\end{widetext}
where ${\cal P}$ is the set of relevant $P$-wave pentaquark channels (including their $(I,S)$ labels),
and $Z_N$ ensures $\langle N_{\rm phys}|N_{\rm phys}\rangle=1$.

\subsection{Perturbative coefficients and energy denominators}

Let the unperturbed nucleon mass be $M_N^{(0)}$ and the unperturbed pentaquark eigenenergies be
$E_n^{(0)}$ (in the same Hamiltonian scheme used to obtain the hyperfine eigenvectors and spectrum
splittings). Denote the energy gaps
\begin{equation}
\Delta_n \equiv E_n^{(0)}-M_N^{(0)}.
\end{equation}
The $E_n^{(0)}$ includes the spectrum generated by the light cone Hamiltonian, together with the hyper-fine splitting arises from the spin color interaction. The later one can be obtained by diagonalizing the hyper-interactions matrix shown in Table~\ref{tab:CS12} and ~\ref{tab:CS32}. The Energy gap between the light cone $S$ wave and $P$ wave state is around 150MeV as shown in the second column in Table~\ref{tab:Mix}. Note that only the transverse $P$ components are related in this work since the sigma and pion interactions couple to these transverse modes only. The mass gap between the longitudinal and transverse component is about 150MeV, reflecting the breaking of $O(3)$ symmetry in the light cone frame. The full spectrum of the P state is summarized in Table~\ref{tab:specanT_consolidated}. The parameter $c_0^p=c_0+150$(MeV) with $c_0=1756$(MeV) is an overall parameter by fixing the S wave spectrum in our previous work~\cite{He:2025dik}, and additional 150(MeV) is mass gap between light cone S wave and transverse P wave component.

To leading order, the admixture coefficients are
\begin{eqnarray}
C^{(\sigma)}_{n}(S_z)
&=&
\frac{\langle P_n,S_z|T_\sigma|N,S_z\rangle}{\Delta_n},
\nonumber\\
C^{(\pi)}_{n}(S_z)
&=&
\frac{\langle P_n,S_z|T_\pi|N,S_z\rangle}{\Delta_n},
\label{eq:Cn_pert_def}
\end{eqnarray}
with $T_\sigma$ and $T_\pi$ defined in Eqs.~\eqref{eq:Tsigma_delta} and \eqref{eq:Tpi_delta}. The
normalization factor is, at the same order,
\begin{equation}
Z_N
=
1-\sum_{n}\Big(|C_n^{(\sigma)}|^2+|C_n^{(\pi)}|^2
+2\,{\rm Re}\big[C_n^{(\sigma)}(C_n^{(\pi)})^\ast\big]\Big)+\cdots,
\label{eq:ZN_def}
\end{equation}
where the interference term is present if both operators connect $|N\rangle$ to the same orthonormal
pentaquark channel $|P_n\rangle$.

\subsection{Explicit $\sigma$-induced and $\pi$-induced admixture in the hyperfine eigenbasis}

For the $\sigma$ and $\pi$ operators, we use the orthonormal hyperfine eigenvectors $|P_n\rangle$ to represent the 27 states shown in Table~\ref{tab:Pwave_counts}.
The explicit mixed nucleon
state with $S_z=\tfrac{1}{2}$ can be written compactly as
\begin{widetext}
\begin{equation}
\begin{aligned}
|N_{\rm phys},\tfrac{1}{2}\rangle
&=
\sqrt{Z_N}\,
\Bigg[
|N,\tfrac{1}{2}\rangle
+\sum_{n=1}^{27}C_n^{(\pi)}(\frac{1}{2})\,
|P_n,\tfrac{1}{2}\rangle
+\sum_{n=1}^{27}C_n^{(\sigma)}(\frac{1}{2})\,
|P_n,\tfrac{1}{2}\rangle
\Bigg],
\end{aligned}
\label{eq:Nphys_final_both}
\end{equation}
\end{widetext}

with
\begin{eqnarray}    
C^{(\sigma)}_{n}\!\left(\tfrac{1}{2}\right)&=&
\frac{\langle P_n,\tfrac{1}{2}|T_\sigma|N,\tfrac{1}{2}\rangle}{\Delta_{n}}=\frac{\langle N,\tfrac{1}{2}|T_\sigma|P_n,\tfrac{1}{2}\rangle^*}{\Delta_{n}},
\nonumber\\
C^{(\pi)}_{n}\!\left(\tfrac{1}{2}\right)&=&
\frac{\langle P_n,\tfrac{1}{2}|T_\pi|N,\tfrac{1}{2}\rangle}{\Delta_{n}}=\frac{\langle N,\tfrac{1}{2}|T_\pi|P_n,\tfrac{1}{2}\rangle^*}{\Delta_{n}}
\label{eq:C_sigma_vi}
\end{eqnarray}
The coefficients $C^{(\sigma)}_{n}$ and $C^{(\pi)}_{n}$ are presented in Table.~\ref{tab:Cn}. They can be obtained using the results presented in Table~.\ref{tab:specanT_consolidated}.

\begin{table*}[htbp]
\centering
\begin{tabular}{|c|c|c|c|c|} 
\hline
 & \makecell{Spectrum (MeV)} & $C_n^{(\sigma)}(\frac{1}{2})$ &  $C_n^{(\pi)}(\frac{1}{2})$  \\ \hline
\multirow{13}{*}{$\left(\tfrac{1}{2}\right)^+$}
& 1243.4  & 0 &0\\
& 1243.4  & 0 &0\\
& 1564.2  & 0 &0\\      
& 1564.2  & $0.0667703 -0.11565i$ &$0.0667703i +0.11565$\\     
& 1787.2  & 0 &0 \\      
& 1787.2  & 0 &0 \\ 
& 1800.4  & 0 &0 \\      
& 1800.4  & 0 &0 \\    
& 1800.4  & $-0.0895611+0.155124 i$  &$-0.0895611i-0.155124 $\\    
& 2109.3  & $-0.0914374+0.158374 i$  &$-0.0914374i-0.158374 $\\ 
& 2109.3  & $0$ &0 \\ 
& 2109.3  & $0$ &0 \\ 
& 2443.2  & $0$ &0 \\
\hline
\multirow{11}{*}{$\left(\tfrac{3}{2}\right)^+$}
& 1686.25  & 0 & 0 \\
& 1686.25  & 0 & 0 \\
& 1686.25  & $0.108715 -0.1883i$  & $0.108715i +0.1883$\\      
& 1783.92  & 0 & 0 \\     
& 1783.92  & 0 & 0 \\       
& 1930.42   & 0 & 0 \\ 
& 1930.42   & $0.119017 - 0.206144i$ &$0.119017i + 0.206144$\\      
& 2150.17  & $-0.0706654 + 0.122396i$ &$-0.0706654i - 0.122396$ \\    
& 2150.17  & 0 & 0  \\
& 2150.17  & 0  & 0 \\
& 2150.17  & 0  & 0 \\
\hline
\multirow{3}{*}{$\left(\tfrac{5}{2}\right)^+$}
& 2150.2  & $0$ & 0 \\
& 2150.2  & $0$ & 0\\
& 2150.2  & $0$ & 0\\
\hline
\end{tabular}
\caption{Spectrum lines and coefficients $C_n^\sigma(\frac{1}{2})$ and $C_n^\pi(\frac{1}{2})$. }
\label{tab:Cn}
\end{table*}

This explicit state is the direct analogue of the MSZ mixed-nucleon wave function: a bare $qqq$
nucleon dressed by a controlled $qqqq\bar q$ cloud, with the dressing amplitudes determined by
microscopic transition operators ($\sigma$ and $\pi$ here) and by the hyperfine-resolved five-quark
spectrum through the energy denominators.

\subsection{Dominant mixing states}
\label{subsec:sigma_histogram}

Using the coefficients $C_n^{(\sigma)}\!\left(\frac12\right)$ listed in Table~\ref{tab:Cn}, we define the probability carried by the $n$-th hyperfine-diagonalized pentaquark eigenchannel in the physical nucleon in Eq.~\eqref{eq:Nphys_final_both},
with the coefficients defined in Eq.~(\ref{eq:C_sigma_vi}). In the present case, only $6$ out of the $27$ P-wave pentastates carry nonzero $\sigma$-probability, while the remaining $21$ states are exactly null within the accuracy of Table~\ref{tab:Cn}. This already shows that the $\sigma$-induced nucleon dressing is highly sparse and strongly constrained by symmetry. The existence of $27$ total P-wave pentastates follows from the counting in Table~I: $13$ for $S=\frac12$, $11$ for $S=\frac32$, and $3$ for $S=\frac52$. The canonical OSFC families are $L[4]\otimes SF[31]$, $L[31]\otimes SF[31]$, $L[31]\otimes SF[22]$, $L[31]\otimes SF[4]$, and $L[31]\otimes SF[211]$. The representative flavor$\otimes$spin irreps relevant for the different total-spin sectors is summrized in Table~\ref{tab:FS_decomp}. 

For the $27$ entries shown in Table~\ref{tab:Cn}, the nonzero values of $P_n^{(\sigma)}$ are
\begin{align}
P_4^{(\sigma)}  &= 0.01782, \nonumber\\
P_9^{(\sigma)}  &= 0.03209, \nonumber\\
P_{10}^{(\sigma)} &= 0.03344, \nonumber\\
P_{16}^{(\sigma)} &= 0.04728, \nonumber\\
P_{20}^{(\sigma)} &= 0.05664, \nonumber\\
P_{21}^{(\sigma)} &= 0.01998 ,
\end{align}
with all other $P_n^{(\sigma)}=0$.
The largest single contribution is therefore carried by the $n=20$ eigenchannel, followed by $n=16$, and then the pair $n=9,10$.

\begin{figure*}[t]
  \centering
  \includegraphics[width=0.72\textwidth]{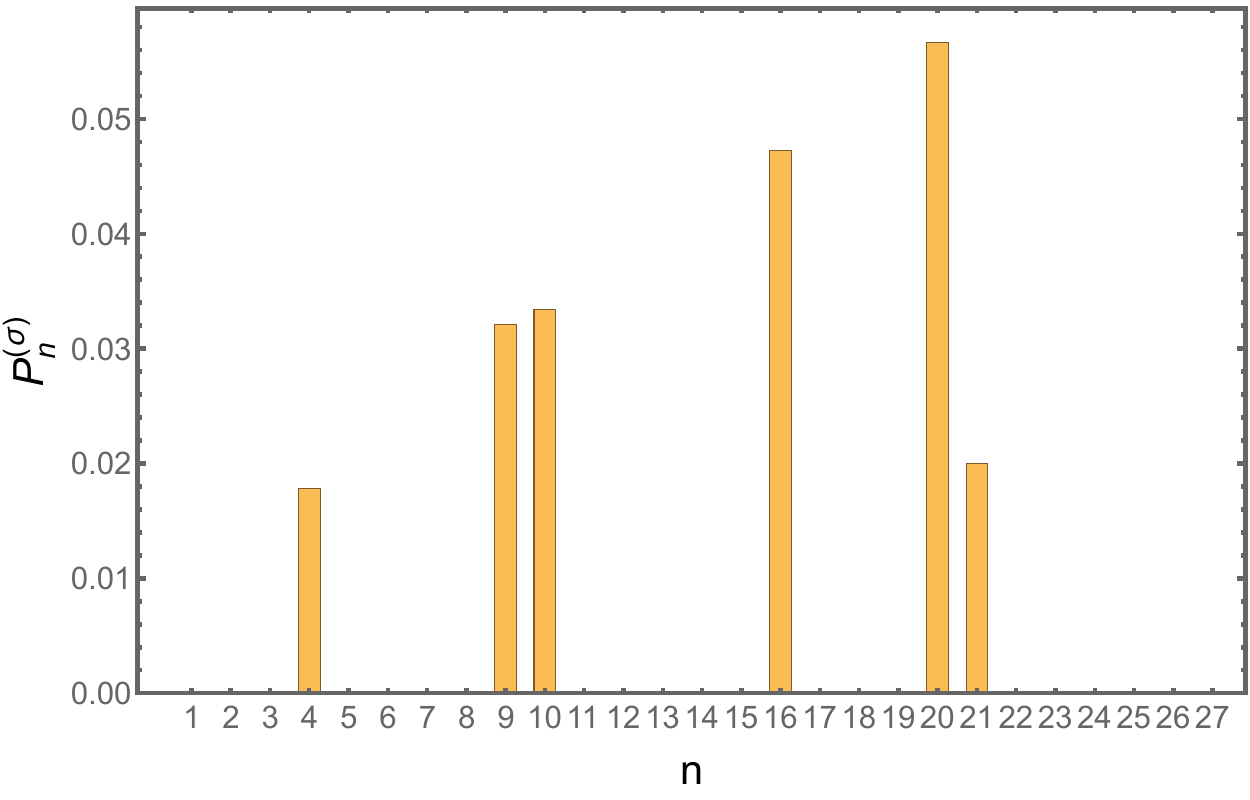}
  \caption{Histogram of the probabilities $P_n^{(\sigma)}=\bigl|C_n^{(\sigma)}\bigr|^2$ for the $27$ hyperfine-diagonalized P-wave pentastates entering Eq.~\eqref{eq:Nphys_general_expansion}. The distribution is highly sparse: $21$ states have vanishing probability, while only $6$ states contribute. }
  \label{fig:sigma_probability_channels}
\end{figure*}

To make the sparse structure more transparent, it is useful to isolate only the nonzero contributions and label them by the corresponding symmetry channels. Adopting the natural ordering implied by Table~\ref{tab:Pwave_counts} together with the symmetry-basis ordering in Tables~\ref{tab:T_S12} and ~\ref{tab:T_S32}, the six nonzero contributions are assigned as
\begin{equation}
\begin{array}{c|c|c}
n & \text{dominant symmetry channel} & P^{(\sigma)}\\
\hline
4   & [4][31]_b~[S=1/2] & 0.01782\\
9   & 0.585[4][31]a+0.811[4][31]c~[S=1/2] & 0.03209\\
10  & -0.811[4][31]a+0.585[4][31]c~[S=1/2] & 0.03344\\
16  & 0.688[4][31]a+0.725[4][31]c~[S=3/2] & 0.04728\\
20  & [4][31]_b~[S=3/2] & 0.05664\\
21  & -0.725[4][31]a+0.688[4][31]c~[S=3/2]& 0.01998
\end{array}
\label{eq:channel_assignment_sigma}
\end{equation}

The first three states are the spin 1/2 states, whereas the last three are spin 3/2 states. This channel assignment should be viewed as the dominant symmetry-channel identification inherited from the basis ordering; after hyperfine diagonalization, each physical eigenstate is in general a linear combination of the underlying OSFC basis states. Nevertheless, the pattern is physically robust: the nonzero strength is concentrated in a very small set of hyperfine-selected channels.

.\subsection{Pion-induced admixture and chiral symmetry}
\label{subsec:pi_chiral}

The analysis of the $\pi$-induced mixing coefficients $C_n^{(\pi)}$ proceeds in complete analogy with the $\sigma$ case discussed above. In particular, the transition amplitudes are defined by Eq.~(\ref{eq:C_sigma_vi}), and the corresponding probabilities are
\begin{equation}
P_n^{(\pi)} \equiv \left|C_n^{(\pi)}\!\left(\tfrac12\right)\right|^2 .
\end{equation}
A direct inspection of Table~VII shows that the pattern of nonzero entries in the $\pi$ channel is identical to that of the $\sigma$ channel: only $6$ out of the $27$ hyperfine eigenstates contribute, while the remaining $21$ entries vanish.

This similarity is not accidental, but follows from the underlying chiral structure of the transition operators. The $\sigma$-type operator corresponds to a scalar $^3P_0$ pair-creation vertex, while the $\pi$-type operator introduces a spin-momentum coupling associated with pseudoscalar emission. Despite their different Dirac structures, both operators couple the nucleon to the same set of Pauli-allowed pentaquark configurations and share the same selection rules at the level of color-spin-flavor symmetry.

In particular, as noted in Sec.~\ref{sec:mixing_consolidated}, the $\pi$-type interaction involves an additional longitudinal spin operator, but its contribution vanishes due to the specific spin structure $S_z^{(4)} + S_z^{(\bar q)}$ acting on the pion state. As a result, the effective transition matrix elements reduce to the same symmetry structure as the $\sigma$-induced ones. Consequently, both operators probe the same restricted subset of OSFC channels, leading to identical support in the space of hyperfine eigenstates. 

From the perspective of chiral symmetry, this reflects the fact that the $\sigma$ and $\pi$ operators form components of a common chiral multiplet. In a light-front Hamiltonian framework where the dominant dynamics is governed by color-spin interactions and transverse orbital structure, the distinction between scalar and pseudoscalar couplings is subleading compared to the symmetry constraints imposed by the Pauli principle and the hyperfine Hamiltonian. The observed equality of the support of $P_n^{(\sigma)}$ and $P_n^{(\pi)}$ is therefore a direct manifestation of this underlying chiral structure.

In summary, while the magnitudes of the coefficients $C_n^{(\sigma)}$ and $C_n^{(\pi)}$ may differ at the level of detailed dynamics, the set of contributing pentaquark eigenchannels is identical. This provides a nontrivial consistency check of the framework and confirms that the dominant five-quark admixture in the nucleon is controlled primarily by symmetry selection rules rather than by the specific Dirac structure of the transition operator.

\subsection{Mixing percentages}

The total five-quark weight induced by the $\sigma$ operator is
\begin{equation}
P_{\rm tot}^{(\sigma)}=\sum_{n=1}^{27}\left|C_n^{(\sigma)}\right|^2=0.20727,
\end{equation}
corresponding to about $20.7\%$ at the level of the unnormalized expansion coefficients. Likewise, the $\pi$-induced contribution satisfies
\begin{equation}
P_{\rm tot}^{(\pi)}=\sum_{n=1}^{27}\left|C_n^{(\pi)}\right|^2=0.20727,
\end{equation}
since the two operators populate the same set of pentaquark eigenchannels with identical magnitudes.

A crucial simplification arises from the fact that, as seen in Table~\ref{tab:Cn}, the $\pi$-induced coefficients are the complex conjugates of the $\sigma$-induced ones,
\begin{equation}
C_n^{(\pi)} = iC_n^{(\sigma)}.
\end{equation}
As a result, the interference term in Eq.~\eqref{eq:Nphys_final_both} vanishes
\begin{equation}
2\,\mathrm{Re}\sum_n C_n^{(\sigma)}\left(C_n^{(\pi)}\right)^*
=0 .
\end{equation}
The total contribution of $\sigma$ and $\pi$ interactions become
\begin{equation}
P_{\rm tot}^{(\sigma+\pi)}
=
P_{\rm tot}^{(\sigma)} + P_{\rm tot}^{(\pi)}
=
0.415.
\end{equation}

The properly normalized five-quark fraction is then
\begin{equation}
\mathcal{P}_{5q}
=
\frac{P_{\rm tot}^{(\sigma+\pi)}}{1+P_{\rm tot}^{(\sigma+\pi)}}
\simeq 0.293,
\end{equation}
corresponding to approximately $29\%$ five-quark content and $71\%$ three-quark core in the physical nucleon.

This result reflects a nontrivial dynamical consequence of the chiral structure of the transition operators: although the $\sigma$ and $\pi$ operators individually generate comparable admixtures, their contributions 
add incoherently.

\subsection{Comparison with recent results}

In this subsection we compare our results for nucleon–pentaquark mixing with the recent analysis of five-quark components in the nucleon presented in Ref.~\cite{Miesch:2025ael}. Both works share the common objective of quantifying the role of $qqqq\bar{q}$ configurations in the nucleon wave function and provide further evidence that higher Fock components beyond the three-quark core are phenomenologically significant. In particular, both approaches identify dynamical mechanisms that generate mixing between $|qqq\rangle$ and $|qqqq\bar{q}\rangle$ sectors and find that the resulting five-quark probability is sizable, at the level of ${\cal O}(10\%-30\%)$. This agreement supports the general picture that nucleon structure is intrinsically multi-partonic.

The two analyses differ qualitatively in how symmetry and dynamics constrain the mixing pattern. A primary  distinction lies in the implementation of the Pauli principle. In the present work, the four-quark core is constructed using a complete S$_4$ permutation-group classification, ensuring that the combined orbital, spin-flavor, and color wave function is fully antisymmetric and free of redundancies. This leads to exact selection rules at the level of the basis itself. As a direct consequence, only $6$ out of the $27$ positive-parity $P$-wave pentastates contribute to nucleon mixing, while the remaining $21$ vanish identically. The resulting sparsity is therefore not a dynamical accident but a symmetry-enforced constraint.

In contrast, Ref.~\cite{Miesch:2025ael} does not organize the five-quark basis in terms of a complete S$_4$ classification, but uses instead a large monom basis ($4\times3^6\times2^5\times2^5$) to diagonalize the S$_4$ permutation group 
with extensive use of Mathematica. As a result, the Pauli constraints are implemented less explicitly, and a larger set of configurations can contribute to the mixing. This leads to a more distributed pattern of amplitudes in which the underlying symmetry selection rules are less transparent, and the effective number of contributing channels is significantly larger.

 Additionally, there is a slight difference in the structure of the transition operators. In our light-front Hamiltonian framework, the $\sigma$- and $\pi$-type operators form a chiral pair and act on the same restricted set of Pauli-allowed configurations. Their amplitudes are related by a fixed phase,
\begin{equation}
C_n^{(\pi)} = i\,C_n^{(\sigma)},
\end{equation}
which implies that the interference term vanishes exactly in the normalization. The two contributions therefore add incoherently, leading to a tightly constrained and correlated pattern of mixing.

Finally, in the present work, hyperfine color-spin interactions are diagonalized explicitly in the full symmetry-adapted basis, yielding orthonormal eigenchannels that reorganize the original configurations into physically meaningful states. The mixing is then expressed in this eigenbasis, which reveals a clear hierarchy in which only a small number of eigenchannels dominate the five-quark admixture. In contrast, analyses that do not perform a full hyperfine diagonalization typically work in a fixed configuration basis, where the strength is distributed more broadly and the identification of dominant structures is less direct.

These differences lead to a qualitatively distinct interpretation of the five-quark content of the nucleon. In the present work, the result
\begin{equation}
P_{5q} \simeq 29\%,
\end{equation}
emerges from a highly sparse and symmetry-selected set of channels, dominated by a few hyperfine eigenstates. This indicates that the five-quark component is not a  superposition of many configurations, but rather a structured admixture controlled by symmetry selection rules and chiral dynamics. The comparison therefore highlights the importance of implementing exact permutation symmetry, together with chiral operator structure and hyperfine diagonalization, in order to reveal the underlying organization of higher Fock components in baryons.

\section{Conclusions}
\label{sec_conclusions}

In this work we have carried out a systematic light-front Hamiltonian analysis of nucleon–pentaquark mixing induced by $\sigma$- and $\pi$-type transition operators in a fully Pauli-consistent five-quark basis. Using a permutation-group construction of the orbital, spin-flavor, and color degrees of freedom, we have built a complete set of positive-parity P-wave pentaquark states and organized them into symmetry-adapted bases suitable for dynamical calculations. The inclusion of hyperfine color-spin interactions and their diagonalization yields a set of orthonormal eigenchannels that provide a physically meaningful description of the five-quark sector.

A central result of this analysis is the emergence of strong symmetry selection rules that drastically reduce the effective number of contributing channels. Among the $27$ possible P-wave pentastates, only $6$ contribute to the physical nucleon state, while the remaining $21$ vanish identically. The resulting five-quark admixture is therefore highly sparse and dominated by a small number of hyperfine-selected channels. Furthermore, we have shown that the $\sigma$- and $\pi$-induced amplitudes populate the same subset of states and are related by complex conjugation, reflecting their common chiral structure. 
After normalization, the nucleon contains a five-quark component of approximately $29\%$, with the remainder residing in the three-quark core.

Beyond these specific results, an important outcome of this work is the explicit construction of the mixed nucleon wave function in a fully symmetry-controlled framework. The expressions given in Sec.~\ref{sec:Nphys_explicit}, together with the tabulated coefficients in Table~\ref{tab:Cn}, provide a concrete and reusable representation of the nucleon state including its five-quark components. In this sense, the present work serves as a repository of the mixed nucleon wave function, suitable for future theoretical and phenomenological applications and for cross-checks with alternative approaches.

The availability of this explicit wave function opens several directions for further study. The five-quark components provide a natural framework for analyzing flavor asymmetries in the nucleon sea, including the $\bar d - \bar u$ imbalance and possible strange-quark contributions. They also encode nontrivial orbital structure, offering a pathway to quantifying orbital angular momentum contributions to the nucleon spin. In addition, the mixed wave function can be used to compute static and dynamical observables such as magnetic moments and transition form factors, as well as more differential quantities including generalized parton distributions (GPDs) and transverse-momentum-dependent distributions (TMDs). In all these cases, the strong symmetry selection rules identified here imply that only a small subset of channels will dominate the physical observables.

More broadly, the framework developed here provides a bridge between symmetry-based constructions of hadronic wave functions and dynamical light-front calculations. The combination of Pauli-consistent basis states, controlled transition operators, and hyperfine dynamics offers a systematic approach to incorporating higher Fock components in baryon structure. The resulting picture is one in which nucleon–pentaquark mixing is governed primarily by symmetry and chiral structure, leading to a reduced and highly organized five-quark content that can be quantitatively explored in future studies.

\section{Acknowledgements}
This work is supported by the Office of Science, U.S. Department of Energy under Contract  No. DE-FG-88ER40388. FH is supported by the U.S. Department of Energy, Office of Science, Office of Nuclear Physics, under Grant No. DE-SC0013065. 
This research is also supported in part within the framework of the Quark-Gluon Tomography (QGT)  Topical Collaboration, under contract no. DE-SC0023646.

\appendix

\section{$S_3$ tensor-product projections and cluster recoupling}
\label{App_3}

Although the pentaquark core is governed by $S_4$, $S_3$ enters whenever one isolates a three-quark
cluster or performs intermediate couplings. The simplest couplings serve as consistency checks and define phase conventions.

Coupling two symmetric irreps yields
\begin{equation}
[3]_A\otimes [3]_B=[3]_{AB},
\qquad
\big([3]_{AB}:[3]_A\otimes[3]_B\big)=A[3]\;B[3].
\end{equation}
Coupling a symmetric with a mixed irrep gives
\begin{eqnarray}
&&[3]_A\otimes [21]_B=[21]_{AB},
\nonumber\\
&&\big([21]_{AB}:[3]_A\otimes[21]_B\big)_{\alpha,\beta}=A[3]\;B[21]_{\alpha,\beta}.
\end{eqnarray}
The nontrivial case is $[21]\otimes[21]$, which contains all three irreps:
\begin{equation}
[21]_A\otimes[21]_B=[3]_{AB}\oplus[21]_{AB}\oplus[111]_{AB}.
\end{equation}
The explicit projections are
\begin{widetext}
\begin{eqnarray}
\big([3]_{AB}:[21]_A\otimes[21]_B\big)
&=&
\frac{1}{\sqrt{2}}\Big(A[21]_\alpha\,B[21]_\alpha+A[21]_\beta\,B[21]_\beta\Big),
\nonumber\\
\big([111]_{AB}:[21]_A\otimes[21]_B\big)
&=&
\frac{1}{\sqrt{2}}\Big(A[21]_\alpha\,B[21]_\beta-A[21]_\beta\,B[21]_\alpha\Big),
\nonumber\\
\big([21]_{AB}:[21]_A\otimes[21]_B\big)_\alpha
&=&
\frac{1}{\sqrt{2}}\Big(A[21]_\alpha\,B[21]_\beta+A[21]_\beta\,B[21]_\alpha\Big),
\nonumber\\
\big([21]_{AB}:[21]_A\otimes[21]_B\big)_\beta
&=&
\frac{1}{\sqrt{2}}\Big(A[21]_\alpha\,B[21]_\alpha-A[21]_\beta\,B[21]_\beta\Big).
\nonumber\\
\end{eqnarray}
\end{widetext}
In physical terms, the first line isolates the exchange-even component of the product, the second 
isolates the exchange-odd component, and the last two span the mixed irrep. These are precisely 
the combinations that appear, for example, when two mixed-symmetry subsystems are coupled to produce 
a symmetric (or antisymmetric) effective cluster state.

\section{$S_4$ tensor-product projections for the four-quark core}
\label{App_4}

For four identical quarks the permutation group is S$_4$, and the relevant irreducible representations 
are $[4]$, $[31]$, $[22]$, $[211]$, and $[1111]$. The Pauli principle requires that the total four-quark 
wave function transform as $[1111]$ under S$_4$. Writing the four-quark wave function as a product of orbital, 
spin-flavor, and color factors, this constraint becomes
\begin{equation}
[f_{\rm orb}] \otimes [f_{\rm SF}] \otimes [f_{\rm color}] \supset [1111].
\end{equation}
This single condition determines which combinations of orbital excitation, color structure, and spin-flavor 
symmetry are allowed.

We now list the $S_4$ projections required to couple symmetry types across different physical factors of 
the four-quark core. The simplest products involve a totally symmetric factor. Since $[4]$ is the identity 
representation of $S_4$ in the sense of tensoring, it leaves the other symmetry type unchanged
\begin{widetext}
\begin{equation}
[4]_A\otimes[4]_B=[4]_{AB},
\qquad
\big([4]_{AB}:[4]_A\otimes[4]_B\big)=A[4]\;B[4],
\end{equation}
\begin{equation}
[4]_A\otimes[31]_B=[31]_{AB},
\qquad
\big([31]_{AB}:[4]_A\otimes[31]_B\big)_{\alpha}=A[4]\;B[31]_{\alpha},
\end{equation}
\begin{equation}
[4]_A\otimes[22]_B=[22]_{AB},
\qquad
\big([22]_{AB}:[4]_A\otimes[22]_B\big)_{\alpha}=A[4]\;B[22]_{\alpha},
\end{equation}
\begin{equation}
[4]_A\otimes[211]_B=[211]_{AB},
\qquad
\big([211]_{AB}:[4]_A\otimes[211]_B\big)_{\alpha}=A[4]\;B[211]_{\alpha}.
\end{equation}
These relations are heavily used when the orbital part is taken in its lowest configuration $[4]$ and one wishes 
to infer that the entire symmetry bookkeeping is carried by color and spin-flavor.

The workhorse decomposition in many pentaquark channels is $[31]\otimes[31]$. It arises, for example, when both 
orbital and spin-flavor parts are of $[31]$ type (a common situation for negative-parity states built from a 
single $P$-wave excitation together with a mixed-symmetry spin-flavor core), or when color is in a mixed irrep 
and must be combined with another mixed factor to reach $[1111]$. The decomposition is
\begin{equation}
[31]_A\otimes[31]_B=[4]_{AB}\oplus[31]_{AB}\oplus[22]_{AB}\oplus[211]_{AB},
\end{equation}
and the explicit normalized projections read
\begin{equation}
\begin{aligned}
\big([4]_{AB}:[31]_A\otimes[31]_B\big)
&=
\frac{1}{\sqrt{3}}\Big(A[31]_\alpha B[31]_\alpha+A[31]_\beta B[31]_\beta+A[31]_\gamma B[31]_\gamma\Big),\\
\big([31]_{AB}:[31]_A\otimes[31]_B\big)_\alpha
&=
\frac{1}{\sqrt{3}}\Big(A[31]_\alpha B[31]_\beta+A[31]_\beta B[31]_\alpha\Big)
+\frac{1}{\sqrt{6}}\Big(A[31]_\alpha B[31]_\gamma+A[31]_\gamma B[31]_\alpha\Big),\\
\big([31]_{AB}:[31]_A\otimes[31]_B\big)_\beta
&=
\frac{1}{\sqrt{3}}\Big(A[31]_\alpha B[31]_\alpha-A[31]_\beta B[31]_\beta\Big)
+\frac{1}{\sqrt{6}}\Big(A[31]_\gamma B[31]_\beta+A[31]_\beta B[31]_\gamma\Big),\\
\big([31]_{AB}:[31]_A\otimes[31]_B\big)_\gamma
&=
\frac{1}{\sqrt{6}}\Big(A[31]_\alpha B[31]_\alpha+A[31]_\beta B[31]_\beta\Big)
-\frac{2}{\sqrt{6}}\,A[31]_\gamma B[31]_\gamma,\\
\big([22]_{AB}:[31]_A\otimes[31]_B\big)_\alpha
&=
\frac{1}{\sqrt{6}}\Big(A[31]_\alpha B[31]_\beta+A[31]_\beta B[31]_\alpha\Big)
-\frac{1}{\sqrt{3}}\Big(A[31]_\alpha B[31]_\gamma+A[31]_\gamma B[31]_\alpha\Big),\\
\big([22]_{AB}:[31]_A\otimes[31]_B\big)_\beta
&=
\frac{1}{\sqrt{6}}\Big(A[31]_\alpha B[31]_\alpha-A[31]_\beta B[31]_\beta\Big)
-\frac{1}{\sqrt{3}}\Big(A[31]_\beta B[31]_\gamma+A[31]_\gamma B[31]_\beta\Big),\\
\big([211]_{AB}:[31]_A\otimes[31]_B\big)_\alpha
&=
\frac{A[31]_\alpha B[31]_\gamma-A[31]_\gamma B[31]_\alpha}{\sqrt{2}},\\
\big([211]_{AB}:[31]_A\otimes[31]_B\big)_\beta
&=
\frac{A[31]_\beta B[31]_\gamma-A[31]_\gamma B[31]_\beta}{\sqrt{2}},\\
\big([211]_{AB}:[31]_A\otimes[31]_B\big)_\gamma
&=
\frac{A[31]_\alpha B[31]_\beta-A[31]_\beta B[31]_\alpha}{\sqrt{2}}.
\end{aligned}
\end{equation}
A useful physical way to read these expressions is the following. The $[4]$ projector extracts the ``aligned'' 
combination where the basis labels match ($\alpha\alpha+\beta\beta+\gamma\gamma$), i.e.\ the maximally symmetric 
component. The $[211]$ projectors are manifestly antisymmetric under interchange of $A\leftrightarrow B$ in the 
paired basis labels, reflecting the deeper antisymmetry content of $[211]$. The remaining $[31]$ and $[22]$ 
combinations interpolate between these extremes and correspond to distinct Pauli-allowed recouplings.

The coupling $[31]\otimes[22]$ arises whenever a pair-symmetry structure ($[22]$) must be combined with a 
one-quark-distinguished structure ($[31]$), as happens in diquark-motivated bases or when mixing channels 
with different internal clusterizations. One has
\begin{equation}
[31]_A\otimes[22]_B=[31]_{AB}\oplus[211]_{AB},
\end{equation}
with
\begin{equation}
\begin{aligned}
\big([31]_{AB}:[31]_A\otimes[22]_B\big)_\alpha
&=
\frac{1}{2}\Big(B[22]_\alpha A[31]_\beta+B[22]_\beta A[31]_\alpha\Big)
-\frac{1}{\sqrt{2}}\,B[22]_\alpha A[31]_\gamma,\\
\big([31]_{AB}:[31]_A\otimes[22]_B\big)_\beta
&=
\frac{1}{2}\Big(B[22]_\alpha A[31]_\alpha-B[22]_\beta A[31]_\beta\Big)
-\frac{1}{\sqrt{2}}\,B[22]_\beta A[31]_\gamma,\\
\big([31]_{AB}:[31]_A\otimes[22]_B\big)_\gamma
&=
-\frac{1}{\sqrt{2}}\Big(B[22]_\alpha A[31]_\alpha+B[22]_\beta A[31]_\beta\Big),\\
\big([211]_{AB}:[31]_A\otimes[22]_B\big)_\alpha
&=
\frac{A[31]_\alpha B[22]_\beta+A[31]_\beta B[22]_\alpha+\sqrt{2}\,A[31]_\gamma B[22]_\alpha}{2},\\
\big([211]_{AB}:[31]_A\otimes[22]_B\big)_\beta
&=
\frac{A[31]_\alpha B[22]_\alpha-A[31]_\beta B[22]_\beta+\sqrt{2}\,A[31]_\gamma B[22]_\beta}{2},\\
\big([211]_{AB}:[31]_A\otimes[22]_B\big)_\gamma
&=
\frac{A[31]_\alpha B[22]_\beta-A[31]_\beta B[22]_\alpha}{\sqrt{2}}.
\end{aligned}
\end{equation}
Here the appearance of both symmetric and antisymmetric combinations is again the group-theoretic manifestation of 
whether the resulting coupling can participate in an overall antisymmetric four-quark state.

The product $[31]\otimes[211]$ is important when one factor already contains substantial antisymmetry and must 
be coupled to a mixed factor. It decomposes as
\begin{equation}
[31]_A\otimes[211]_B=[31]_{AB}\oplus[22]_{AB}\oplus[211]_{AB}\oplus[1111]_{AB}.
\end{equation}
In pentaquark applications, the explicit appearance of $[1111]$ in this product is especially valuable: it 
provides a direct route to a Pauli-allowed core when the remaining factor is symmetric. The projections are
\begin{equation}
\begin{aligned}
\big([31]_{AB}:[31]_A\otimes[211]_B\big)_\alpha
&=
\frac{A[31]_\gamma B[211]_\alpha+A[31]_\beta B[211]_\gamma}{\sqrt{2}},\\
\big([31]_{AB}:[31]_A\otimes[211]_B\big)_\beta
&=
\frac{A[31]_\gamma B[211]_\beta-A[31]_\alpha B[211]_\gamma}{\sqrt{2}},\\
\big([31]_{AB}:[31]_A\otimes[211]_B\big)_\gamma
&=
\frac{-A[31]_\alpha B[211]_\alpha-A[31]_\beta B[211]_\beta}{\sqrt{2}},
\\[4pt]
\big([22]_{AB}:[31]_A\otimes[211]_B\big)_\alpha
&=
\frac{1}{\sqrt{6}}\Big(A[31]_\alpha B[211]_\beta+A[31]_\beta B[211]_\alpha\Big)
-\frac{1}{\sqrt{3}}\Big(A[31]_\beta B[211]_\gamma-A[31]_\gamma B[211]_\alpha\Big),\\
\big([22]_{AB}:[31]_A\otimes[211]_B\big)_\beta
&=
\frac{1}{\sqrt{6}}\Big(A[31]_\alpha B[211]_\alpha-A[31]_\beta B[211]_\beta\Big)
+\frac{1}{\sqrt{3}}\Big(A[31]_\alpha B[211]_\gamma+A[31]_\gamma B[211]_\beta\Big),
\\[4pt]
\big([211]_{AB}:[31]_A\otimes[211]_B\big)_\alpha
&=
\frac{1}{\sqrt{3}}\Big(A[31]_\alpha B[211]_\beta+A[31]_\beta B[211]_\alpha\Big)
+\frac{1}{\sqrt{6}}\Big(A[31]_\beta B[211]_\gamma-A[31]_\gamma B[211]_\alpha\Big),\\
\big([211]_{AB}:[31]_A\otimes[211]_B\big)_\beta
&=
\frac{1}{\sqrt{3}}\Big(A[31]_\alpha B[211]_\alpha-A[31]_\beta B[211]_\beta\Big)
-\frac{1}{\sqrt{6}}\Big(A[31]_\alpha B[211]_\gamma+A[31]_\gamma B[211]_\beta\Big),\\
\big([211]_{AB}:[31]_A\otimes[211]_B\big)_\gamma
&=
-\frac{1}{\sqrt{6}}\Big(A[31]_\alpha B[211]_\beta-A[31]_\beta B[211]_\alpha\Big)
+\frac{2}{\sqrt{6}}\,A[31]_\gamma B[211]_\gamma,
\\[4pt]
\big([1111]_{AB}:[31]_A\otimes[211]_B\big)
&=
\frac{1}{\sqrt{3}}\Big(A[31]_\alpha B[211]_\beta-A[31]_\beta B[211]_\alpha+A[31]_\gamma B[211]_\gamma\Big).
\end{aligned}
\end{equation}

The remaining products are included for completeness and for later use in channel mixing, where operators 
can connect basis states built from different internal symmetry couplings.

The product $[22]\otimes[22]$ decomposes as
\begin{equation}
[22]_A\otimes[22]_B=[4]_{AB}\oplus[22]_{AB}\oplus[1111]_{AB},
\end{equation}
with projections
\begin{equation}
\begin{aligned}
\big([4]_{AB}:[22]_A\otimes[22]_B\big)
&=
\frac{1}{\sqrt{2}}\Big(A[22]_\alpha B[22]_\alpha+A[22]_\beta B[22]_\beta\Big),\\
\big([22]_{AB}:[22]_A\otimes[22]_B\big)_\alpha
&=
\frac{1}{\sqrt{2}}\Big(A[22]_\alpha B[22]_\beta+A[22]_\beta B[22]_\alpha\Big),\\
\big([22]_{AB}:[22]_A\otimes[22]_B\big)_\beta
&=
\frac{1}{\sqrt{2}}\Big(A[22]_\alpha B[22]_\alpha-A[22]_\beta B[22]_\beta\Big),\\
\big([1111]_{AB}:[22]_A\otimes[22]_B\big)
&=
\frac{1}{\sqrt{2}}\Big(A[22]_\alpha B[22]_\beta-A[22]_\beta B[22]_\alpha\Big).
\end{aligned}
\end{equation}
The appearance of $[1111]$ here is another direct mechanism to build an antisymmetric four-quark state 
from pairwise correlated structures, which is why $[22]$-based diquark pictures can be compatible with 
Pauli constraints if the remaining factors are arranged appropriately.

The product $[22]\otimes[211]$ yields
\begin{equation}
[22]_A\otimes[211]_B=[31]_{AB}\oplus[211]_{AB},
\end{equation}
with
\begin{equation}
\begin{aligned}
\big([31]_{AB}:[22]_A\otimes[211]_B\big)_\alpha
&=
\frac{1}{2}\Big(A[22]_\alpha B[211]_\beta+A[22]_\beta B[211]_\alpha\Big)
+\frac{1}{\sqrt{2}}\,A[22]_\beta B[211]_\gamma,\\
\big([31]_{AB}:[22]_A\otimes[211]_B\big)_\beta
&=
\frac{1}{2}\Big(A[22]_\alpha B[211]_\alpha-A[22]_\beta B[211]_\beta\Big)
-\frac{1}{\sqrt{2}}\,A[22]_\alpha B[211]_\gamma,\\
\big([31]_{AB}:[22]_A\otimes[211]_B\big)_\gamma
&=
\frac{1}{\sqrt{2}}\Big(A[22]_\alpha B[211]_\alpha+A[22]_\beta B[211]_\beta\Big),
\\[4pt]
\big([211]_{AB}:[22]_A\otimes[211]_B\big)_\alpha
&=
\frac{1}{2}\Big(A[22]_\alpha B[211]_\beta+A[22]_\beta B[211]_\alpha\Big)
-\frac{1}{\sqrt{2}}\,A[22]_\beta B[211]_\gamma,\\
\big([211]_{AB}:[22]_A\otimes[211]_B\big)_\beta
&=
\frac{1}{2}\Big(A[22]_\alpha B[211]_\alpha-A[22]_\beta B[211]_\beta\Big)
+\frac{1}{\sqrt{2}}\,A[22]_\alpha B[211]_\gamma,\\
\big([211]_{AB}:[22]_A\otimes[211]_B\big)_\gamma
&=
\frac{1}{\sqrt{2}}\Big(A[22]_\alpha B[211]_\beta-A[22]_\beta B[211]_\alpha\Big).
\end{aligned}
\end{equation}
In later use, these relations control how pairwise symmetry in one factor (e.g.\ orbital) correlates 
with antisymmetry content in another (e.g.\ color).

Finally, the product $[211]\otimes[211]$ decomposes as
\begin{equation}
[211]_A\otimes[211]_B=[4]_{AB}\oplus[31]_{AB}\oplus[22]_{AB}\oplus[211]_{AB},
\end{equation}
with
\begin{equation}
\begin{aligned}
\big([4]_{AB}:[211]_A\otimes[211]_B\big)
&=
\frac{1}{\sqrt{3}}\Big(A[211]_\alpha B[211]_\alpha+A[211]_\beta B[211]_\beta+A[211]_\gamma B[211]_\gamma\Big),\\
\big([31]_{AB}:[211]_A\otimes[211]_B\big)_\alpha
&=
\frac{1}{\sqrt{6}}\Big(A[211]_\alpha B[211]_\beta+A[211]_\beta B[211]_\alpha\Big)
-\frac{1}{\sqrt{3}}\Big(A[211]_\beta B[211]_\gamma+A[211]_\gamma B[211]_\beta\Big),\\
\big([31]_{AB}:[211]_A\otimes[211]_B\big)_\beta
&=
\frac{1}{\sqrt{6}}\Big(A[211]_\alpha B[211]_\alpha-A[211]_\beta B[211]_\beta\Big)
+\frac{1}{\sqrt{3}}\Big(A[211]_\alpha B[211]_\gamma+A[211]_\gamma B[211]_\alpha\Big),\\
\big([31]_{AB}:[211]_A\otimes[211]_B\big)_\gamma
&=
-\frac{1}{\sqrt{6}}\Big(A[211]_\alpha B[211]_\alpha+A[211]_\beta B[211]_\beta\Big)
+\frac{2}{\sqrt{6}}\,A[211]_\gamma B[211]_\gamma,
\\[4pt]
\big([22]_{AB}:[211]_A\otimes[211]_B\big)_\alpha
&=
\frac{1}{\sqrt{6}}\Big(A[211]_\alpha B[211]_\beta+A[211]_\beta B[211]_\alpha\Big)
+\frac{1}{\sqrt{3}}\Big(A[211]_\beta B[211]_\gamma+A[211]_\gamma B[211]_\beta\Big),\\
\big([22]_{AB}:[211]_A\otimes[211]_B\big)_\beta
&=
\frac{1}{\sqrt{6}}\Big(A[211]_\alpha B[211]_\alpha-A[211]_\beta B[211]_\beta\Big)
-\frac{1}{\sqrt{3}}\Big(A[211]_\alpha B[211]_\gamma+A[211]_\gamma B[211]_\alpha\Big),
\\[4pt]
\big([211]_{AB}:[211]_A\otimes[211]_B\big)_\alpha
&=
\frac{1}{\sqrt{2}}\Big(A[211]_\beta B[211]_\gamma-A[211]_\gamma B[211]_\beta\Big),\\
\big([211]_{AB}:[211]_A\otimes[211]_B\big)_\beta
&=-
\frac{1}{\sqrt{2}}\Big(A[211]_\alpha B[211]_\gamma-A[211]_\gamma B[211]_\alpha\Big)
,\\
\big([211]_{AB}:[211]_A\otimes[211]_B\big)_\gamma
&=
\frac{1}{\sqrt{2}}\Big(A[211]_\alpha B[211]_\beta-A[211]_\beta B[211]_\alpha\Big).
\end{aligned}
\end{equation}
\end{widetext}

\section{Light-front $P$-wave pentaquark orbital states}
\label{sec:lf_pwave_consolidated}

A five-parton $qqqq\bar q$ light-front state is described by longitudinal momentum fractions $x_i$
and transverse momenta $\boldsymbol{k}_{i\perp}$ with
\begin{equation}
\sum_{i=1}^{5}x_i=1,\qquad x_i\ge0,\qquad \sum_{i=1}^{5}\boldsymbol{k}_{i\perp}=0,
\end{equation}
together with internal spin-flavor-color quantum numbers.  The orbital content on the light front
is organized by the conserved projection $L_z$ generated by transverse rotations,
\begin{equation}
L_z=-\,i\sum_{i=1}^{5}\frac{\partial}{\partial\varphi_i},
\end{equation}
where $\varphi_i$ is the azimuthal angle of $\boldsymbol{k}_{i\perp}$.  It is convenient to use
$k_\perp^\pm=k_x\pm i k_y\propto e^{\pm i\varphi}$ so that a single factor of $k_\perp^\pm$ carries
$L_z=\pm1$ and therefore implements the transverse members of a $P$-wave multiplet.

To separate the internal motion we introduce four longitudinal momenta using Jacobi momenta
$x_{\alpha,\beta,\gamma,\delta}$ in the center-of-momentum frame, chosen so that
$\alpha,\beta,\gamma$ resolve the internal motion of the $q^4$ core while $\delta$ resolves the
relative motion of the $q^4$ subsystem with respect to the antiquark.  A convenient explicit choice for the longitudinal part,
\bea\label{eq:jacobi_penta_consolidated}
x_\alpha&=&\frac{x_1-x_2}{\sqrt{2}},\nonumber\\
x_\beta&=&\frac{x_1+x_2-2x_3}{\sqrt{6}} ,\nonumber\\
x_\gamma&=&\frac{x_1+x_2+x_3-3x_4}{\sqrt{12}} ,\nonumber\\
x_\delta&=&\frac{x_1+x_2+x_3+x_4-4x_5}{\sqrt{20}}
\eea

where $1,2,3,4$ label the identical quarks and $5$ labels the antiquark. The transverse Jacobi coordinates can be defined similarly.
The final expression for the light front Hamiltonian, free of CM, is then given by~\cite{He:2025dik}
\bea\label{eq:LFHamiJob}
H_{LF}&\equiv&\sum_{i=1}^5\frac{k^2_{i\perp}+m_Q^2}{x_i} + 5\sigma_Ta 
\nonumber\\
&-&\frac{\sigma_T}{a}\sum_{\xi=\alpha,\beta,\gamma,\delta}((\partial/\partial x_\xi)^2+M^2 (\partial/\partial_{\vec{k}_{\xi\perp}})^2) \nonumber\\
\eea
where $a=7.59$ has been determined by minimizing the ground state mass~\cite{He:2025dik}.
To diagonalize the Light front Hamiltonian in Eq.~(\ref{eq:LFHamiJob}), one needs to find the proper basis. For the longitudinal part, we use the complete and orthogonal basis constructed using Slater determinants~\cite{He:2025dik},
\begin{widetext} 
\bea
\label{SLAT4}
\varphi_{\tilde{n}}[N=5]=\frac {1}{5^{3/4}}
\begin{vmatrix}
   1& 1  & \dots  & 1\\
    e^{i\tilde  n_{21}\tilde s_1}& e^{i\tilde  n_{21}\tilde s_2}  & \dots  & e^{i\tilde  n_{21}\tilde s_5} \\
    \vdots & \vdots & \ddots & \vdots &  \\
      e^{i\tilde  n_{51}\tilde s_1}& e^{i\tilde  n_{51}\tilde s_2}  & \dots  & e^{i\tilde  n_{51}\tilde s_5} 
\end{vmatrix}
\eea
with    
\bea
\label{MAPS}
\tilde{s}_1&=&\frac{1}{10} \left(2\sqrt{5} x_{\delta }+2\sqrt{3} x_{\gamma }+4+\sqrt{2} x_{\alpha }+\sqrt{6} x_{\beta }\right)
\nonumber\\
\tilde{s}_2&=&\frac{1}{30} \left(-12 \sqrt{2} x_{\alpha }-2 \sqrt{6} x_{\beta }+3 \sqrt{5} x_{\delta }+\sqrt{3} x_{\gamma }+6\right)
\nonumber\\
\tilde{s}_3&=&\frac{1}{30} \left(3 \sqrt{2} x_{\alpha }-7 \sqrt{6} x_{\beta }-4 \sqrt{3} x_{\gamma }\right)
\nonumber\\
\tilde{s}_4&=&\frac{1}{10} \left(\sqrt{2} x_{\alpha }+\sqrt{6} x_{\beta }-\sqrt{5} x_{\delta }-3 \sqrt{3} x_{\gamma }-2\right)
\nonumber\\
\tilde{s}_5&=&\frac{1}{10} \left(-2\sqrt{5} x_{\delta }+2\sqrt{3} x_{\gamma }-4+\sqrt{2} x_{\alpha }+\sqrt{6} x_{\beta }\right)\nonumber\\
\eea
\end{widetext}
Note that the correct normalization factor is $\frac {1}{5^{3/4}}$,  the factor $\frac {1}{5^{4/3}}$ given in our previous paper~\cite{He:2025dik} was a typo and does not effect any numerical results presented in ~\cite{He:2025dik}. The coefficients $\tilde n_{i1}$ are ordered as $0<\tilde n_{21}<\tilde n_{31}<\tilde n_{41}...$ to avoid repeated counting of the same state.
The eigenfunctions used for the each transverse component are of generalized Laguerre polynomials, 
\begin{widetext}
\ba  
\label{eq:lagre}
\psi^\perp_{n,m}(k_{\xi\perp},\beta) &=&{1\over \sqrt{\pi}} \beta^{1/4} \sqrt{n! \over (n+|m|)!} e^{-\frac{k_{\xi\perp}^2\beta^{1/2}}{2}+i m \phi}  
\,(k_{\xi\perp}  \beta^{1/4})^{|m|}L_n^{|m|}\big( \beta^{1/2} k_{\xi\perp}^2\big)
\ea
\end{widetext}
with $\beta=5a/(\sigma_T M^2)$ and and eigenvalues
$$E_{n,m}=2\sqrt{5}(\sigma_T M^2/a)^{1/2}(2n+|m|+1)\,.$$ 
Note that the basis in Eq.~(\ref{eq:lagre}). $\xi=\alpha,\beta,\gamma,\delta$ represent the different Jacobi components, the total transverse eigenfunctions are the product of $\psi^\perp_{n,m}(k_{\xi\perp},\beta)$ over four Jacobi components.

In the rest frame, a
minimal $P$-shell orbital basis with definite $S_4$ symmetry for the four identical quarks is
spanned by the totally symmetric irrep $L[4]$ and the mixed irrep $L[31]$, expressed on one Jacobi direction,
\begin{widetext}
\bea
&&L[4](L_z=\pm 1)\propto e^{\pm i\phi_\delta},\qquad
L[31]_\alpha(L_z=\pm 1)\propto e^{\pm i\phi_\alpha},\qquad
L[31]_\beta(L_z=\pm 1)\propto e^{\pm i\phi_\beta},\qquad
L[31]_\gamma(L_z=\pm 1)\propto e^{\pm i\phi_\gamma},
\nonumber\\
&&L[4](L_z=0)\propto x_\delta,\qquad
L[31]_\alpha(L_z=0)\propto x_\alpha,\qquad
L[31]_\beta(L_z=0)\propto x_\beta,\qquad
L[31]_\gamma(L_z=0)\propto x_\gamma,
\label{eq:restframe_irreps_compact}
\eea
\end{widetext}
which furnish the $[4]$ and $[31]$ irreps of $S_4$ when combined with the appropriate $S_4$
projection relations among the $[31]$ basis vectors. The components with $L_z=\pm 1$ are related to the rotation angular in the transverse direction. The component with $L_z=0$ is related to the longitudinal momentum. 

The numbers of eigenfunction used to diagonalize the light cone Hamiltonian are very huge since we need multiply the longitudinal basis and  transverse basis over four Jacobi coordinates once we combine different choices of $(n,m)$ satisfies the truncation condition. In this calculation, we set the index $m=\pm 1, 0$  in Eq.~(\ref{eq:lagre}) for the $\delta$ coordinate, and  $m=0$ for other Jacobi coordinates since the sigma- and pion-type interactions shown in Eq.~(\ref{eq:Tsigma_delta}) and Eq.~(\ref{eq:Tpi_delta}) contain only the $e^{\pm i\phi_\delta}$ component.
The longitudinal and transverse eigen-sets are truncated using the following constraints
\bea
E_L&=&\sum_{i=2}^{5}\tilde n_{i1}^2-\frac 15\bigg(\sum_{i=2}^{5} \tilde n_{i1}\bigg)^2\leq \frac{1184\pi^2}{5}, \nonumber\\
E_T&=&\sum_{i=1}^{4}(2n_i+|m|_i+1)\leq 10,
\eea

In practical LF Hamiltonian diagonalizations one often obtains orthonormal five-body eigenstates
$\psi_n(x,\boldsymbol{k}_\perp)$ and then reconstructs the symmetry-adapted orbital content by
projection.  For the $L[4]$ transverse sector one
projects onto the azimuthal phase of the $\delta$ Jacobi transverse momentum, while for the $L[31]$
longitudinal sector one projects onto the longitudinal Jacobi coordinates, as shown in Eq.~(\ref{eq:restframe_irreps_compact}), 
enforcing orthogonality condition
between the different mixed-symmetry directions, .
This yields the expansion coefficients
\bea
&&\int [d^2\vec{k}_{\perp,_\xi}][dx_\xi ]\psi^{m}_{L[4]}(x,k_\perp)e^{\pm i\phi_{\hat{\delta}}}=C^\perp_{L[4]}\delta_{m,\mp 1}, 
\nonumber\\
&&\int [d^2\vec{k}_{\perp,_\xi}][dx_\xi ]\psi^{0}_{L[4]}(x,k_\perp)x_j=C^{||}_{L[4]}\delta_{j,\hat{\delta}}, 
\nonumber\\
&&\int [d^2\vec{k}_{\perp,_\xi}][dx_\xi ]\psi^{0}_{L[31]_i}(x,k_\perp)x_j=C^{||}_{L[31]}\delta_{ij}
\eea
As we have shown in Sec.~\ref{sec:construct_P}, the light cone P wave for different representation can be expressed as
\bea
\psi^{+1}_{L[4]}(x,k_\perp)&=&(0.394653 +0.29994i) \psi_2(x,k_\perp)
\nonumber\\
&+&(-0.24334 + 0.931223i) \psi_3(x,k_\perp)\nonumber\\
\psi^{-1}_{L[4]}(x,k_\perp)&=&(0.641508 -0.618858i)  \psi_2(x,k_\perp)
\nonumber\\
&-&(0.334356+0.045252i) \psi_3(x,k_\perp),
\nonumber\\
\psi^{0}_{L[31]_\alpha}(x,k_\perp)&=&\psi_8(x,k_\perp),
\nonumber\\
\psi^{0}_{L[31]_\beta}(x,k_\perp)&=&(0.053i+0.526)\psi_6(x,k_\perp)
\nonumber\\
&+&(-0.580+0.620i)\psi_9(x,k_\perp) \nonumber\\
\psi^{0}_{L[31]_\gamma}(x,k_\perp)&=&(0.475-0.372i)\psi_5(x,k_\perp)
\nonumber\\
&+&(0.291i-0.742)\psi_{10}(x,k_\perp) \nonumber\\
\psi^{0}_{L[4]_\delta}(x,k_\perp)&=&(-0.521+0.352i)\psi_4(x,k_\perp)
\nonumber\\
&+&(0.725+0.280i)\psi_{11}(x,k_\perp) \nonumber\\
\eea
The subscript represent $L_z$ component. 
These wave functions are then coupled to the spin-flavor and color
blocks of the OSFC basis and projected onto the fully antisymmetric $[1111]$ irrep of $S_4$ in the
$q^4$ subsystem.  This produces the complete set of light-front $P$-wave pentaquark basis states
with definite $(I,S,J^P)$ used throughout the manuscript and makes explicit that the algebraic OSFC
construction and the projection-based implementation differ only by a unitary rotation
within degenerate symmetry sectors.

\section{Light-front $S$-wave three quark orbital states}
\label{sec:lf_swave_consolidated}
For the spin orbital interaction, the mixing between S wave three quark state and P wave penta state is generated by the transverse part, the light cone wave function of three quark state can be diagonalized by the full Hamiltonian,
\bea\label{eq:LFHamiJobN3}
H^N_{LF}&\equiv&\sum_{i=1}^3\frac{k^2_{i\perp}+m_Q^2}{x_i} + 3\sigma_Ta_N 
\nonumber\\
&-&\frac{\sigma_T}{a_N}\sum_{\xi=\alpha,\beta}((\partial/\partial x_\xi)^2+M^2 (\partial/\partial_{\vec{k}_{\xi\perp}})^2) 
\eea
To diagonalize the Hamiltonian, the longitudinal basis are obtained using Slater determinant with $N=3$.
\bea
\label{SLAT4}
\varphi_{\tilde{n}}[N=3]=\frac {1}{3^{3/4}}
\begin{vmatrix}
   1& 1    & 1\\
    e^{i\tilde  n_{21}\tilde s'_1}& e^{i\tilde  n_{21}\tilde s'_2}    & e^{i\tilde  n_{21}\tilde s'_3} \\
      e^{i\tilde  n_{31}\tilde s'_1}& e^{i\tilde  n_{31}\tilde s'_2}  & e^{i\tilde  n_{31}\tilde s'_3} 
\end{vmatrix}
\eea
with
\bea
\tilde s'_1&=&\frac{1}{6} \left(\sqrt{2} x_\alpha+\sqrt{6} x_\beta+2\right),
\nonumber\\
\tilde s'_2&=&-\frac{1}{3} \left(\sqrt{2} x_\alpha\right),
\nonumber\\
\tilde s'_2&=&\frac{1}{6} \left(\sqrt{2} x_\alpha-\sqrt{6} x_\beta-2\right),
\eea
The transverse basis for $\xi=\alpha,\beta$ are chosen to be generalized Laguerre
polynomials as given in Eq.~(\ref{eq:lagre}).
Since we are interested in the S wave three quark state, we set $m_\alpha=m_\beta=0$, and impose the truncation condition $2n_\alpha+2n_\beta+2\leq 10$. By minimizing the ground state mass, we find $a_N=4.35$. The ground state of light cone Hamiltonian defined in Eq.~(\ref{eq:LFHamiJobN3}) corresponds to the wave function $\psi_S(x,k_\perp)$ given in Eq.~(\ref{eq:trans}).

\section{Explicit wavefunctions}
\label{sec:full_pwavefunction}

\subsection{Nucleon OSFC}
The bare nucleon state $|N ⟩$ is a 
three-quark state in the S-wave,
\begin{equation}
    |N\rangle=L[3]SF_3[3]C_3[111]
\end{equation}
with
\bea
&&C_3[111]=\frac{1}{\sqrt{6}}(|R G B\rangle-|R B G\rangle\nonumber\\
&&+|G B R\rangle-|G R B\rangle+|B R G\rangle-|B G R\rangle)
\eea
and 
\begin{equation}
\begin{aligned}
 &SF_3[3]=\frac{1}{\sqrt{2}}(S_3[21]_\alpha F_3[21]_\alpha+S_3[21]_\beta F_3[21]_\beta)=\\
 &\frac{1}{\sqrt{18}}[2|u \uparrow u \uparrow d \downarrow\rangle-|u \uparrow u \downarrow d \uparrow\rangle-|u \downarrow u \uparrow d \uparrow\rangle \\
& \quad+2|u \uparrow d \downarrow u \uparrow\rangle-|u \uparrow d \uparrow u \downarrow\rangle-|u \downarrow d \uparrow u \uparrow\rangle \\
& \quad+2|d \downarrow u \uparrow u \uparrow\rangle-|d \uparrow u \uparrow u \downarrow\rangle-|d \uparrow u \downarrow u \uparrow\rangle]
\end{aligned}
\end{equation}
this is the wavefunction of proton and for neutron we just need replace $u$ with $d$.

\subsection{Pentastates OSFC}\label{sec:OSFC}
For the 4 quarks in the pentaquarkwavefunction the color partition must be C[211] and we choose the Young tableaux and wave function to be
\begin{widetext}
\begin{equation}
\begin{aligned}
C[211]_\alpha=\ytableaushort{1 3, 2, 4}\quad \ytableaushort{R C, G, B}\hspace{5em}
C[211]_\beta=\ytableaushort{1 2, 3, 4}\quad \ytableaushort{R C, G, B}\hspace{5em}
[211]_\gamma=\ytableaushort{1 4, 2, 3}\quad \ytableaushort{R C, G, B}
\end{aligned}
\end{equation}
the $C=R,G,B$ can be three color and the explicit form are
\begin{subequations}
\label{eq:colorR}
\begin{align}
C[211]_\alpha(R) &= \frac{1}{\sqrt{48}} \bigl( 3|R G R B\rangle - 3|G R R B\rangle - 3|R B R G\rangle + 3|B R R G\rangle  - |R G B R\rangle + |G R B R\rangle + |R B G R\rangle - |B R G R\rangle \nonumber \\
&\qquad + 2|G B R R\rangle - 2|B G R R\rangle \bigr) \\
C[211]_\beta(R)  &= \frac{1}{\sqrt{16}} \bigl( 2|R R G B\rangle - 2|R R B G\rangle - |R G R B\rangle - |G R R B\rangle  + |R B R G\rangle + |B R R G\rangle + |R G B R\rangle + |G R B R\rangle \nonumber \\
&\qquad - |R B G R\rangle - |B R G R\rangle \bigr) \\
C[211]_\gamma(R) &= \frac{1}{\sqrt{6}}  \bigl( |B R G R\rangle + |R G B R\rangle + |G B R R\rangle - |R B G R\rangle - |G R B R\rangle - |B G R R\rangle \bigr)
\end{align}
\end{subequations}
and $C=G,B$ will be
\begin{subequations}
\label{eq:colorG}
\begin{align}
C[211]_\alpha(G) &= \frac{1}{\sqrt{48}} \bigl( 3|R G G B\rangle - 3|G R G B\rangle - 3|B G G R\rangle + 3|G B G R\rangle  - |R G B G\rangle + |G R B G\rangle - |G B R G\rangle + |B G R G\rangle \nonumber \\
&\qquad + 2|B R G G\rangle - 2|R B G G\rangle \bigr) \\
C[211]_\beta(G)  &= \frac{1}{\sqrt{16}} \bigl( 2|G G B R\rangle - 2|G G R B\rangle + |R G G B\rangle + |G R G B\rangle - |R G B G\rangle - |G R B G\rangle + |G B R G\rangle + |B G R G\rangle \nonumber \\
&\qquad - |G B G R\rangle - |B G G R\rangle \bigr) \\
C[211]_\gamma(G) &= \frac{1}{\sqrt{6}}  \bigl( |R G B G\rangle - |G R B G\rangle - |R B G G\rangle + |B R G G\rangle  + |G B R G\rangle - |B G R G\rangle \bigr)
\end{align}
\end{subequations}
\begin{subequations}
\label{eq:colorB}
\begin{align}
C[211]_\alpha(B) &= \frac{1}{\sqrt{48}} \bigl( 3|B R B G\rangle - 3|R B B G\rangle + 3|G B B R\rangle - 3|B G B R\rangle  + |R B G B\rangle - |B R G B\rangle - |G B R B\rangle + |B G R B\rangle \nonumber \\
&\qquad + 2|R G B B\rangle - 2|G R B B\rangle \bigr) \\
C[211]_\beta(B)  &= \frac{1}{\sqrt{16}} \bigl( 2|B B R G\rangle - 2|B B G R\rangle + |R B G B\rangle + |B R G B\rangle  - |G B R B\rangle - |B G R B\rangle - |R B B G\rangle - |B R B G\rangle \nonumber \\
&\qquad + |G B B R\rangle + |B G B R\rangle \bigr) \\
C[211]_\gamma(B) &= \frac{1}{\sqrt{6}}  \bigl( |R G B B\rangle - |G R B B\rangle - |R B G B\rangle + |B R G B\rangle + |G B R B\rangle - |B G R B\rangle \bigr)
\end{align}
\end{subequations}
The five quark wave function is
\begin{equation}
C[211]_{\xi}=\frac{1}{\sqrt{3}}\bigl(C[211]_{\xi}(R) \bar{R}+C[211]_{\xi}(G) \bar{G}+C[211]_{\xi}(B) \bar{B}\bigr)
\end{equation}
where $\bar{C}=\bar{R},\bar{G},\bar{B}$ is the antiquark colors. The full OSFC wavefunction is then 
\begin{equation}
\Psi=\frac{1}{\sqrt{3}}\bigl[C[211]_\beta\left(LSF[31]\right)_\alpha-C[211]_\alpha\left(LSF[31]\right)_\beta+C[211]_\gamma\left(LSF[31]\right)_\gamma\bigr] 
\end{equation}
The $LSF[31]$ wavefunction can be generated by the group decomposition rule\eqref{App_4},
the S wave function for five quarks can be generated by recoupling through pertinent Clebsch-Gordon coefficients
\begin{equation}
S[4]\left[\frac{5}{2} \frac{5}{2}\right]=|\uparrow \uparrow \uparrow \uparrow \bar{\uparrow}\rangle
\end{equation}
and
\begin{equation}
S[4]\left[\frac{3}{2} \frac{3}{2}\right]=\sqrt{\frac{4}{5}}|\uparrow \uparrow \uparrow \uparrow \bar{\downarrow}\rangle-\sqrt{\frac{1}{20}}(|\uparrow \uparrow \uparrow \downarrow \bar{\uparrow}\rangle+|\uparrow \uparrow \downarrow \uparrow \bar{\uparrow}\rangle+|\uparrow \downarrow \uparrow \uparrow \bar{\uparrow}\rangle+|\downarrow \uparrow \uparrow \uparrow \bar{\uparrow}\rangle)
\end{equation}
and 
\begin{equation}
\begin{aligned}
S[22]_\alpha\left[\frac{1}{2} \frac{1}{2}\right] & =\sqrt{\frac{1}{4}}(|\uparrow \downarrow \uparrow \downarrow \bar{\uparrow}\rangle-|\downarrow \uparrow \uparrow \downarrow \bar{\uparrow}\rangle-|\uparrow \downarrow \downarrow \uparrow \bar{\uparrow}\rangle+|\downarrow \uparrow \downarrow \uparrow \bar{\uparrow}\rangle) \\
S[22]_\beta\left[\frac{1}{2} \frac{1}{2}\right] & =\sqrt{\frac{1}{12}}(2|\uparrow \uparrow \downarrow \downarrow \bar{\uparrow}\rangle-|\uparrow \downarrow \uparrow \downarrow \bar{\uparrow}\rangle-|\downarrow \uparrow \uparrow \downarrow \bar{\uparrow}\rangle-|\uparrow \downarrow \downarrow \uparrow \bar{\uparrow}\rangle-|\downarrow \uparrow \downarrow \uparrow \bar{\uparrow}\rangle+2|\downarrow \downarrow \uparrow \uparrow \bar{\uparrow}\rangle)
\end{aligned}
\end{equation}
and 
    \begin{equation}
\begin{aligned}
S[31]_\alpha\left[\frac{3}{2} \frac{3}{2}\right] & =\sqrt{\frac{1}{2}}(|\uparrow \downarrow \uparrow \uparrow \bar{\uparrow}\rangle-|\downarrow \uparrow \uparrow \uparrow \bar{\uparrow}\rangle) \\
S[31]_\beta\left[\frac{3}{2} \frac{3}{2}\right] & =\sqrt{\frac{1}{6}}(2|\uparrow \uparrow \downarrow \uparrow \bar{\uparrow}\rangle-|\uparrow \downarrow \uparrow \uparrow \bar{\uparrow}\rangle-|\downarrow \uparrow \uparrow \uparrow \bar{\uparrow}\rangle) \\
S[31]_\gamma\left[\frac{3}{2} \frac{3}{2}\right] & =\sqrt{\frac{1}{12}}(3|\uparrow \uparrow \uparrow \downarrow \bar{\uparrow}\rangle-|\uparrow \uparrow \downarrow \uparrow \bar{\uparrow}\rangle-|\uparrow \downarrow \uparrow \uparrow \bar{\uparrow}\rangle-|\downarrow \uparrow \uparrow \uparrow \bar{\uparrow}\rangle)
\end{aligned}
\end{equation}
and
    \begin{equation}
\begin{aligned}
& S[31]_\alpha\left[\frac{1}{2} \frac{1}{2}\right]=\sqrt{\frac{1}{3}}(|\uparrow \downarrow \uparrow \uparrow \bar{\downarrow}\rangle-|\downarrow \uparrow \uparrow \uparrow \bar{\downarrow}\rangle)-\sqrt{\frac{1}{12}}(|\uparrow \downarrow \uparrow \downarrow \bar{\uparrow}\rangle+|\uparrow \downarrow \downarrow \uparrow \bar{\uparrow}\rangle-|\downarrow \uparrow \uparrow \downarrow \bar{\uparrow}\rangle-|\downarrow \uparrow \downarrow \uparrow \bar{\uparrow}\rangle) \\
& S[31]_\beta\left[\frac{1}{2} \frac{1}{2}\right]=\frac{1}{3}(2|\uparrow \uparrow \downarrow \uparrow \bar{\downarrow}\rangle-|\uparrow \downarrow \uparrow \uparrow \bar{\downarrow}\rangle-|\downarrow \uparrow \uparrow \uparrow \bar{\downarrow}\rangle) \\
& \left.-\frac{1}{6}(2|\uparrow \uparrow \downarrow \downarrow \bar{\uparrow}\rangle-|\uparrow \downarrow \uparrow \downarrow \bar{\uparrow}\rangle-|\downarrow \uparrow \uparrow \downarrow \bar{\uparrow}\rangle+|\uparrow \downarrow \downarrow \uparrow \bar{\uparrow}\rangle+|\downarrow \uparrow \downarrow \uparrow \bar{\uparrow}\rangle-2| \downarrow \downarrow \uparrow \uparrow \bar{\uparrow}\rangle\right) \\
& S[31]_\gamma\left[\frac{1}{2} \frac{1}{2}\right]=\sqrt{\frac{1}{18}}(3|\uparrow \downarrow \uparrow \uparrow \bar{\downarrow}\rangle-|\downarrow \uparrow \uparrow \uparrow \bar{\downarrow}\rangle-|\uparrow \uparrow \downarrow \uparrow \bar{\downarrow}\rangle-|\uparrow \uparrow \uparrow \downarrow \bar{\downarrow}\rangle) \\
& -\sqrt{\frac{1}{18}}(|\uparrow \uparrow \downarrow \downarrow \bar{\uparrow}\rangle+|\uparrow \downarrow \uparrow \downarrow \bar{\uparrow}\rangle+|\downarrow \uparrow \uparrow \downarrow \bar{\uparrow}\rangle-|\uparrow \downarrow \downarrow \uparrow \bar{\uparrow}\rangle-|\downarrow \uparrow \downarrow \uparrow \bar{\uparrow}\rangle-|\downarrow \downarrow \uparrow \uparrow \bar{\uparrow}\rangle)
\end{aligned}
\end{equation}

Note that the anti-quark spin satisfies $\overline{\uparrow}=-i \sigma^2 \uparrow=\downarrow$ and $\overline{\downarrow}=-i \sigma^2 \downarrow=-\uparrow$, and the flavor function follow the similar rules, we just need to replace the $\uparrow \rightarrow u, \downarrow \rightarrow d$ and $\overline{\uparrow} \rightarrow \overline{d}, \overline{\downarrow} \rightarrow \overline{u}$, the isospin of antiquark is $I_z(\overline{u})=-\frac{1}{2}, I_z(\overline{d})=+\frac{1}{2}$
for example, the full pentaquark wavefunction for isospin $1/2$ and spin $5/2$ can be written as 
\begin{align}
\varphi_{S[4]F[31]}^{L[4]SF[31]}\left[\tfrac{1}{2}\tfrac{1}{2},\tfrac{5}{2}\tfrac{5}{2}\right]
&= \frac{1}{\sqrt{3}}L[4]S[4] \bigl( C[211]_\beta F[31]_\alpha-C[211]_\alpha F[31]_\beta+C[211]_\gamma F[31]_\gamma \bigr) \nonumber\\
&=\frac{1}{\sqrt{3}}L[4]|\uparrow \uparrow \uparrow \uparrow \overline{\uparrow}\rangle \Bigl( C[211]_\beta (\sqrt{\frac{1}{3}}(|uduu \bar{u}\rangle-|duuu \bar{u}\rangle)-\sqrt{\frac{1}{12}}(|udud  \bar{d}\rangle+|uddu\bar{d}\rangle-|duud \bar{d}\rangle-|dudu \bar{d}\rangle)) \nonumber\\
&-C[211]_\alpha (
\frac{1}{3}(2|uudu \bar{u}\rangle-|uduu \bar{u}\rangle-|duuu \bar{u}\rangle)-\frac{1}{6}(2|uudd \bar{d}\rangle-|udud \bar{d}\rangle-|duud \bar{d}\rangle+|uddu \bar{d}\rangle \nonumber\\
&+|dudu \bar{d}\rangle-2| dduu \bar{d}\rangle)  +C[211]_\gamma (\sqrt{\frac{1}{18}}(3|uduu \bar{u}\rangle-|duuu \bar{u}\rangle-|uudu \bar{u}\rangle-|uuud \bar{u}\rangle)  \nonumber\\
& -\sqrt{\frac{1}{18}}(|uudd \bar{d}\rangle+|udud \bar{d}\rangle+|duud \bar{d}\rangle-|uddu \bar{d}\rangle-|dudu \bar{d}\rangle-|dduu \bar{d}\rangle))) \Bigr) \nonumber\\
\varphi_{S[4]F[31]}^{L[31]SF[31]}\left[\tfrac{1}{2}\tfrac{1}{2},\tfrac{5}{2}\tfrac{5}{2}\right]
&= \frac{1}{\sqrt{3}}S[4] \Bigl( C[211]_\beta \bigl[ \tfrac{1}{\sqrt{3}}(L_5[31]_\alpha F[31]_\beta+\alpha \leftrightarrow \beta)  +\tfrac{1}{\sqrt{6}}(L[31]_\alpha F[31]_\gamma+\alpha \leftrightarrow \gamma) \bigr]  \nonumber\\
&\quad -C[211]_\alpha \bigl[ \tfrac{1}{\sqrt{3}}(L[31]_\alpha F[31]_\alpha-\alpha \rightarrow \beta) +\tfrac{1}{\sqrt{6}}(L[31]_\gamma F[31]_\beta+\gamma \leftrightarrow \beta) \bigr]  \nonumber\\
&\quad +C[211]_\gamma \bigl[ \tfrac{1}{\sqrt{6}}(L[31]_\alpha F[31]_\alpha+\alpha \rightarrow \beta) -\tfrac{2}{\sqrt{6}} L[31]_\gamma F[31]_\gamma \bigr] \Bigr) \\
\varphi_{S[4]F[22]}^{L[31]SF[22]}\left[\tfrac{1}{2}\tfrac{1}{2},\tfrac{5}{2}\tfrac{5}{2}\right]
&= \frac{1}{\sqrt{3}}S[4] \Bigl( C[211]_\beta \bigl[\tfrac{1}{2}\bigl(F[22]_\alpha L[31]_\beta+\alpha \leftrightarrow \beta\bigr)  -\tfrac{1}{\sqrt{2}} F[22]_\alpha L[31]_\gamma\bigl]  \nonumber\\
&\quad -C[211]_\alpha \bigl[\tfrac{1}{2}\bigl(F[22]_\alpha L[31]_\alpha-\alpha \rightarrow \beta\bigr)  -\tfrac{1}{\sqrt{2}} F[22]_\beta L[31]_\gamma\bigl]  \nonumber\\
&\quad -C[211]_\gamma \bigl[\tfrac{1}{\sqrt{2}}\bigl(F[22]_\alpha L[31]_\alpha+\alpha \rightarrow \beta\bigr] \Bigr)
\end{align}
We only show the maximum weight representations, the state with lower $S_z$ component can be obtained by applying the spin lowing operator.
The wavefunction for isospin $1/2$ and spin $3/2$ can be written as 
\begin{align}
\varphi_{S[31]F[31]}^{L[4]SF[31]_a}\left[\tfrac{1}{2}\tfrac{1}{2},\tfrac{3}{2}\tfrac{3}{2}\right]
&= \frac{1}{\sqrt{3}}L[4] \Bigl( C[211]_\beta\bigl[ \frac{1}{\sqrt{3}}\left(S[31]_\alpha F[31]_\beta+\alpha \leftrightarrow \beta\right)+\frac{1}{\sqrt{6}}\left(S[31]_\alpha F[31]_\gamma+\alpha \leftrightarrow \gamma\right)\bigr] \nonumber\\
&\quad -C[211]_\alpha \bigl[\frac{1}{\sqrt{3}}\left(S[31]_\alpha F[31]_\alpha-\alpha \to \beta\right)+\frac{1}{\sqrt{6}}\left(S[31]_\gamma F[31]_\beta+\gamma \leftrightarrow \beta\right) \bigr] \nonumber\\
&\quad +C[211]_\gamma \bigl[\frac{1}{\sqrt{6}}\left(S[31]_\alpha F[31]_\alpha+\alpha \to \beta\right)-\frac{2}{\sqrt{6}} S[31]_\gamma F[31]_\gamma \bigr]\Bigr)\\
\varphi_{S[31]F[22]}^{L[4]SF[31]_b}\left[\tfrac{1}{2} \tfrac{1}{2}, \tfrac{3}{2} \tfrac{3}{2}\right]
&=\frac{1}{\sqrt{3}}L[4]\left(C[211]_\beta \bigl[\frac{1}{2}\left(F[22]_\alpha S[31]_\beta+\alpha \leftrightarrow \beta\right)-\frac{1}{\sqrt{2}} F[22]_\alpha S[31]_\gamma\bigr]\right.  \nonumber\\
& \quad-C[211]_\alpha\bigl[ \frac{1}{2}\left(F[22]_\alpha S[31]_\alpha-\alpha \rightarrow \beta\right)-\frac{1}{\sqrt{2}} F[22]_\beta S[31]_\gamma\bigr]  \nonumber\\
& \left.\quad-C[211]_\gamma\bigl[ \frac{1}{\sqrt{2}}\left(F[22]_\alpha S[31]_\alpha+\alpha \rightarrow \beta\right)\bigr]\right)\\
\varphi_{S[4]F[31]}^{L[4]SF[31]_c}\left[\tfrac{1}{2} \tfrac{1}{2}, \tfrac{3}{2} \tfrac{3}{2}\right]
&=\frac{1}{\sqrt{3}}L[4]S[4]\left(C[211]_\beta F[31]_\alpha-C[211]_\alpha F[31]_\beta +C[211]_\gamma F[31]_\gamma\right)\\
\varphi_{S[31]F[31]}^{L[31]SF[31]_a}\left[\tfrac{1}{2} \tfrac{1}{2}, \tfrac{3}{2} \tfrac{3}{2}\right]
&=\frac{1}{\sqrt{3}}\left(C[211]_\beta \bigl[ \frac{1}{\sqrt{3}}\left(L[31]_\alpha SF[31]_\beta+\alpha \leftrightarrow \beta\right)+\frac{1}{\sqrt{6}}\left(L[31]_\alpha SF[31]_\gamma+\alpha \leftrightarrow \gamma\right)\bigr]\right. \nonumber\\
& \quad-C[211]_\alpha\bigl[\frac{1}{\sqrt{3}}\left(L[31]_\alpha SF[31]_\alpha-\alpha \to \beta\right)+\frac{1}{\sqrt{6}}\left(L[31]_\gamma SF[31]_\beta+\gamma \leftrightarrow \beta\right) \bigr] \nonumber\\
& \left.\quad+C[211]_\gamma\bigl[\frac{1}{\sqrt{6}}\left(L[31]_\alpha SF[31]_\alpha+\alpha \to \beta\right)-\frac{2}{\sqrt{6}} L[31]_\gamma SF[31]_\gamma \bigr]\right) \\
&\left\{\begin{aligned}  
&SF[31]_\alpha=\bigl[ \frac{1}{\sqrt{3}}\left(S[31]_\alpha F[31]_\beta+\alpha \leftrightarrow \beta\right)+\frac{1}{\sqrt{6}}\left(S[31]_\alpha F[31]_\gamma+\alpha \leftrightarrow \gamma\right)\bigr]\nonumber\\
&SF[31]_\beta=\bigl[\frac{1}{\sqrt{3}}\left(S[31]_\alpha F[31]_\alpha-\alpha \to \beta\right)+\frac{1}{\sqrt{6}}\left(S[31]_\gamma F[31]_\beta+\gamma \leftrightarrow \beta\right) \bigr]\\
&SF[31]_\gamma=\bigl[\frac{1}{\sqrt{6}}\left(S[31]_\alpha F[31]_\alpha+\alpha \to \beta\right)-\frac{2}{\sqrt{6}} S[31]_\gamma F[31]_\gamma \bigr]
\end{aligned}
\right.\\
\varphi_{S[31]F[22]}^{L[31]SF[31]_b}\left[\tfrac{1}{2} \tfrac{1}{2}, \tfrac{3}{2} \tfrac{3}{2}\right]
&=\frac{1}{\sqrt{3}}\left(C[211]_\beta \bigl[ \frac{1}{\sqrt{3}}\left(L[31]_\alpha SF[31]_\beta+\alpha \leftrightarrow \beta\right)+\frac{1}{\sqrt{6}}\left(L[31]_\alpha SF[31]_\gamma+\alpha \leftrightarrow \gamma\right)\bigr]\right. \nonumber\\
& \quad-C[211]_\alpha\bigl[\frac{1}{\sqrt{3}}\left(L[31]_\alpha SF[31]_\alpha-\alpha \to \beta\right)+\frac{1}{\sqrt{6}}\left(L[31]_\gamma SF[31]_\beta+\gamma \leftrightarrow \beta\right) \bigr] \nonumber\\
& \left.\quad+C[211]_\gamma\bigl[\frac{1}{\sqrt{6}}\left(L[31]_\alpha SF[31]_\alpha+\alpha \to \beta\right)-\frac{2}{\sqrt{6}} L[31]_\gamma SF[31]_\gamma \bigr]\right)\\
&\left\{\begin{aligned}  
&SF[31]_\alpha=\bigl[\frac{1}{2}\left(F[22]_\alpha S[31]_\beta+\alpha \leftrightarrow \beta\right)-\frac{1}{\sqrt{2}} F[22]_\alpha S[31]_\gamma\bigr]\nonumber\\
&SF[31]_\beta=\bigl[ \frac{1}{2}\left(F[22]_\alpha S[31]_\alpha-\alpha \rightarrow \beta\right)-\frac{1}{\sqrt{2}} F[22]_\beta S[31]_\gamma\bigr]\nonumber\\
&SF[31]_\gamma=-\bigl[ \frac{1}{\sqrt{2}}\left(F[22]_\alpha S[31]_\alpha+\alpha \rightarrow \beta\right)\bigr]
\end{aligned}
\right.\\
\varphi_{S[4]F[31]}^{L[31]SF[31]_c}\left[\tfrac{1}{2} \tfrac{1}{2}, \tfrac{3}{2} \tfrac{3}{2}\right]
&=\frac{1}{\sqrt{3}}S[4]\left(C[211]_\beta \bigl[ \frac{1}{\sqrt{3}}\left(L[31]_\alpha F[31]_\beta+\alpha \leftrightarrow \beta\right)+\frac{1}{\sqrt{6}}\left(L[31]_\alpha F[31]_\gamma+\alpha \leftrightarrow \gamma\right)\bigr]\right. \nonumber\\
& \quad-C[211]_\alpha\bigl[\frac{1}{\sqrt{3}}\left(L[31]_\alpha F[31]_\alpha-\alpha \to \beta\right)+\frac{1}{\sqrt{6}}\left(L[31]_\gamma F[31]_\beta+\gamma \leftrightarrow \beta\right) \bigr] \nonumber\\
& \left.\quad+C[211]_\gamma\bigl[\frac{1}{\sqrt{6}}\left(L[31]_\alpha F[31]_\alpha+\alpha \to \beta\right)-\frac{2}{\sqrt{6}} L[31]_\gamma F[31]_\gamma \bigr]\right)\\
\varphi_{S[31]F[31]}^{L[31]SF[22]_a}\left[\tfrac{1}{2} \tfrac{1}{2}, \tfrac{3}{2} \tfrac{3}{2}\right]
&=\frac{1}{\sqrt{3}}\left(C[211]_\beta \bigl[\frac{1}{2}\left(SF[22]_\alpha L[31]_\beta+\alpha \leftrightarrow \beta\right)-\frac{1}{\sqrt{2}} SF[22]_\alpha L[31]_\gamma\bigr]\right. \nonumber\\
& \quad-C[211]_\alpha\bigl[ \frac{1}{2}\left(SF[22]_\alpha L[31]_\alpha-\alpha \rightarrow \beta\right)-\frac{1}{\sqrt{2}} SF[22]_\beta L[31]_\gamma\bigr] \nonumber\\
& \left.\quad-C[211]_\gamma\bigl[ \frac{1}{\sqrt{2}}\left(SF[22]_\alpha L[31]_\alpha+\alpha \rightarrow \beta\right)\bigr]\right)\\
&\left\{\begin{aligned}  
&SF[22]_\alpha=\bigl[\frac{1}{\sqrt{6}}\left(S[31]_\alpha F[31]_\beta+\alpha \leftrightarrow \beta\right)-\frac{1}{\sqrt{3}}\left(S[31]_\alpha F[31]_\gamma+\alpha \leftrightarrow \gamma\right)\bigr]\nonumber\\
&SF[22]_\beta=\bigl[ \frac{1}{\sqrt{6}}\left(S[31]_\alpha F[31]_\alpha-\alpha \leftrightarrow \beta\right)-\frac{1}{\sqrt{3}}\left(S[31]_\beta F[31]_\gamma+\beta \leftrightarrow \gamma\right)\bigr]
\end{aligned}
\right.\\
\varphi_{S[4]F[22]}^{L[31]SF[22]_b}\left[\tfrac{1}{2} \tfrac{1}{2}, \tfrac{3}{2} \tfrac{3}{2}\right]
&=\frac{1}{\sqrt{3}}S[4]\left(C[211]_\beta \bigl[\frac{1}{2}\left(F[22]_\alpha L[31]_\beta+\alpha \leftrightarrow \beta\right)-\frac{1}{\sqrt{2}} F[22]_\alpha L[31]_\gamma\bigr]\right. \nonumber\\
& \quad-C[211]_\alpha\bigl[ \frac{1}{2}\left(F[22]_\alpha L[31]_\alpha-\alpha \rightarrow \beta\right)-\frac{1}{\sqrt{2}} F[22]_\beta L[31]_\gamma\bigr] \nonumber\\
& \left.\quad-C[211]_\gamma\bigl[ \frac{1}{\sqrt{2}}\left(F[22]_\alpha L[31]_\alpha+\alpha \rightarrow \beta\right)\bigr]\right)\nonumber\\
\varphi_{S[31]F[31]}^{L[31]SF[4]}\left[\tfrac{1}{2} \tfrac{1}{2}, \tfrac{3}{2} \tfrac{3}{2}\right]
&=\frac{1}{3}\left(C[211]_\beta L[31]_\alpha-C[211]_\alpha L[31]_\beta+C[211]_\gamma L[31]_\gamma\right)\nonumber\\
&\left(S[31]_\alpha F[31]_\alpha+S[31]_\beta F[31]_\beta+S[31]_\gamma F[31]_\gamma\right)\\
\varphi_{S[31]F[31]}^{L[31]SF[211]_a}\left[\tfrac{1}{2} \tfrac{1}{2}, \tfrac{3}{2} \tfrac{3}{2}\right]
&=\frac{1}{\sqrt{3}}\left(C[211]_\beta \frac{1}{\sqrt{2}}\bigl[L[31]_\gamma SF[211]_\alpha+L[31]_\beta SF[211]_\gamma\bigr]\right. \nonumber\\
& \quad-C[211]_\alpha\frac{1}{\sqrt{2}}\bigl[L[31]_\gamma SF[211]_\beta-L[31]_\alpha SF[211]_\gamma\bigr]\nonumber\\
& \left.\quad-C[211]_\gamma\frac{1}{\sqrt{2}}\bigl[L[31]_\alpha SF[211]_\alpha+L[31]_\beta SF[211]_\beta\bigr]\right)\\
&\left\{\begin{aligned}  
&SF[211]_\alpha=\frac{1}{\sqrt{2}}\bigl[S[31]_\alpha F[31]_\gamma-S[31]_\gamma F[31]_\alpha\bigr]\nonumber\\
&SF[211]_\beta=\frac{1}{\sqrt{2}}\bigl[S[31]_\beta F[31]_\gamma-S[31]_\gamma F[31]_\beta\bigr]\nonumber\\
&SF[211]_\gamma=\frac{1}{\sqrt{2}}\bigl[S[31]_\alpha F[31]_\beta-S[31]_\beta F[31]_\alpha\bigr]
\end{aligned}
\right.\\
\varphi_{S[31]F[22]}^{L[31]SF[211]_b}\left[\tfrac{1}{2} \tfrac{1}{2}, \tfrac{3}{2} \tfrac{3}{2}\right]
&=\frac{1}{\sqrt{3}}\left(C[211]_\beta \frac{1}{\sqrt{2}}\bigl[L[31]_\gamma SF[211]_\alpha+L[31]_\beta SF[211]_\gamma\bigr]\right. \nonumber\\
& \quad-C[211]_\alpha\frac{1}{\sqrt{2}}\bigl[L[31]_\gamma SF[211]_\beta-L[31]_\alpha SF[211]_\gamma\bigr]\nonumber\\
& \left.\quad-C[211]_\gamma\frac{1}{\sqrt{2}}\bigl[L[31]_\alpha SF[211]_\alpha+L[31]_\beta SF[211]_\beta\bigr]\right)\\
&\left\{\begin{aligned}  
&SF[211]_\alpha=\frac{1}{2}\bigl[S[31]_\alpha F[22]_\beta+S[31]_\beta F[22]_\alpha+\sqrt{2} S[31]_\gamma F[22]_\alpha\bigr]\nonumber\\
&SF[211]_\beta=\frac{1}{2}\bigl[S[31]_\alpha F[22]_\alpha-S[31]_\beta F[22]_\beta+\sqrt{2} S[31]_\gamma F[22]_\beta\bigr]\nonumber\\
&SF[211]_\gamma=\frac{1}{\sqrt{2}}\bigl[S[31]_\alpha F[22]_\beta-S[31]_\beta F[22]_\alpha\bigr]
\end{aligned}
\right.
\end{align}
The wavefunction for isospin $1/2$ and spin $1/2$ can be written as 
\begin{align}
\varphi_{S[31]F[31]}^{L[4]SF[31]_a}\left[\tfrac{1}{2} \tfrac{1}{2}, \tfrac{1}{2} \tfrac{1}{2}\right]
&=\frac{1}{\sqrt{3}}L[4]\left(C[211]_\beta \bigl[ \frac{1}{\sqrt{3}}\left(S[31]_\alpha F[31]_\beta+\alpha \leftrightarrow \beta\right)+\frac{1}{\sqrt{6}}\left(S[31]_\alpha F[31]_\gamma+\alpha \leftrightarrow \gamma\right)\bigr]\right. \nonumber\\
& \quad-C[211]_\alpha\bigl[\frac{1}{\sqrt{3}}\left(S[31]_\alpha F[31]_\alpha-\alpha \to \beta\right)+\frac{1}{\sqrt{6}}\left(S[31]_\gamma F[31]_\beta+\gamma \leftrightarrow \beta\right) \bigr]\nonumber\\
& \left.\quad+C[211]_\gamma\bigl[\frac{1}{\sqrt{6}}\left(S[31]_\alpha F[31]_\alpha+\alpha \to \beta\right)-\frac{2}{\sqrt{6}} S[31]_\gamma F[31]_\gamma \bigr]\right)\\
\varphi_{S[31]F[22]}^{L[4]SF[31]_b}\left[\tfrac{1}{2} \tfrac{1}{2}, \tfrac{1}{2} \tfrac{1}{2}\right]
&=\frac{1}{\sqrt{3}}L[4]\left(C[211]_\beta\bigl[\frac{1}{2}\left(F[22]_\alpha S[31]_\beta+\alpha \leftrightarrow \beta\right)-\frac{1}{\sqrt{2}} F[22]_\alpha S[31]_\gamma\bigr]\right. \nonumber\\
& \quad-C[211]_\alpha\bigl[ \frac{1}{2}\left(F[22]_\alpha S[31]_\alpha-\alpha \rightarrow \beta\right)-\frac{1}{\sqrt{2}} F[22]_\beta S[31]_\gamma\bigr]\nonumber\\
& \left.\quad-C[211]_\gamma\bigl[ \frac{1}{\sqrt{2}}\left(F[22]_\alpha S[31]_\alpha+\alpha \rightarrow \beta\right)\bigr]\right)\\
\varphi_{S[22]F[31]}^{L[4]SF[31]_c}\left[\tfrac{1}{2} \tfrac{1}{2}, \tfrac{1}{2} \tfrac{1}{2}\right]
&=\frac{1}{\sqrt{3}}L[4]\left(C[211]_\beta\bigl[\frac{1}{2}\left(S[22]_\alpha F[31]_\beta+\alpha \leftrightarrow \beta\right)-\frac{1}{\sqrt{2}} S[22]_\alpha F[31]_\gamma\bigr]\right. \nonumber\\
& \quad-C[211]_\alpha\bigl[ \frac{1}{2}\left(S[22]_\alpha F[31]_\alpha-\alpha \rightarrow \beta\right)-\frac{1}{\sqrt{2}} S[22]_\beta F[31]_\gamma\bigr]\nonumber\\
& \left.\quad-C[211]_\gamma\bigl[ \frac{1}{\sqrt{2}}\left(S[22]_\alpha F[31]_\alpha+\alpha \rightarrow \beta\right)\bigr]\right)\\
\varphi_{S[31]F[31]}^{L[31]SF[31]_a}\left[\tfrac{1}{2} \tfrac{1}{2}, \tfrac{1}{2} \tfrac{1}{2}\right]
&=\frac{1}{\sqrt{3}}\left(C[211]_\beta \bigl[ \frac{1}{\sqrt{3}}\left(L[31]_\alpha SF[31]_\beta+\alpha \leftrightarrow \beta\right)+\frac{1}{\sqrt{6}}\left(L[31]_\alpha SF[31]_\gamma+\alpha \leftrightarrow \gamma\right)\bigr]\right. \nonumber\\
& \quad-C[211]_\alpha\bigl[\frac{1}{\sqrt{3}}\left(L[31]_\alpha SF[31]_\alpha-\alpha \to \beta\right)+\frac{1}{\sqrt{6}}\left(L[31]_\gamma SF[31]_\beta+\gamma \leftrightarrow \beta\right) \bigr] \nonumber\\
& \left.\quad+C[211]_\gamma\bigl[\frac{1}{\sqrt{6}}\left(L[31]_\alpha SF[31]_\alpha+\alpha \to \beta\right)-\frac{2}{\sqrt{6}} L[31]_\gamma SF[31]_\gamma \bigr]\right)\\
&\left\{\begin{aligned}  
&SF[31]_\alpha=\bigl[ \frac{1}{\sqrt{3}}\left(S[31]_\alpha F[31]_\beta+\alpha \leftrightarrow \beta\right)+\frac{1}{\sqrt{6}}\left(S[31]_\alpha F[31]_\gamma+\alpha \leftrightarrow \gamma\right)\bigr]\nonumber\\
&SF[31]_\beta=\bigl[\frac{1}{\sqrt{3}}\left(S[31]_\alpha F[31]_\alpha-\alpha \to \beta\right)+\frac{1}{\sqrt{6}}\left(S[31]_\gamma F[31]_\beta+\gamma \leftrightarrow \beta\right) \bigr]\nonumber\\
&SF[31]_\gamma=\bigl[\frac{1}{\sqrt{6}}\left(S[31]_\alpha F[31]_\alpha+\alpha \to \beta\right)-\frac{2}{\sqrt{6}} S[31]_\gamma F[31]_\gamma \bigr]
\end{aligned}
\right.   \\
\varphi_{S[31]F[22]}^{L[31]SF[31]_b}\left[\tfrac{1}{2} \tfrac{1}{2}, \tfrac{1}{2} \tfrac{1}{2}\right]
&=\frac{1}{\sqrt{3}}\left(C[211]_\beta \bigl[ \frac{1}{\sqrt{3}}\left(L[31]_\alpha SF[31]_\beta+\alpha \leftrightarrow \beta\right)+\frac{1}{\sqrt{6}}\left(L[31]_\alpha SF[31]_\gamma+\alpha \leftrightarrow \gamma\right)\bigr]\right. \nonumber\\
& \quad-C[211]_\alpha\bigl[\frac{1}{\sqrt{3}}\left(L[31]_\alpha SF[31]_\alpha-\alpha \to \beta\right)+\frac{1}{\sqrt{6}}\left(L[31]_\gamma SF[31]_\beta+\gamma \leftrightarrow \beta\right) \bigr] \nonumber\\
& \left.\quad+C[211]_\gamma\bigl[\frac{1}{\sqrt{6}}\left(L[31]_\alpha SF[31]_\alpha+\alpha \to \beta\right)-\frac{2}{\sqrt{6}} L[31]_\gamma SF[31]_\gamma \bigr]\right)\\
&\left\{\begin{aligned}  
&SF[31]_\alpha=\bigl[\frac{1}{2}\left(F[22]_\alpha S[31]_\beta+\alpha \leftrightarrow \beta\right)-\frac{1}{\sqrt{2}} F[22]_\alpha S[31]_\gamma\bigr]\nonumber\\
&SF[31]_\beta=\bigl[ \frac{1}{2}\left(F[22]_\alpha S[31]_\alpha-\alpha \rightarrow \beta\right)-\frac{1}{\sqrt{2}} F[22]_\beta S[31]_\gamma\bigr]\nonumber\\
&SF[31]_\gamma=-\bigl[ \frac{1}{\sqrt{2}}\left(F[22]_\alpha S[31]_\alpha+\alpha \rightarrow \beta\right)\bigr]
\end{aligned}
\right.\\
\varphi_{S[22]F[31]}^{L[31]SF[31]_c}\left[\tfrac{1}{2} \tfrac{1}{2}, \tfrac{1}{2} \tfrac{1}{2}\right]
&=\frac{1}{\sqrt{3}}\left(C[211]_\beta \bigl[ \frac{1}{\sqrt{3}}\left(L[31]_\alpha SF[31]_\beta+\alpha \leftrightarrow \beta\right)+\frac{1}{\sqrt{6}}\left(L[31]_\alpha SF[31]_\gamma+\alpha \leftrightarrow \gamma\right)\bigr]\right. \nonumber\\
& \quad-C[211]_\alpha\bigl[\frac{1}{\sqrt{3}}\left(L[31]_\alpha SF[31]_\alpha-\alpha \to \beta\right)+\frac{1}{\sqrt{6}}\left(L[31]_\gamma SF[31]_\beta+\gamma \leftrightarrow \beta\right) \bigr] \nonumber\\
& \left.\quad+C[211]_\gamma\bigl[\frac{1}{\sqrt{6}}\left(L[31]_\alpha SF[31]_\alpha+\alpha \to \beta\right)-\frac{2}{\sqrt{6}} L[31]_\gamma SF[31]_\gamma \bigr]\right)\\
&\left\{\begin{aligned}  
&SF[31]_\alpha=\bigl[\frac{1}{2}\left(S[22]_\alpha F[31]_\beta+\alpha \leftrightarrow \beta\right)-\frac{1}{\sqrt{2}} S[22]_\alpha F[31]_\gamma\bigr]\nonumber\\
&SF[31]_\beta=\bigl[ \frac{1}{2}\left(S[22]_\alpha F[31]_\alpha-\alpha \rightarrow \beta\right)-\frac{1}{\sqrt{2}} S[22]_\beta F[31]_\gamma\bigr]\nonumber\\
&SF[31]_\gamma=-\bigl[ \frac{1}{\sqrt{2}}\left(S[22]_\alpha F[31]_\alpha+\alpha \rightarrow \beta\right)\bigr]
\end{aligned}
\right.\\
\varphi_{S[31]F[31]}^{L[31]SF[22]_a}\left[\tfrac{1}{2} \tfrac{1}{2}, \tfrac{1}{2} \tfrac{1}{2}\right]
&=\frac{1}{\sqrt{3}}\left(C[211]_\beta \bigl[\frac{1}{2}\left(SF[22]_\alpha L[31]_\beta+\alpha \leftrightarrow \beta\right)-\frac{1}{\sqrt{2}} SF[22]_\alpha L[31]_\gamma\bigr]\right. \nonumber\\
& \quad-C[211]_\alpha\bigl[ \frac{1}{2}\left(SF[22]_\alpha L[31]_\alpha-\alpha \rightarrow \beta\right)-\frac{1}{\sqrt{2}} SF[22]_\beta L[31]_\gamma\bigr] \nonumber\\
& \left.\quad-C[211]_\gamma\bigl[ \frac{1}{\sqrt{2}}\left(SF[22]_\alpha L[31]_\alpha+\alpha \rightarrow \beta\right)\bigr]\right)\\
&\left\{\begin{aligned}  
&SF[22]_\alpha=\bigl[\frac{1}{\sqrt{6}}\left(S[31]_\alpha F[31]_\beta+\alpha \leftrightarrow \beta\right)-\frac{1}{\sqrt{3}}\left(S[31]_\alpha F[31]_\gamma+\alpha \leftrightarrow \gamma\right)\bigr]\nonumber\\
&SF[22]_\beta=\bigl[ \frac{1}{\sqrt{6}}\left(S[31]_\alpha F[31]_\alpha-\alpha \leftrightarrow \beta\right)-\frac{1}{\sqrt{3}}\left(S[31]_\beta F[31]_\gamma+\beta \leftrightarrow \gamma\right)\bigr]
\end{aligned}
\right.\\
\varphi_{S[22]F[22]}^{L[31]SF[22]_b}\left[\tfrac{1}{2} \tfrac{1}{2}, \tfrac{1}{2} \tfrac{1}{2}\right]
&=\frac{1}{\sqrt{3}}\left(C[211]_\beta \bigl[\frac{1}{2}\left(SF[22]_\alpha L[31]_\beta+\alpha \leftrightarrow \beta\right)-\frac{1}{\sqrt{2}} SF[22]_\alpha L[31]_\gamma\bigr]\right. \nonumber\\
& \quad-C[211]_\alpha\bigl[ \frac{1}{2}\left(SF[22]_\alpha L[31]_\alpha-\alpha \rightarrow \beta\right)-\frac{1}{\sqrt{2}} SF[22]_\beta L[31]_\gamma\bigr] \nonumber\\
& \left.\quad-C[211]_\gamma\bigl[ \frac{1}{\sqrt{2}}\left(SF[22]_\alpha L[31]_\alpha+\alpha \rightarrow \beta\right)\bigr]\right)\\
&\left\{\begin{aligned}  
&SF[22]_\alpha=\frac{1}{\sqrt{2}}\bigl[S[22]_\alpha F[22]_\beta+\alpha \leftrightarrow \beta\bigl]\nonumber\\
&SF[22]_\beta=\frac{1}{\sqrt{2}}\bigl[S[22]_\alpha F[22]_\alpha-\alpha \leftrightarrow \beta\bigl]
\end{aligned}
\right.\\
\varphi_{S[31]F[31]}^{L[31]SF[4]_a}\left[\tfrac{1}{2} \tfrac{1}{2}, \tfrac{1}{2} \tfrac{1}{2}\right]
&=\frac{1}{3}\left(C[211]_\beta L[31]_\alpha-C[211]_\alpha L[31]_\beta+C[211]_\gamma L[31]_\gamma\right)\nonumber\\
&\left(S[31]_\alpha F[31]_\alpha+S[31]_\beta F[31]_\beta+S[31]_\gamma F[31]_\gamma\right)\\
\varphi_{S[22]F[22]}^{L[31]SF[4]_b}\left[\tfrac{1}{2} \tfrac{1}{2}, \tfrac{1}{2} \tfrac{1}{2}\right]
&=\frac{1}{\sqrt{6}}\left(C[211]_\beta L[31]_\alpha-C[211]_\alpha L[31]_\beta+C[211]_\gamma L[31]_\gamma\right)\nonumber\\
&\left(S[22]_\alpha F[22]_\alpha+S[22]_\beta F[22]_\beta\right)\\
\varphi_{S[31]F[31]}^{L[31]SF[211]_a}\left[\tfrac{1}{2} \tfrac{1}{2}, \tfrac{1}{2} \tfrac{1}{2}\right]
&=\frac{1}{\sqrt{3}}\left(C[211]_\beta \frac{1}{\sqrt{2}}\bigl[L[31]_\gamma SF[211]_\alpha+L[31]_\beta SF[211]_\gamma\bigr]\right. \nonumber\\
& \quad-C[211]_\alpha\frac{1}{\sqrt{2}}\bigl[L[31]_\gamma SF[211]_\beta-L[31]_\alpha SF[211]_\gamma\bigr]\nonumber\\
& \left.\quad-C[211]_\gamma\frac{1}{\sqrt{2}}\bigl[L[31]_\alpha SF[211]_\alpha+L[31]_\beta SF[211]_\beta\bigr]\right)\\
&\left\{\begin{aligned}  
&SF[211]_\alpha=\frac{1}{\sqrt{2}}\bigl[S[31]_\alpha F[31]_\gamma-S[31]_\gamma F[31]_\alpha\bigr]\nonumber\\
&SF[211]_\beta=\frac{1}{\sqrt{2}}\bigl[S[31]_\beta F[31]_\gamma-S[31]_\gamma F[31]_\beta\bigr]\nonumber\\
&SF[211]_\gamma=\frac{1}{\sqrt{2}}\bigl[S[31]_\alpha F[31]_\beta-S[31]_\beta F[31]_\alpha\bigr]
\end{aligned}
\right.\\
\varphi_{S[31]F[22]}^{L[31]SF[211]_b}\left[\tfrac{1}{2} \tfrac{1}{2}, \tfrac{1}{2} \tfrac{1}{2}\right]
&=\frac{1}{\sqrt{3}}\left(C[211]_\beta \frac{1}{\sqrt{2}}\bigl[L[31]_\gamma SF[211]_\alpha+L[31]_\beta SF[211]_\gamma\bigr]\right. \nonumber\\
& \quad-C[211]_\alpha\frac{1}{\sqrt{2}}\bigl[L[31]_\gamma SF[211]_\beta-L[31]_\alpha SF[211]_\gamma\bigr]\nonumber\\
& \left.\quad-C[211]_\gamma\frac{1}{\sqrt{2}}\bigl[L[31]_\alpha SF[211]_\alpha+L[31]_\beta SF[211]_\beta\bigr]\right)\\
&\left\{\begin{aligned}  
&SF[211]_\alpha=\frac{1}{2}\bigl[S[31]_\alpha F[22]_\beta+S[31]_\beta F[22]_\alpha+\sqrt{2} S[31]_\gamma F[22]_\alpha\bigr]\nonumber\\
&SF[211]_\beta=\frac{1}{2}\bigl[S[31]_\alpha F[22]_\alpha-S[31]_\beta F[22]_\beta+\sqrt{2} S[31]_\gamma F[22]_\beta\bigr]\nonumber\\
&SF[211]_\gamma=\frac{1}{\sqrt{2}}\bigl[S[31]_\alpha F[22]_\beta-S[31]_\beta F[22]_\alpha\bigr]
\end{aligned}
\right.\\
\varphi_{S[22]F[31]}^{L[31]SF[211]_c}\left[\tfrac{1}{2} \tfrac{1}{2}, \tfrac{1}{2} \tfrac{1}{2}\right]
&=\frac{1}{\sqrt{3}}\left(C[211]_\beta \frac{1}{\sqrt{2}}\bigl[L[31]_\gamma SF[211]_\alpha+L[31]_\beta SF[211]_\gamma\bigr]\right. \nonumber\\
& \quad-C[211]_\alpha\frac{1}{\sqrt{2}}\bigl[L[31]_\gamma SF[211]_\beta-L[31]_\alpha SF[211]_\gamma\bigr]\nonumber\\
& \left.\quad-C[211]_\gamma\frac{1}{\sqrt{2}}\bigl[L[31]_\alpha SF[211]_\alpha+L[31]_\beta SF[211]_\beta\bigr]\right)\\
&\left\{\begin{aligned}  
&SF[211]_\alpha=\frac{1}{2}\bigl[F[31]_\alpha S[22]_\beta+F[31]_\beta S[22]_\alpha+\sqrt{2} F[31]_\gamma S[22]_\alpha\bigr]\nonumber\\
&SF[211]_\beta=\frac{1}{2}\bigl[F[31]_\alpha S[22]_\alpha-F[31]_\beta S[22]_\beta+\sqrt{2} F[31]_\gamma S[22]_\beta\bigr]\nonumber\\
&SF[211]_\gamma=\frac{1}{\sqrt{2}}\bigl[F[31]_\alpha S[22]_\beta-F[31]_\beta S[22]_\alpha\bigr]
\end{aligned}
\right.
\end{align}
\end{widetext}

\section{Chiral relation between $T_\sigma$ and $T_\pi$ in the Foldy-Wouthuysen reduction}
\label{sec_FW}
The operators in \eqref{eq:Tsigma_delta} and \eqref{eq:Tpi_delta} originate from the relativistic Yukawa couplings
\begin{equation}
\mathcal L_{\rm int}
= g_\sigma\,\bar q q\,\sigma
+ i g_\pi\,\bar q \gamma_5 \tau^a q\,\pi^a .
\label{eq:Yukawa}
\end{equation}
Under an infinitesimal axial (chiral) transformation
\begin{equation}
q \rightarrow \left(1+\frac{i}{2}\epsilon^a \tau^a \gamma_5\right) q ,
\qquad
\bar q \rightarrow 
\bar q \left(1+\frac{i}{2}\epsilon^a \tau^a \gamma_5\right),
\end{equation}
the bilinears mix as
\begin{equation}
\delta(\bar q q)= i\epsilon^a \bar q \gamma_5 \tau^a q,
\qquad
\delta(\bar q i\gamma_5 \tau^a q)= - i\epsilon^a \bar q q ,
\end{equation}
so the scalar and pseudoscalar vertices are chiral partners at the relativistic level.

To connect this to \eqref{eq:Tsigma_delta} and \eqref{eq:Tpi_delta}, we perform a Foldy-Wouthuysen (FW) reduction of the Dirac Hamiltonian
\begin{equation}
H = \boldsymbol{\alpha}\!\cdot\!\mathbf p + \beta m
+ \beta g_\sigma \sigma
+ i \beta g_\pi \gamma_5 \tau^a \pi^a .
\end{equation}
Decomposing $H=\beta m + \mathcal E + \mathcal O$ into even and odd parts,
\begin{equation}
\mathcal E = \beta g_\sigma \sigma ,
\qquad
\mathcal O = \boldsymbol{\alpha}\!\cdot\!\mathbf p
+ i \beta g_\pi \gamma_5 \tau^a \pi^a ,
\end{equation}
The $\sigma$-interaction is even
($[\beta,\mathcal E]=0$), whereas the $\pi$-interaction is odd
($\{\beta,\mathcal O\}=0$).  This grading determines the order at which spin-dependent operators appear in the $1/m$ expansion.
Keeping terms linear in the meson fields and expanding to the first
nontrivial spin-dependent order, the FW Hamiltonian yields:

\paragraph{Pion vertex (odd operator).}
From the interference term in $\mathcal O^2$ one obtains, for positive-energy components,
\begin{equation}
H^{(\pi)}_{\rm FW}
= -\frac{ig_\pi}{2m}\,
\boldsymbol{\sigma}\!\cdot\!\boldsymbol{\nabla}(\tau^a\pi^a)
+ \mathcal O\!\left(\frac{1}{m^2}\right).
\end{equation}
In momentum space ($\mathbf q$ the momentum transfer),
\begin{equation}
V_\pi = -\frac{g_\pi}{2m}\,
\boldsymbol{\sigma}\!\cdot\!\mathbf q \; \tau^a .
\label{eq:Vpi_NR}
\end{equation}
Thus the pseudoscalar coupling produces the leading nonrelativistic
spin-momentum structure $\mathbf S\!\cdot\!\mathbf q\,\tau^a$.

\paragraph{Sigma vertex (even operator).}
The spin-dependent contribution arises from the double commutator term
\begin{equation}
-\frac{1}{8m^2}
\big[\boldsymbol{\alpha}\!\cdot\!\mathbf p,
[\boldsymbol{\alpha}\!\cdot\!\mathbf p,
\beta g_\sigma \sigma]\big] ,
\end{equation}
which yields
\begin{equation}
V_\sigma
= -\,\frac{i g_\sigma}{2m^2}\,
\boldsymbol{\sigma}\!\cdot\!(\mathbf p' \times \mathbf p)
+ \mathcal O\!\left(\frac{1}{m^3}\right).
\label{eq:Vsigma_NR}
\end{equation}
Hence the scalar coupling generates the non-relativistic
spin-orbit structure
$\mathbf S\!\cdot\!(\mathbf p' \times \mathbf p)$.

\paragraph{Light-front kinematics.}
Defining  $K=(\mathbf p'+\mathbf p)/2$ and
$\mathbf q=\mathbf p'-\mathbf p$, one has
$\mathbf p'\times \mathbf p = - K \times \mathbf q$.
In the light-front
$K \simeq K_z \hat{\mathbf z}$, so
\begin{equation}
\boldsymbol{\sigma}\!\cdot\!(\mathbf p' \times \mathbf p)
\;\rightarrow\;
\boldsymbol{\sigma}\!\cdot\!(\hat{\mathbf z}\times \mathbf q).
\end{equation}
Projecting the momentum transfer onto the intrinsic Jacobi coordinate,
$\mathbf q \rightarrow \mathbf p_\delta$, the transition operators become
\begin{equation}
T_\sigma \propto
\mathbf S \!\cdot\!
(\hat{\mathbf z}\times \mathbf p_\delta),
\qquad
T_\pi \propto
(\mathbf S \!\cdot\! \mathbf p_\delta)\,\tau^a ,
\end{equation}
which are~\eqref{eq:Tsigma_delta} and \eqref{eq:Tpi_delta}

Chiral symmetry relates the scalar and pseudoscalar
quark bilinears at the relativistic level.
After FW reduction, the even/odd grading of the
Dirac operators causes the chiral partners to appear at different
orders in the $1/m$ expansion.
The pseudoscalar coupling produces the leading
spin-momentum operator,
while the scalar coupling produces the spin-orbit operator.
These reduce in the light-front intrinsic basis preciselylimit to $T_\sigma$ and $T_\pi$ as given by \eqref{eq:Tsigma_delta} and \eqref{eq:Tpi_delta}

\bibliography{NP}
\end{document}